\documentclass[notitlepage,11pt,amsmath,preprintnumbers,nofootinbib,aps] 
{revtex4-1}

\usepackage{amssymb,url,bm}
\usepackage{amsfonts,multirow}
\usepackage{pifont}
\usepackage{amsmath}
\usepackage{graphicx}
\usepackage{hyperref}
\usepackage{color}
\usepackage{xcolor,colortbl}
\usepackage{array,mathtools,amssymb,booktabs}
\usepackage{float}
\usepackage{setspace}

\renewcommand{\baselinestretch}{1.}
\AtBeginDocument{
\heavyrulewidth=.16em
\lightrulewidth=.1em
\cmidrulewidth=.03em
\belowrulesep=.4ex
\belowbottomsep=0pt
\aboverulesep=.4ex
\abovetopsep=0pt
\cmidrulesep=\doublerulesep
\cmidrulekern=.5em
\defaultaddspace=.5em
}

\renewcommand{\thesubsection}{{\Alph{subsection}}}
\renewcommand{\baselinestretch}{1.1}
\long\def\del #1 \enddel { }

\renewcommand{\baselinestretch}{1.}

\makeatletter
    \def\CT@@do@color{%
      \global\let\CT@do@color\relax
            \@tempdima\wd\z@
            \advance\@tempdima\@tempdimb
            \advance\@tempdima\@tempdimc
    \advance\@tempdimb\tabcolsep
    \advance\@tempdimc\tabcolsep
    \advance\@tempdima2\tabcolsep
            \kern-\@tempdimb
            \leaders\vrule
                    \hskip\@tempdima\@plus  1fill
            \kern-\@tempdimc
            \hskip-\wd\z@ \@plus -1fill }

\definecolor{Gray}{gray}{0.85}
\definecolor{LightGray}{gray}{0.93}
\definecolor{LightGreen}{rgb}{0.88, 1, 0.88}
\definecolor{LightCyan}{rgb}{0.88,1,1}
\definecolor{LightRed}{rgb}{1, 0.85, 0.85}
\definecolor{LightYellow}{rgb}{1, 1, 0.85}
\definecolor{Yellow}{rgb}{1,1,0.05}
\definecolor{LightBlue}{rgb}{0.87, 0.94, 1}
\definecolor{white}{gray}{1}

\newcolumntype{G}{>{\columncolor{LightGray}}c}
\newcolumntype{C}{>{$}c<{$}}
\newcolumntype{?}{!{\vrule width 1pt}}
\newcolumntype{`}{!{\vrule width 1.5pt}}

\newcommand{\eqn}[1]{Eq.\,(\ref{#1})}
\newcommand{\eq}[1]{(\ref{#1})}

\newcommand{\fig}[1]{Fig.\,\ref{#1}}

\newcommand{\tab}[1]{Tab.\,\ref{#1}}
\newcommand{\tabs}[1]{Tabs.\,\ref{#1}}
\newcommand{\sct}[1]{Sect.\,\ref{#1}}

\newcounter{notecount}

\renewcommand{\Re}{{\rm Re}}

\newcommand{\beq}{\begin{equation}}
\newcommand{\eeq}{\end{equation}}
\newcommand{\bsp}{\begin{split}}
\newcommand{\esp}{\end{split}}
\newcommand{\bfi}{\begin{figure}}
\newcommand{\efi}{\end{figure}}
\def\bea{\arraycolsep .1em \begin{eqnarray}}
\def\eea{\end{eqnarray}}
\def\bwt{\begin{widetext}}
\def\ewt{\end{widetext}}

\newcommand{\grdint}[2]{\int \text{d}^{#1} #2 \, \sqrt{g} \,}

\begin{document}

\title{Fixed Points of Quantum Gravity \\ and the Dimensionality of the UV Critical Surface}
\author{Yannick Kluth}
\author{Daniel F. Litim}
\affiliation{Department of Physics and Astronomy, University of Sussex, Brighton, BN1 9QH, U.K.}
\begin{abstract}
We study quantum effects in higher curvature extensions of general relativity using the functional renormalisation group.
New flow equations are derived for general classes of models involving Ricci scalar, Ricci tensor, and Riemann tensor interactions.
Our method is applied to test the asymptotic safety conjecture 
for quantum gravity with polynomial Riemann tensor interactions
of the form $\sim\int \sqrt{g} \,(R_{\mu\nu\sigma\tau}R^{\mu\nu\sigma\tau})^n$ and 
$\sim\int \sqrt{g} \, R\cdot(R_{\mu\nu\sigma\tau}R^{\mu\nu\sigma\tau})^n$, and functions thereof. 
Interacting fixed points, universal scaling dimensions,  gaps in  eigenvalue spectra, quantum equations of motion, 
and de Sitter solutions are identified  by combining high order polynomial approximations,
 Pad\'e resummations, and full numerical integration. 
 Most notably, we discover that quantum-induced shifts of scaling dimensions can lead to a four-dimensional  ultraviolet critical surface.
Increasingly higher-dimensional interactions remain  irrelevant and show near-Gaussian scaling and signatures of weak coupling.
Moreover, a new equal weight condition is put forward to identify  stable eigenvectors  to all orders in the expansion. 
  Similarities and differences with results from the Einstein-Hilbert approximation, $f(R)$ approximations, and $f(R,{\rm Ric}^2)$ models  
 are highlighted and the relevance  of findings for quantum gravity and  the asymptotic safety conjecture  is discussed.

\end{abstract}
\maketitle
\newpage
\onecolumngrid
\begin{spacing}{.95}
\tableofcontents
\end{spacing}
\renewcommand{\baselinestretch}{.8}
\setlength{\parskip}{7pt}

\section{\bf Introduction}\label{sec:intro}
Even more than a century after the discoveries of quantum mechanics and  general relativity, an understanding  of  gravity as a fundamental quantum field theory of the metric field continues to offer challenges. Many proposals have been put forward to combine the principles of quantum physics with general relativity within and beyond quantum field-theoretical settings and include string theory, loop quantum gravity, asymptotically safe gravity, dynamical triangulations,  matrix models, group field theory,  causal sets,   and  more. 

The asymptotic safety conjecture  for gravity  \cite{Weinberg:1980gg} stipulates the existence of an interacting high-energy fixed point 
under the renormalisation group (RG) evolution of couplings \cite{Wilson:1971bg,Wilson:1971dh,Litim:2006dx,Niedermaier:2006ns,Litim:2011cp}.
Quantum effects turn gravitational couplings into  running ones with  a  short-distance fixed point  triggered by the anti-screening nature of gravity.  Renormalisation group  trajectories connect the short-distance quantum regime with  the large-distance regime of classical general relativity,  and a small set of  fundamentally free parameters   ensures predictivity. As such, asymptotic safety bears the promise to overcome the  non-renormalisability  of Einstein's theory  \cite{Goroff:1985sz,Goroff:1985th,Gomis:1995jp} and the absence  of  asymptotic freedom in renormalisable fourth order extensions   \cite{Stelle:1976gc,Anselmi:2018ibi}. 

Proofs for asymptotic safety, however, are hard to come by. 
In four space-time dimensions, and without gravity,  asymptotic safety for gauge-matter theories  has recently  been established rigorously both at weak \cite{Litim:2014uca,Bond:2016dvk,Buyukbese:2017ehm,Bond:2017suy,Bond:2017tbw,Bond:2019npq} and at strong coupling \cite{Bond:2022xvr} using RG methods, also offering new directions for model building. 
Similarly, for certain perturbatively non-renromalisable theories- such as 3D theories with elementary four- or  six-fermion interactions, asymptotic safety has also been established rigorously    \cite{Rosenstein:1988pt,deCalan:1991km,Jakovac:2014lqa,Cresswell-Hogg:2022lgg} using large-$N$ techniques.
For 4D quantum gravity, however, anomalous dimensions are expected to be large and a fixed point search requires  non-perturbative tools. 
Still,  canonical power counting can be used as a guiding principle for a  bootstrap   search \cite{Falls:2013bv}, together with  functional renormalisation \cite{Polchinski:1983gv,Wetterich:1992yh,Morris:1993qb,Reuter:1996cp,Freire:2000bq,Litim:2001up}. By now, strong circumstantial 
hints for asymptotic safety have been found in the Einstein Hilbert theory
\cite{
Souma:1999at,
Souma:2000vs,
 Reuter:2001ag,
 Lauscher:2001ya,
 Litim:2003vp,
 Bonanno:2004sy,
 Fischer:2006fz,
 Litim:2008tt,   
  Eichhorn:2009ah,
 Manrique:2009uh,
  Eichhorn:2010tb,
  Manrique:2010am,
  Manrique:2011jc,
  Litim:2012vz,
  Donkin:2012ud,
  Christiansen:2012rx,
  Codello:2013fpa,
  Christiansen:2014raa,
  Becker:2014qya,
  Falls:2014zba,
  Falls:2015qga,
  Falls:2015cta,
  Christiansen:2015rva,
  Gies:2015tca,
  Benedetti:2015zsw,
  Biemans:2016rvp,
  Pagani:2016dof,
  Falls:2017cze,
  Houthoff:2017oam,
  Knorr:2017fus}
  and higher curvature extensions 
  \cite{
Lauscher:2002sq,
  Codello:2006in,
  Codello:2007bd,
  Machado:2007ea,
  Codello:2008vh,
Benedetti:2009rx,
Benedetti:2009gn,
Benedetti:2010nr,
  Niedermaier:2011zz,
 Niedermaier:2009zz,
 Niedermaier:2010zz,
Groh:2011vn,
  Benedetti:2012dx,
  Dietz:2012ic,
  Ohta:2013uca,
 Benedetti:2013jk,
 Dietz:2013sba,
  Falls:2014tra,
Saltas:2014cta,
Demmel:2014sga,
   Eichhorn:2015bna,
  Ohta:2015efa,
  Ohta:2015fcu,
Demmel:2015oqa,
Falls:2016wsa,
  Falls:2016msz,
  Gies:2016con,
  Christiansen:2016sjn,
  Gonzalez-Martin:2017gza,
  Becker:2017tcx,
  Falls:2017lst,
 Falls:2018ylp,
  deBrito:2018jxt,
  Falls:2020qhj}.
Intriguingly, high-order studies  also  observed  that most quantum scaling dimensions become near-Gaussian despite of  residual interactions,  except for a few dominant  ones, indicating  that asymptotically safe quantum gravity becomes ``as Gaussian as it gets''   \cite{Falls:2013bv,Falls:2014tra,Falls:2017lst,Falls:2018ylp}.

In this paper, we investigate the impact of  higher order  curvature interactions for quantum gravity. 
There are several  motivations for this. Firstly, higher curvature extensions have become of interest as extensions of 
classical  general relativity, often   in view of cosmology and inflation \cite{Sotiriou:2008rp,DeFelice:2010aj,Clifton:2011jh,Capozziello:2011et}, 
and holography   \cite{Hung:2011xb}. It is then natural to also investigate  the impact of quantum fluctuations. 
Secondly, from the viewpoint of the asymptotic safety conjecture, higher order curvature interactions 
will be present rather naturally  owing to residual interactions in the UV, and should therefore be taken 
into consideration. In particular, results in the Einstein-Hilbert or fourth order approximations 
require validation beyond the lowest orders of interaction monomials.
Similarly, 
model studies which  besides 
Einstein Hilbert terms only retain individual higher curvature invariants,
often lead to very large scaling dimensions 
and 
should be revisited, e.g.~\cite{Lauscher:2002sq,Codello:2007bd,Falls:2014tra,Gies:2016con}.

\begin{figure}
	\includegraphics[width=.5\linewidth]{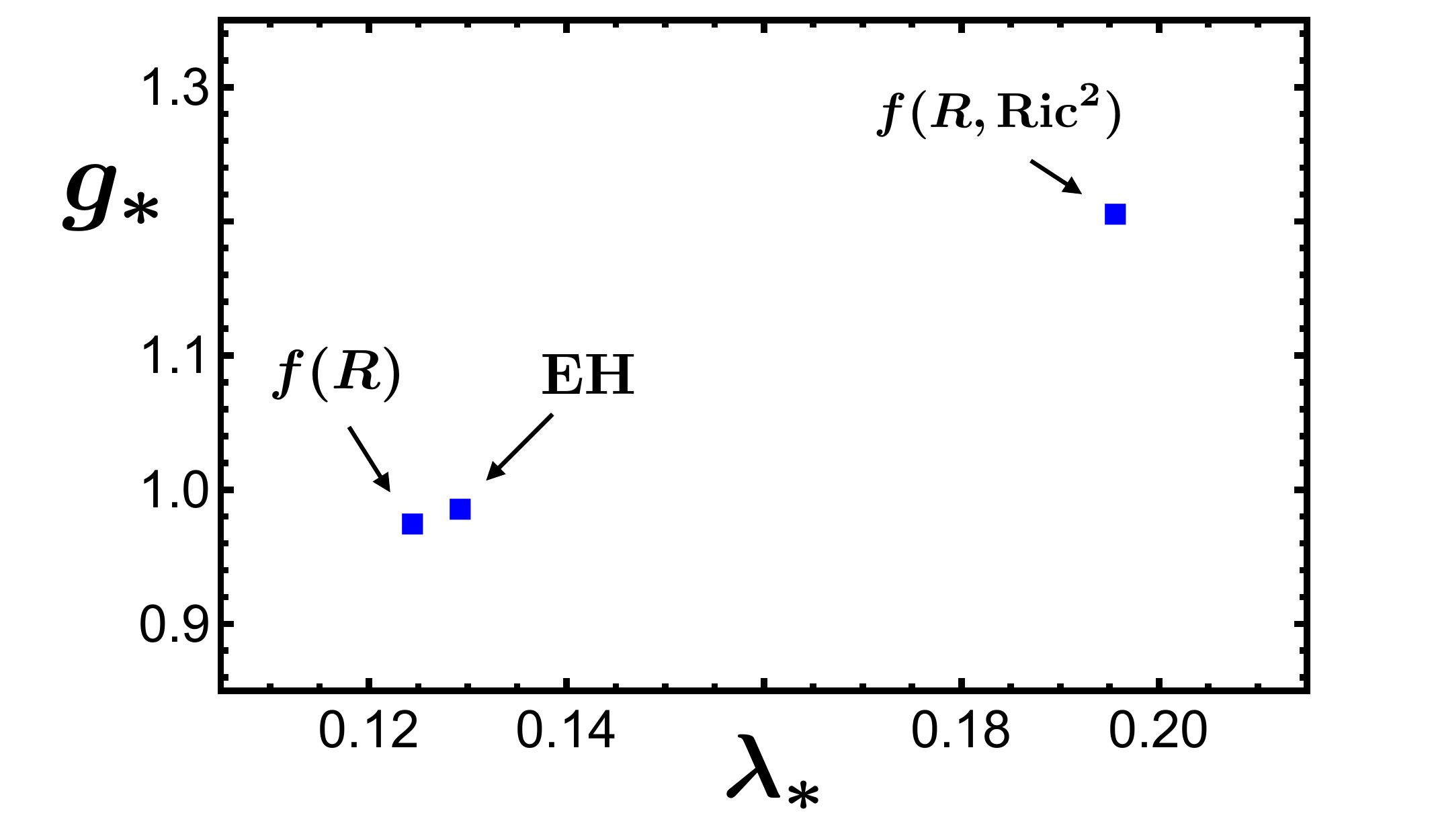}
	\caption{Impact of higher curvature invariants beyond Einstein-Hilbert.}	\label{pLaG}
\end{figure}

To analyse higher curvature invariants systematically, a bootstrap search strategy  was introduced in \cite{Falls:2013bv}. It relies on the  hypothesis  that canonical mass dimension remains a good ordering principle even at  interacting fixed points, which is known to be true for  fixed point systems observed in Nature. 
The virtue of this setup is that the validity of the hypothesis  can be verified a posteriori, order by order in the approximation: If an additional operator with higher mass dimension is added, the newly introduced quantum scaling dimension should be more irrelevant than those found at the previous approximation order, while the remaining scaling dimensions are, at best, shifted by a small amount. 
The bootstrap  strategy has first been used in  
$f(R)$ models   \cite{Falls:2014tra}  
retaining monomials $\int\sqrt{g}R^n$ up to $n=70$ powers in curvature \cite{Falls:2018ylp}. Most notably, the fixed point for  Newton's coupling  and the cosmological constant come out close to the Einstein-Hilbert results, 
supporting the view 
that the latter is a faithful image of the fixed point in the full theory.  On the other hand, the 
inclusion of high order Ricci tensor interactions \cite{Falls:2017lst} 
has lead to more substantial shifts away from Einstein Hilbert 
results. As such,  Fig.~\ref{pLaG} illustrates clearly  that the impact of higher order curvature interactions
requires further scrutiny.

A third motivation for our work is the number of fundamentally free parameters of asymptotically safe gravity, also known as the dimension of the UV critical surface.
In particular, one may wonder whether classically marginal or irrelevant interactions may become relevant quantum-mechanically.
Thus far, the observed quantum shifts in the eigenvalue spectra compared to canonical values are typically of the order of a few.
In principle,
this pattern implies that  some dim-6 invariants may very well become relevant in the quantum world. It also suggests 
 that the number of fundamentally free parameters
  cannot be larger than the number of operators with mass dimension $\leq 6$; see  \tab{tab:dim246operators} for a full list. 
  Therefore, interaction monomials up to dim-6 are of particular interest if we want to understand the dimensionality of the UV critical surface. In practice, however,
 examples for any dim-$6$ interaction term to become relevant in quantum gravity have not yet been observed, either because
 quantum corrections are  not large enough, 
 or because they point into the irrelevant direction  \cite{Falls:2014tra,Falls:2017lst,Falls:2018ylp}. 
 It is therefore important to check 
 whether this remains true for approximations involving other dim-$6$ curvature invariants. 
 
 With these aims in mind,  we derive  new functional renormalisation group equations for models of quantum gravity involving Ricci scalar, Ricci tensor, and Riemann tensor interactions. 
   The setup is rather general, and
  covers  many  higher curvature extensions of classical general relativity \cite{Sotiriou:2008rp,DeFelice:2010aj,Clifton:2011jh,Capozziello:2011et}.
  Our three-parameter families of  quantum effective actions, once Taylor-expanded to high order,
  allow for systematic  fixed point searches beyond Einstein Hilbert gravity.
 A key focus of this work are fixed points in new models of quantum gravity  with high order Riemann tensor interactions, which are studied in depth. 
   Results include   fixed points,  scaling dimensions, gaps in eigenvalue spectra, quantum equations of motion, de Sitter solutions,
  and the impact of Riemann  interactions on the location of fixed points, complementing~Fig.~\ref{pLaG}. 
  We also study  the new dim-6 curvature invariant 
  $\int\sqrt{g}R\, {\rm Riem}^2$ which can become relevant  at shortest distances. This  feature is qualitatively  new and different from the previously studied dim-6 invariants $\int\sqrt{g}R^3$, $\int\sqrt{g}R\, {\rm Ric}^2$, and the Goroff-Sagniotti term $\int \sqrt{g} R^{\rho \sigma}_{\ \ \mu \nu} R^{\mu \nu}_{\ \ \alpha \beta} R^{\alpha \beta}_{\ \ \rho \sigma}$, and the implications are elaborated in  detail. 
 
  Another novelty of our study  is a first complete order-by-order analysis of eigenperturbations at fixed points. 
  With the help of a new equal weight condition,
we find that eigenperturbations are genuinely stable, and in accord with expectations from eigenvalue spectra.
We also provide a detailed comparison of results with previous work and highlight similarities and differences
due to different types of higher curvature interactions.
  On the technical side, we expand beyond previous efforts and use
  polynomial approximations of the action up to 144 orders in curvature, alongside Pad\'e resummations and  full numerical integration, to achieve our results.

\aboverulesep = 0mm
\belowrulesep = 0mm

\begin{table*}[t]
	\centering
	\addtolength{\tabcolsep}{10pt}
	\setlength{\extrarowheight}{8pt}
\scalebox{0.8}{
	\begin{tabular}{`c?c`}
		\toprule
		\rowcolor{Yellow} & \bf Curvature Invariants \\[1ex] \midrule
\rowcolor{LightGray}			\bf dim-$\bm0$ & $1$ \\[1ex]
	\bf dim-$\bm2$ & $R$ \\[1ex]
\rowcolor{LightGray}			 \bf dim-$\bm4$ & $R^2, R_{\mu \nu} R^{\mu \nu}, R_{\rho \sigma \mu \nu} R^{\rho \sigma \mu \nu}, \Box R$ \\[1ex]
\multirow{2}{*}{\bf dim-$\bm6$} & 
$R^3\,, R R_{\mu \nu} R^{\mu \nu},\,
R R_{\rho \sigma \mu \nu} R^{\rho \sigma \mu \nu},\,
R_{\mu \nu} R^{\nu \rho} R_\rho^\mu,\,
R_{\mu \rho} R_{\nu \sigma} R^{\mu \nu \rho \sigma},\,
R^\mu_\nu R_{\nu \alpha \beta \gamma} R^{\mu \alpha \beta \gamma},$ \\
	& 
	$R^{\mu \nu}_{\,\,\,\,\,\, \rho \sigma} R^{\rho \sigma}_{\,\,\,\,\,\, \alpha \beta} R^{\alpha \beta}_{\,\,\,\,\,\, \mu \nu},\,
	R_{\mu \nu \rho \sigma} R^{\mu \,\,\, \rho}_{\,\,\, \alpha \,\,\, \beta} R^{\nu \alpha \sigma \beta},\nabla_\mu R \nabla^\mu R,\,
	\nabla_\rho R_{\mu \nu} \nabla^\rho R^{\mu \nu}$ \\[1ex]
		\bottomrule
	\end{tabular}
}
	\caption{List of  curvature invariants up to canonical mass dimension six.}
	\label{tab:dim246operators}
\end{table*}

The remainder of this work is organised as follows. 
In \sct{setup} we derive the main equations using functional renormalisation with a particular emphasis on Riemann tensor interactions, and lay out our fixed point search strategy.
In \sct{sec:mainfp} we discuss the main Riemann fixed point, its fixed point effective action, de~Sitter solutions, and residuals.
In \sct{universality}, we  find the universal scaling exponents and the UV critical surface. We also perform a bootstrap test and observe signatures of weak coupling.
In \sct{sec:2ndphyfp} we detail our results for a pair of weakly coupled Riemann fixed points.
In \sct{misc}, we identify all eigenvectors and eigenperturbations for all fixed points by removing a rescaling ambiguity through a new equal weight condition. 
In \sct{sec:discus} we discuss the main features of our results and compare with earlier studies in different approximations.
Finally, \sct{sec:conclusions} gives a brief summary of our results and conclusions. 
Three appendices summarise technicalities such as   first and second variations (\sct{AppA}), Hessians (\sct{AppB}), and  the gravitational renormalisation group equations (\sct{sec:floweq}).

\section{\bf Renormalisation Group for Gravity}
\label{setup}
In this section, we derive new functional renormalisation group equations for quantum gravity involving Riemann tensor interactions using heat kernel methods on spherical backgrounds. Our work completes a line of research initiated in \cite{Falls:2014tra,Falls:2016wsa,Falls:2017lst}.

\subsection{Action}
\label{sec:RGgravity}
We are interested in gravitational actions of the form
\beq
	\begin{split}
		\overline{\Gamma}_k [ g_{\mu \nu} ] =& \int \text{d}^4 x \, \sqrt{g} \left[ F_k (X) + R \cdot Z_k (X) \right] \,, 
	\end{split}
	\label{eqn:effansatz}
\eeq
with $X$ being an arbitrary dim-4 curvature invariant. $F$ and $Z$ are {\it a priori} unspecified functions. The index $k$ indicates the Wilsonian RG scale at which the quantum effective action is evaluated. Ultimately, we are interested in the quantum effects induced by fluctuations of the metric fields, and the RG flow $k\partial_k$ of the quantum effective action and the gravitational couplings contained in it, and its fixed points in the high energy limit. The most general form of $X$ can be written as
\beq \label{X}
X=a\, R^2 +b\, {\rm Ric}^2+c\, {\rm Riem}^2\,.
\eeq
Since the overall normalisation of $X$ is irrelevant in that it can always be absorbed into a redefinition of coupling constants, only two out of the three parameters $\{a,b,c\}$ are truly independent.\footnote {A simple choice of independent parameters are the surface angles $\{\theta,\phi\}$ of the three-dimensional unit sphere spanned by the parameters $\{a,b,c\}$.} This slight redundancy is of no relevance for the present work and we will therefore stick to the parameters $\{a,b,c\}$. 
The family of theories described by \eqref{eqn:effansatz} with \eqref{X} contains the Einstein-Hilbert theory with scale-dependent cosmological constant $\Lambda_k$ and Newton's coupling $G_k$ as the leading order coefficients 
\beq\label{LaG}
F_k(0)=\frac{\Lambda_k}{{8\pi\,G_k}}\,,\quad Z_k(0)=-\frac{1}{16 \pi G_k}
\eeq
 of a Taylor expansion in curvature. In the infrared limit $(k\to 0)$ the running couplings will have to agree with data, $i.e.$~with Newton's constant $G_{0} \approx 6.67 \times 10^{-11} {\rm m}^3/({\rm kg\, s}^2)$ and with the cosmological constant $\Lambda_{0} \approx 1.1\times 10^{-52}\,{\text{m}}^{-2}$.
Besides being toy models for asymptotically safe gravity, actions of this type also arise as higher curvature extensions of classical general relativity, e.g.~\cite{Sotiriou:2008rp,DeFelice:2010aj,Clifton:2011jh,Capozziello:2011et}. 

 A key feature of the models \eq{eqn:effansatz} with \eq{X} is that a polynomial expansion of the action retains only a single  curvature invariant with canonical mass dimension $(-d+2n)$ for any $n\ge 0$, whose RG flows are found unambiguously on  spherical backgrounds \cite{Falls:2017lst}. To make this statement more explicit, we construct a complete operator basis at any mass dimension containing the terms $X^n$ and $R \, X^n$ together with operators vanishing on spherical backgrounds. One way to construct such a basis consists in using the scalar Ricci curvature together with the traceless Ricci tensor and the Weyl tensor. Assuming $X$ to be non-vanishing on the sphere, all operators of this basis except for $X^n$ and $R \, X^n$ can be chosen to be proportional to at least one traceless Ricci or the Weyl tensor. Thus, all operators in this basis except for $X^n$ and $R \, X^n$ vanish on a spherical geometry. In this way, spherical backgrounds can be used to project onto actions of the form \eq{eqn:effansatz} with \eq{X} provided $X$ does not vanish on spheres (see also \cite{Kluth:2022vnq}). Effectively, in this setting the terms 
 $X^{n}$ and $R\, X^{n}$ act as representatives for higher order curvature invariants of dimension $4n$ and $4n+2$, respectively. Hence, the corresponding actions on spheres takes the form
\beq
	\begin{split}
		\overline{\Gamma}_k \Large|_{\rm sphere}=& \int \text{d}^4 x \, \sqrt{g} \cdot \bar f_{a,b,c}(R) \,, 
	\end{split}
	\label{rays}
\eeq
similar to the local potential approximation for $f(R)$-type gravity advocated in \cite{Benedetti:2012dx}. In this light, our models are generalisations of the  local potential approximation on maximally symmetric backgrounds, the main addition being   that the underlying dynamics not only involves the Ricci scalar, but also  the fluctuations due to Ricci and Riemann tensor interactions. We will sometimes refer to the functions $\bar f_{a,b,c}(R)$ as ``rays'' in the space of curvature invariants. In these conventions,  the ray  $\bar f_{1,0,0}(R)$  correspond to the conventional, $f(R)$-type, local potential approximation \cite{Benedetti:2012dx} (see also \cite{Falls:2013bv, Falls:2014tra}). The ray $\bar f_{0,1,0}(R)$  which involve   Ricci tensor interactions  have previously been investigated in \cite{Falls:2017lst}. The ray $\bar f_{0,0,1}(R)$  which are sensitive to   Riemann tensor interactions are the central subject of this study.

\subsection{Functional Renormalisation}
Next, we derive the renormalisation group flow for general gravitational actions \eq{eqn:effansatz} with  \eq{X} using the method of functional renormalisation  \cite{Wetterich:1992yh,Reuter:1996cp} in the context of gravity \cite{Litim:2008tt,Litim:2011cp} (see \cite{Percacci:2017fkn,Reuter:2019byg} for recent textbooks),
\begin{equation}
	\partial_t \Gamma_k = \frac{1}{2} \text{Tr} \left\{ \Big( \partial_t \mathcal{R}_k \Big) \left( \Gamma_k^{(2)} + \mathcal{R}_k \right)^{-1} \right\} \, ,
	\label{eqn:wetterich}
\end{equation}
where $\Gamma_k$ is the gauge fixed effective average action, $\Gamma_k^{(2)}$ its second variation matrix and $\mathcal{R}_k$ the regulator. 
A path integral representation for these classes of theories has been given in \cite{Falls:2017lst}. Note that, within the background field method, this functional differential equation gives rise to a functional $\Gamma_k [\overline{g}_{\mu \nu}, \phi_i]$, depending on a background metric $\overline{g}_{\mu \nu}$ and some quantum fluctuation fields $\phi_i$. We can rewrite this as
\beq
	\Gamma_k [\overline{g}_{\mu \nu}, \phi_i] = \overline{\Gamma}_k [g_{\mu \nu}] + \hat{\Gamma}_k [\overline{g}_{\mu \nu}, \phi_i] \, ,
\eeq
with
\beq
	\hat{\Gamma}_k [\overline{g}_{\mu \nu}, 0] = 0 \,.
\eeq
We then simplify this ansatz using the so called ``single field'' or ``background field'' approximation. This means, we impose that $\hat{\Gamma}_k$ includes only terms of the bare action arising from the functional measure. Explicitly, this includes gauge fixing terms $\Gamma_{\text{gf}, k}$ and ghost terms $\Gamma_{\text{gh}, k}$ whose RG-running is neglected. Thus, they only enter on the right-hand side of \eq{eqn:wetterich} through the Hessian $\Gamma_k^{(2)}$. Since the left-hand side then depends only on $\overline{\Gamma}_k[g_{\mu \nu}]$ it is sufficient to evaluate the right-hand side at vanishing fluctuation fields $\phi_i = 0$, including $h_{\mu \nu} = 0$.
In addition, we choose the gauge fixing to be
\beq
	\Gamma_{\text{gf}, k} = \frac{1}{2 \alpha} \grdint{d}{x} \mathcal{F}_\mu \mathcal{F}^\mu \, ,
\eeq
with
\beq
	\mathcal{F}_\mu = \sqrt{2} \kappa \left( \nabla^\nu h_{\mu \nu} - \frac{1}{d} \nabla_\mu h \right) \, .
\eeq
Here, $\kappa = (32 \pi G_N)^{-1/2}$ and the gauge parameter is chosen to be $\alpha \rightarrow 0$.
Further, we split the background metric field from the quantum fields using a linear split,
\beq
	g_{\mu \nu} = \overline{g}_{\mu \nu} + h_{\mu \nu} \, ,
	\label{eqn:linsplit}
\eeq
where $h_{\mu \nu}$ denotes the metric quantum fluctuation.
Alternatively, one may  use an exponential split as advocated in \cite{Kawai:1992np,Kawai:1993mb}  (see also \cite{Nink:2014yya,Falls:2015qga,Falls:2015cta}) where the background field is split off the quantum field multiplicatively according to 
\beq
g_{\mu\nu} =   \bar{g}_{\mu\lambda} 
(\exp {{h}})^{\lambda}{}_\nu\equiv  \bar{g}_{\mu\nu}+ h_{\mu \nu} + \frac12 h_{\mu\tau} h^{\tau}_{\ \nu} +{\cal O}(h^3)\,.
\eeq
In this case quadratic terms in the fluctuation field ${h}_{\mu\nu}$ are retained and contribute both to the Hessians in \eq{eqn:wetterich}
 and  the flow of couplings. 
In this work, and to make contact with previous studies, we adopt to the linear split \eq{eqn:linsplit} throughout.
Moreover, the York decomposition is used to decompose the metric fluctuations according to
\beq
	\begin{split}
		h_{\mu \nu} =& \, h^T_{\mu \nu} + \nabla_\mu \xi_\nu + \nabla_\nu \xi_\mu
		+ \left( \nabla_\mu \nabla_\nu - \frac{g_{\mu \nu}}{d} \nabla^2 \right) \sigma + \frac{g_{\mu \nu}}{d} h \, .
	\end{split}
\eeq
This ensures that no non-minimal differential operator will be present in the second variation matrix $\Gamma_k^{(2)}$ when evaluated on the sphere. Ghost contributions coming from this field redefinition as well as the gauge fixing term are denoted as $\Gamma_{\text{gh}, k}$ such that
\beq
	\Gamma_k = \overline{\Gamma}_k + \Gamma_{\text{gf}, k} + \Gamma_{\text{gh}, k} \, ,
	\label{eqn:fullgamma}
\eeq
within our approximation.
With these definitions we find for the flow equation
\beq\label{fRG}
	\begin{split}
		\partial_t \Gamma_k =& \, \frac{1}{2} \text{Tr}_{2} \frac{\partial_t \mathcal{R}_k^{h^T h^T} \left( - \nabla^2 \right)}{\Gamma^{(2)}_{h^T h^T} \left( - \nabla^2 \right) + \mathcal{R}_k^{h^T h^T} \left( - \nabla^2 \right)} + \frac{1}{2} \text{Tr}_{0} \frac{\partial_t \mathcal{R}_k^{h h} \left( - \nabla^2 \right)}{\Gamma^{(2)}_{h h} \left( - \nabla^2 \right) + \mathcal{R}_k^{h h} \left( - \nabla^2 \right)} \\
		& - \frac{1}{2} \text{Tr}_{0}'' \frac{\partial_t R_k ( - \nabla^2  - \frac{R}{d - 1})}{- \nabla^2 - \frac{R}{d - 1} + R_k ( - \nabla^2 - \frac{R}{d - 1})} - \frac{1}{2} \text{Tr}_{1}' \frac{\partial_t R_k \left( -\nabla^2 - \frac{R}{d} \right)}{-\nabla^2 - \frac{R}{d} + R_k \left( - \nabla^2 - \frac{R}{d} \right)} \,.
	\end{split}
\eeq
Here, traces w.r.t. spin $2$ fields (transverse traceless tensors) are denoted by $\text{Tr}_{2}$, traces of spin $1$ fields (transverse vectors) by $\text{Tr}_{1}$, and traces of scalar modes by $\text{Tr}_{0}$. Primes are used to denote the exclusion of lowest modes in the calculation of these traces. 
At one loop accuracy, the flow equation can be integrated in closed form for any regulator to provide a regularised form of the full one loop effective action.

For the present purposes, we are interested in quantum effects beyond one loop. To that end, we choose the regulator $\mathcal{R}_k$ for each Hessian according to the replacement rule
\beq
	\Gamma_k \left( - \nabla^2 \right) + \mathcal{R}_k \left( - \nabla^2 \right) \rightarrow \Gamma_k \left( - \nabla^2 + R_k \left( - \nabla^2 \right) \right) \, ,
\eeq
and $R_k \left( z \right)$ denotes the momentum cutoff which we take to be the optimised cut-off \cite{Litim:2000ci,
Litim:2001up,Litim:2003vp},
\beq\label{opt}
	R_k \left( z \right) = \left( k^2 - z \right) \theta \left( k^2 - z \right) \, .
\eeq
The optimised cutoff function is key to finding explicit analytical expressions of flows  \cite{Litim:2003vp}. Cutoff functions $R_k(z)$ where $z=- \nabla^2$ are termed type I, and those where $z=- \nabla^2+a_\phi R$ are termed type II  \cite{Codello:2008vh}. 
For the flow equation \eq{fRG} we use a type I cut-off for the fluctuations arising from the physical transverse and scalar modes of the graviton. However, using a type I cutoff for the terms arising from the gauge fixing and ghost contributions leads to spurious, convergence-limiting unphysical poles in the flow equation, for any real Ricci curvature \cite{Falls:2018ylp}. Therefore, following \cite{Falls:2018ylp}, these have been removed  by using  suitable type II cut-offs for the terms arising from the gauge fixing and ghost contributions, as can be seen explicitly from the two last traces in \eq{fRG}.

The first and second variations of $X$ as defined in \eq{X} are given in App.~\ref{AppA}, see \eq{eqn:xvar1} and \eq{eqn:xvar2}. The variations of the Ricci scalar curvature can be found in \cite{Falls:2017lst,Percacci:2017fkn}, for example, and the Hessians $\overline{\Gamma}_k^{(2)}$ for $h^T h^T$ and $hh$ are  given in   App,~\ref{AppB}, see \eq{eqn:hessianhtht} and \eq{eqn:hessianhh}, respectively.

To obtain the final form of the flow equation, we define dimensionless functions and parameters through
\beq\label{fz}
	\begin{split}
		f (x) =& \frac{1}{16 \pi} k^{-d} F_k (X) \, , \quad\quad
		z (x) = \frac{1}{16 \pi} k^{-d + 2} Z_k (X) \, ,
	\end{split}
\eeq
with dimensionless variables and fields given by
	$x = k^{-4} X$ 
	and $r = k^{-2} R$, 
and calculate the traces using usual heat kernel techniques \cite{Avramidi:2000bm,Falls:2017lst,Kluth:2019vkg}. 
Owing to the optimised cutoff only a finite set of heat kernel coefficients is required.
More recently, all heat kernel coefficients on spheres in any dimension have been derived \cite{Kluth:2019vkg}, which enables the use of more general cutoff functions. The resulting flow equation for general $(a, b, c)$, which is the main new result of this section,  takes the form
\bea	 \partial_t f + r \,\partial_t z =-4 f - 2 rz 
+ (4a + b + \frac23 c) \,r^2 \,(f' + rz') \label{floweqnabc} 
+ \frac{1}{24\pi}\, I [f, z;a,b,c] (r) \,.	
\label{eqn:floweqngen2}
\eea
Terms on the RHS arise due to the canonical scaling of fields and variables, and due to quantum fluctuations. The explicit form of the fluctuation integrals $ I [f, z;a,b,c] (r)$ is given  in App.~\ref{sec:floweq}. 
It is worth pointing out that due to the freedom of an overall rescaling of all couplings,  models with parameters $(a,b,c)$ are physically equivalent to models with parameters $(a',b',c')$, provided that $(a,b,c)=\lambda\cdot  (a',b',c')$, and $\lambda\neq0$  a real number. We indicate this equivalence with the symbol ``$\simeq$''.
 Flow equations in the special cases $(a,b,c)\simeq (1,0,0)$ representing $f(R)$ gravity, or $(a,b,c)\simeq(0,1,0)$ representing certain types of higher order Ricci tensor interactions, have been given previously  in  \cite{Falls:2014tra,Falls:2017lst,Falls:2018ylp}. 
For a path integral representation of the theory for all scales, see \cite{Falls:2017lst}.

We remark that \eqref{eqn:floweqngen2} is a coupled set of partial differential equations for  two unknown functions $f$ and $z$.  The flow is partly disentangled by projecting onto the even and odd parts with respect to background curvature, leading to
\beq\label{FlowEvenOdd}
\begin{array}{rcl}
 \partial_t f  &=&
 \displaystyle
 -4 f  
+ (4a + b + \frac23 c) \,r^2 \, f'  
+ \frac{I [f, z] (r)+I [f, z] (-r)}{48\pi}\,.	
\\[3ex]
  \partial_t z &=&
\displaystyle
 - 2 z 
+ (4a + b + \frac23 c) \,r^2 \, z' 
+ \frac{I [f, z] (r) - I [f, z] (-r)}{48\,\pi\,r}\,.	
\end{array}\eeq
The representation \eq{FlowEvenOdd} is very convenient for numerical integrations of the flow.
The  flow equations \eq{eqn:floweqngen2} and \eq{FlowEvenOdd} are a central new result of this study. They are put to work for Riemann tensor interactions in the following sections.

\subsection{Emergence of General Relativity}\label{GR}
Prior to searching for quantum fixed points, it is interesting to show how classical general relativity emerges in the classical limit, following  \cite{Falls:2014tra, Falls:2017lst}.  
Most importantly, while the renormalisation flows of the functions $f$ and $z$ mix on the quantum level, \eq{FlowEvenOdd}, the mixing is absent  in the classical limit where the fluctuation integrals can be neglected $I\to 0$,
\beq\label{classical}
\begin{array}{rcl}
 \partial_t f  &=&
 \displaystyle
 -4 f  
+ (4a + b + \frac23 c) \,r^2 \, f'  
\\[2ex]
  \partial_t z &=&
\displaystyle
 - 2 z 
+ (4a + b + \frac23 c) \,r^2 \, z' \,.
\end{array}
\eeq
In this limit, the flow \eq{classical} is further simplified by introducing the new function
\begin{equation}
\label{fbar}
\bar f(r)= f({x})+r\cdot z(x)\,,
\end{equation}
where $x\equiv X/k^4=(a+\frac{b}{4}+\frac{c}{6})r^2$ on spheres, leading to
\begin{equation}
\left(\partial_t+ 4 -2 r\,\partial_r\right)\,\bar f=0 \,.
\label{classical2}
\end{equation}
It states that all dimensionful couplings in the classical theory are independent of the energy scale. 
The flow \eq{classical2} 
has the general solution
\beq\label{FPclassical}
\bar f(r,t)=r^2 \cdot H\left( r\cdot e^{2t}\right)
\eeq
where the function $H(z)$  is determined by the initial conditions for couplings at the reference scale $t=0$. 
Classical fixed points are the  $t$-independent solutions of \eq{FPclassical}. A trivially $t$-independent solution is achieved via the boundary condition $H(z)=$~const. It leads to a line of  fixed points for classical theories of gravity with actions $\sim\lambda_2\int \sqrt{g}X$ where $X=a\,R^2+b\,{\rm Ric}^2+c\,{\rm Riem}^2$, parametrized by the free parameter  $\lambda_2$ which  in four space-time dimensions is a marginal coupling. 
The linearity of \eq{classical2} also  implies the existence of a Gaussian fixed point $\bar{f}_*\equiv 0$\,. Moreover, from the flow for the inverse,
\begin{equation}
\left(\partial_t- 4 -2 r\,\partial_r\right)\,(\bar f^{-1})=0 \,,
\label{classical3}
\end{equation}
we observe the existence of an infinite Gaussian fixed point  \cite{Nicoll:1974zza} 
\beq
\label{infG}
1/\bar f_*\equiv 0\,.
\eeq
The Gaussian and infinite Gaussian fixed points arise from \eq{FPclassical} in asymptotic UV and IR  limits where $t\to\pm\infty$, respectively \cite{Falls:2014tra, Falls:2017lst}. 

We continue with the scaling analysis for monomials of degree $n$ in curvature. For these, the result \eq{FPclassical}  states that the corresponding
coupling $\lambda_n$  scales canonically with Gaussian exponents $\vartheta_{{\rm G}}$,
\beq\label{classicalscaling}
\begin{array}{rl}
\lambda_n(t)&=\lambda_n(0)\exp(\vartheta_{{\rm G},n} t)\\[1ex]
\vartheta_{{\rm G},n}&=2n-4\,.
\end{array}
\eeq
Hence, the dimensionless vacuum energy term $(n=0)$ and the dimensionless Ricci scalar coupling $(n=1)$ are relevant operators, and their dimensionless couplings diverge towards the IR, leading to the infinite Gaussian fixed point \eq{infG} in the IR. Using  \eq{LaG} and $G_k=g/k^2$ and $\Lambda_k=\lambda\, k^2$, we can relate the IR diverging couplings $\lambda_0$ and $\lambda_1$ to the dimensionless Newton coupling $g$  and the dimensionless cosmological constant $\lambda$ as
\beq\label{gla}
g=-\frac1{\lambda_1}\,,\quad \lambda=- \frac{\lambda_0}{2\lambda_1}\,.
\eeq
In terms of these, the infinite Gaussian fixed point implies
\beq\label{GaussLambda}
1/\lambda\to 0\,,\quad g\to 0\,,
\eeq
in such a way that $\Lambda_k$ and $G_k$~are constants in the IR limit $k\to 0$. We conclude that general relativity with positive or negative vacuum energy  corresponds to the infinite Gaussian  IR fixed point \eq{GaussLambda} provided that $\lambda$ is positive  or negative, respectively. Moreover, the infinite Gaussian fixed point is IR attractive in both couplings. The theory also displays an IR fixed point corresponding to a vanishing vacuum energy,
\beq\label{Gauss0}
\lambda\to 0\,,\quad g\to 0\,.
\eeq
This fixed point is IR attractive in $g$ and IR repulsive in $\lambda$. Classically, it can only be achieved by fine-tuning the vacuum energy to zero through the boundary condition.
This analysis can straightforwardly be extended to higher order monomials. 
According to \eq{classicalscaling}, for all couplings with $n>2$ $(n<2)$ the Gaussian fixed point $\lambda_n \to 0$ is IR attractive (repulsive) and therefore approached in the IR limit (UV limit), whereas the infinite Gaussian fixed point $1/\lambda_n \to 0$ is IR repulsive (attractive) and therefore approached in the UV limit (IR limit). 

Finally we emphasize that the emergence of classical general relativity as the infinite Gaussian fixed point  \eq{infG} with \eq{GaussLambda} or \eq{Gauss0} is  consistent with the non-perturbative RG flow \eq{FlowEvenOdd}. The reason for this is that the fluctuation-induced contributions  $I[f,z]$ in \eq{FlowEvenOdd} are parametrically suppressed  over the classical contributions owing to $I[f,z]/\bar f \to 0$ in the limit $1/\bar f\to 0$, also recalling \eq{fbar}.   Therefore, the limit  \eq{infG}  collapses the full RG flow onto the classical flow, which has  general relativity with positive, negative, or vanishing cosmological constant amongst its low energy solutions, possibly amended by higher curvature interactions. The result is  independent of the parameters $(a,b,c)$. Previously, the emergence of classical general relativity as the infinite Gaussian fixed point  of the functional RG flow has been shown for $f(R)$ and $f(R,{\rm Ric}^2)$ models  \cite{Falls:2014tra, Falls:2017lst}.

\subsection{Riemann Tensor Interactions}
\label{sec:riemtenint}
For the rest of this work, we fix the free parameters $(a,b,c)$ of the flow equation \eq{floweqnabc}  to be
\beq
	(a, b, c) \simeq (0, 0, 1) \,.
	\label{Riemann}
\eeq
We recall that  \eq{Riemann} represents the   class of equivalent models with $(a, b, c)=(0,0,\lambda)$ where $\lambda\neq0$ is any real number. In terms of \eq{X} this corresponds to the choice
\beq\label{XRiem}
	X = \text{Riem}^2 \,.
\eeq
For this setting, the first few leading terms of the gravitational effective action \eq{eqn:effansatz} in a polynomial expansion read 
\bea\label{model}
\overline{\Gamma}_k [ g_{\mu \nu} ] &=& \int d^4x \sqrt{g} \Big[ \bar{\lambda}_0 + \bar{\lambda}_1 R  + \bar{\lambda}_{2} R_{\rho \sigma \mu \nu} R^{\rho \sigma \mu \nu} 
+  \bar{\lambda}_{3} R \cdot R_{\rho \sigma \mu \nu} R^{\rho \sigma \mu \nu} + \bar{\lambda}_4  (R_{\rho \sigma \mu \nu} R^{\rho \sigma \mu \nu})^2+ ...    \Big]\ \ \  \ \ 
\eea
up to higher powers in  curvature ${\cal O}(R\cdot  {\rm Riem}^4, {\rm Riem}^6)$. For the fixed point search below, we retain interaction terms up to including  $\sim \bar\lambda_{142}\,(R_{\rho \sigma \mu \nu} R^{\rho \sigma \mu \nu} )^{71}$ and $\sim \bar\lambda_{143} \,R\cdot (R_{\rho \sigma \mu \nu} R^{\rho \sigma \mu \nu} )^{71}$ monomials, corresponding to a total of $N_{\rm max}=144$ interaction terms in the effective action \eq{model}. 

Most notably, this action involves new Riemann curvature invariants which have not been studied previously within the non-perturbative RG for gravity. 
The main purpose of this study is to establish whether the  dynamics due to Riemann tensor interactions  is going to affect the UV behaviour of the theory in any significant manner.  Using a polynomial ansatz for \eq{fz} of the form
\begin{equation}\label{series}
	f (x) = \sum_{\ell = 0}^{\left\lfloor \frac{N - 1}{2} \right\rfloor} \lambda_{2 \ell} x^{2 \ell} \, , \qquad z (x) = \sum_{\ell = 0}^{\left\lfloor \frac{N - 2}{2} \right\rfloor} \lambda_{2 \ell + 1} x^{2 \ell} \, ,
\end{equation}
and evaluating everything on a spherical background, a series expansion in small dimensionless Ricci scalar curvatures is performed and equations containing the beta functions and the couplings of the theory are found. The parameter $N$ determines the number of different operators included in the effective average action \eq{model}.

At the fixed point, the beta functions vanish and the remainder of the flow equation \eq{floweqnabc} using \eq{Riemann} is given by
\begin{equation}
	4 f - \frac{2}{3} r^2 f' + r \left( 2 z - \frac{2}{3} r^2 z' \right) = \frac{1}{24\pi} \,I_0 [f, z] (r) \, .
	\label{eqn:floweqnriemfp}
\end{equation}
Performing a series expansion in $r$ we can bring the flow equation into the form
\begin{equation}
	0 = \sum_{\ell = 0}^\infty r^\ell P_\ell (\lambda_0, ..., \lambda_{\ell + 2}) \, .
	\label{eqn:flowexpansion}
\end{equation}
To fulfil this at any approximation order $N$, all coefficients $P_\ell$ with $\ell < N$ must vanish. Hence, we are left with a system of algebraic equations which can be solved to find fixed points. In $f(R)$ theories of gravity, the polynomial conditions can be solved recursively starting with a linear equation for $\lambda_2$ from $P_0$, and leading to unique expressions for all couplings in terms of two parameters ($\lambda_0,\lambda_1)$ which then need to be determined by other means \cite{Falls:2014tra}. In our models, and ultimately due to the  presence of fourth order propagating degrees of freedoms, it turns out that $P_0$ is quadratic in the highest coupling \cite{Falls:2017lst}. In consequence, a recursive algebraic solution is more cumbersome  wherefore we resort to a numerical method instead. 

\subsection{Search Algorithm and Fixed Point Scan}
\label{sec:boundarycond}

After having derived the expressions $P_\ell \left( \lambda_0, \dots, \lambda_{\ell + 2} \right)$, we need to solve them to find fixed points of the system. At any approximation order $N$, we are given $N$ equations 
\beq\label{Pell}
	P_\ell (\lambda_0, \dots, \lambda_{\ell + 2}) = 0 \qquad \text{with} \quad \ell < N \, ,
\eeq
depending on $N + 2$ couplings $\lambda_0, \dots, \lambda_{N + 1}$. Thus, there are more open couplings than equations to solve and we need to impose boundary conditions on two of them. The effect of these boundary conditions on the fixed point solutions has been studied in \cite{Falls:2014tra}. In our fixed point search algorithm, we follow the choice of standard boundary conditions given by
\begin{equation}\label{free}
	\lambda_{N} = \lambda_{N + 1} = 0 \, ,
\end{equation}
at each approximation order $N$. Sometimes we refer to the choice \eq{free} as  "free" boundary conditions for the fixed point search. 

A recursive solution strategy to find the exact expressions for the fixed point couplings in gravity has been put forward in \cite{Falls:2014tra}. Solving \eq{Pell} provides us with expressions for the couplings $\lambda_{\ell+2}$ in terms of the lower order couplings $\lambda_{n}$ with $n<\ell+2$. In $f(R)$ theories, these lead to explicit   expressions for couplings which reduce to algebraic expressions for all couplings $\lambda_{n}(\lambda_0,\lambda_1)$  with $n\ge 2$ upon iteration  (level 2) \cite{Falls:2013bv,Falls:2014tra,Falls:2018ylp}. We emphasize that the expressions are algebraic throughout, the reason for this being that the first nontrivial  equation is  linear in the unknown quantity ($i.e.$~the coupling $\lambda_2$). 
For theories with general Ricci tensor or Riemann tensor interactions  such as here this is no longer the case and the leading equation becomes quadratic. One may still solve the recursive relations for each branch separately, though the expressions become impractical very fast. Alternatively, 
one may  start with the next coupling in line, $\lambda_3$, which again obeys a linear equation. Upon iteration, we  find closed algebraic expressions $\lambda_{n}(\lambda_0,\lambda_1,\lambda_2)$  starting with $n\ge 3$ (level 3). The quadratic condition giving $\lambda_2(\lambda_0,\lambda_1)$ is then exploited at the very end.

In this work, we perform a brute force numerical fixed point search.
This can be  initiated by using the fixed point in the Einstein-Hilbert approximation, the so-called Reuter fixed point,  as a starting point  \cite{Reuter:1996cp,
Souma:1999at,Lauscher:2001ya,Litim:2003vp}. Then, we enhance the operator base and include the one with the smallest larger canonical mass dimension into the system, search for solutions in the vicinity of the previous fixed point coordinates, and repeat in the form of a bootstrap $N\to N+1$  \cite{Falls:2013bv} until some maximum order which we take to be $N_\text{max} = 144$. This strategy has worked well for $f(R)$ gravity \cite{Falls:2013bv,Falls:2014tra,Falls:2018ylp} and $f(R,{\rm Ric}^2)$ gravity \cite{Falls:2017lst}, where the fixed point coupling $g_*$ remains close to its Einstein Hilbert value. In this sense, these $f(R)$  and $f(R,{\rm Ric}^2)$ fixed points 
may be viewed as the higher order extension of the Reuter fixed point. 

On the other hand, the system may very well have global fixed point solutions where Newton's coupling and the cosmological constant differ sizeably from their  Einstein-Hilbert values. We explore the latter possibility by modifying the search algorithm and adding random perturbations to the fixed point couplings at some order. By this token, we explore the possibility whether additional stable fixed points are available. If so, these new fixed point candidates are then further extended to higher orders, as before, using the bootstrap strategy. We also check that the thereby-found solutions, if any, do not depend on the order at which  random perturbations in the initial values were added initially. Overall, these strategies allow us to scan a substantial portion of parameter space. 
As a word of caution, however, we stress that our  search is not exhaustive. Due to the large parameter space there may still exist fixed points which have gone unnoticed in the numerical search, or fixed point solutions which defy the ordering principle set by canonical mass dimension \cite{Falls:2013bv}.

 \begin{figure}
	\includegraphics[width=.5\linewidth]{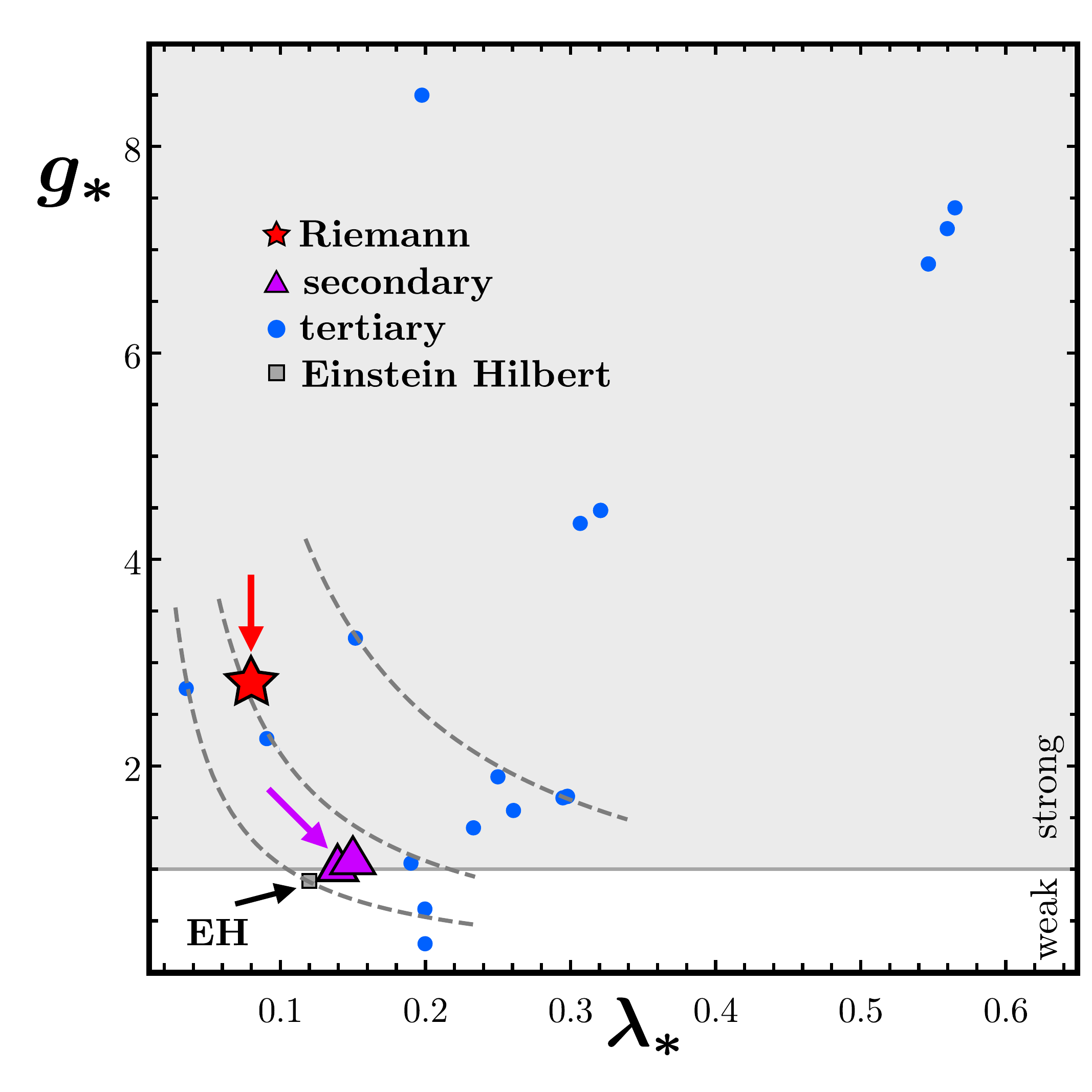}
	\caption{Shown are the values for the cosmological constant  and Newton's coupling ($\lambda,g)_*$ of 23 Riemann fixed point candidates whose coordinates are determined up to order $N=36$ in the expansion. 
	The colour coding differentiates  primary (red), secondary (magenta) or tertiary (blue) Riemann fixed points, and, for comparison, the Einstein Hilbert fixed point (gray box). Gray dashed lines correspond to constant $\lambda_*\,g_*$, and the gray shaded area indicates regions of strong coupling $g_*>1$. Further viability tests select the primary and secondary fixed point candidates (red and magenta arrow) and eliminate all tertiary candidates (see main text).}
		\label{scatter}
\end{figure}

The result of our scan gives 23 fixed point candidates all of which have been computed consistently up to order $N = 36$ in a first run.\footnote{We refer to these fixed points as ``fixed point candidates'' since they may contain spurious fixed point solutions. Thus, some of candidates may show instabilities in coupling coordinates.}
To illustrate the parameter range considered by our search algorithm we show these candidates in \fig{scatter}.
As a point of reference, we also show the result in the Einstein Hilbert approximation, which in our conventions is given by
\beq\label{EH-data}
g_*|_{\rm EH}=0.943\,,\quad \lambda_*|_{\rm EH}=0.119\,,\quad \lambda_*g_*|_{\rm EH}=0.112\,.
\eeq
In \fig{scatter}, we observe a fair spread in the  values for $g_*$. Most fixed point candidates except two are similarly or more strongly coupled than the Einstein-Hilbert one, with  $g_*>g_*|_{\rm EH}\approx 1$ indicated by the gray shaded area.
 Results show isolated fixed point candidates as well as  small clusters of two, three or four candidates close to each other. A cluster of four candidates 
 is close to the Einstein Hilbert solution. Two of the fixed points are termed  "secondary" and two further ones are too close to be visible in Fig.~\ref{scatter}. Further small clusters are in regions with larger gravitational coupling. We also find three subsets of candidates which have very similar values for the scale-invariant product of couplings $\lambda_*\cdot g_*$. In Fig.~\ref{scatter}, this is highlighted by the three (dashed gray) lines of constant $\lambda_*\cdot g_*$. There are four fixed point candidates  with couplings very close to $g_*|_{\rm EH}$ and $\lambda_*|_{\rm EH}$, and six candidates whose product is close to $(\lambda_*\cdot g_*)|_{\rm EH}$.

\begin{figure}
\begin{minipage}{\linewidth}
\begin{minipage}{.5\textwidth}
  \centering
  \includegraphics[width=.8\linewidth]{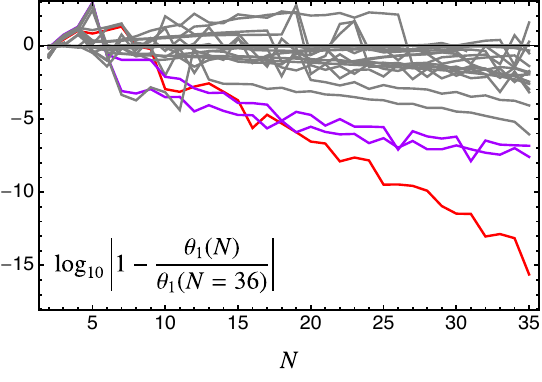}
\end{minipage}%
\begin{minipage}{.5\textwidth}
  \centering
  \includegraphics[width=.8\linewidth]{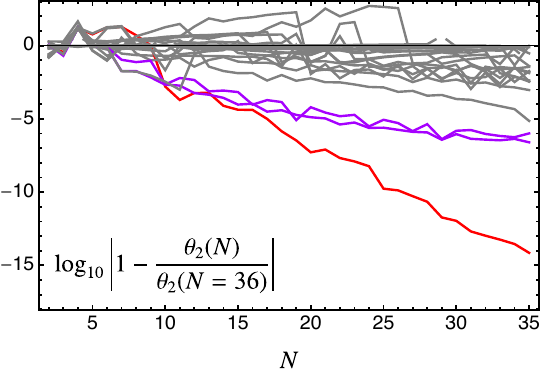}
\end{minipage}
\begin{minipage}{.5\textwidth}
  \centering
  \includegraphics[width=.8\linewidth]{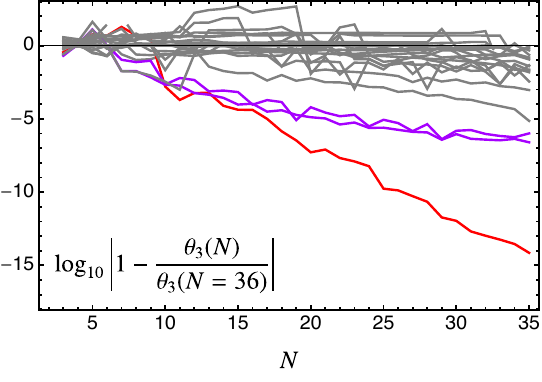}
\end{minipage}%
\begin{minipage}{.5\textwidth}
  \centering
  \includegraphics[width=.8\linewidth]{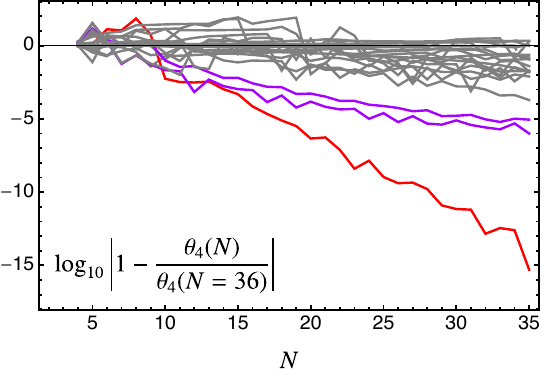}
\end{minipage}
\end{minipage}
\caption{Convergence of the four most relevant eigenvalues (top left to bottom right) for all fixed point candidates shown in \fig{scatter} towards their values at $N = 36$. Tertiary fixed points are illustrated by grey lines, the two secondary fixed points by purple and the primary fixed point by red lines. The quantity on the y-axis gives minus the number of relevant digits, $i.e.$~the number of digits which no longer change when extending the approximation order. Two fixed point candidates, deemed unreliable due to very large relevant eigenvalues, have been excluded (see main text).}
\label{fig:Compare_Conv}
\end{figure}

Once a fixed point candidate has been found consistently up to order $N = 36$, we further check its viability by studying convergence of couplings, and the stability and convergence of scaling exponents, and discard those which turn out to be unstable, have unnaturally large relevant or irrelevant eigenvalues, or show other inconsistencies. To explain this step, we show the convergence of the four most relevant eigenvalues for the fixed point candidates in \fig{fig:Compare_Conv}.\footnote{Two spurious candidates have been excluded from the plot as they display unphysically large relevant eigenvalues of the order of $-10^{13}$ (at $N = 36$), and increasingly negative with increasing approximation orders thereafter.} According to this figure, we select the three fixed points with the best observed convergence properties up to order $N = 36$. Note that some of the tertiary fixed points show a mild convergence in the first few scaling exponents which weakens, however, when going to higher orders. A similar picture is established in the convergence of fixed point coordinates. We emphasize that our criterion of convergence does not preselect the size of the gravitational scaling dimensions in any way.

This second step eliminates the 20 fixed point candidates termed "tertiary", including two fixed point candidates very close to the  ``secondary" ones which are not visible in Fig.~\ref{scatter}. Hence, we are left with the primary Riemann fixed point and two secondary fixed points, highlighted by a red and a magenta arrow in  Fig.~\ref{scatter}. These fixed points are then computed to even higher approximations as discussed below.
For ease of reference, we denote the three remaining fixed points  as FP$_{4s}$,  FP$_4$, and FP$_3$, where the integer states the number of relevant eigendirections, and the subscript ``$s$''  distinguishes the more strongly coupled fixed point of the two  with four relevant directions.
Occasionally, we also refer to FP$_{4s}$ as the ``primary'' Riemann fixed point whose properties are analysed in more depth in Secs.~\ref{sec:mainfp} and~\ref{universality}.  The analysis and discussion of the  secondary Riemann fixed points   FP$_4$ and FP$_3$ is relegated to Sec.\ref{sec:2ndphyfp}.

A further consistency check of these fixed points found with standard boundary conditions is given by the use of non-perturbative boundary conditions as introduced in \cite{Falls:2014tra}. Once a fixed point is found at all orders up to $N_\text{max}$, the idea of improved boundary conditions at any lower  approximation order $N < N_\text{max}$  is to  fix the two highest couplings according to the  values already found at the higher order approximations,
\beq\label{improved}
	\lambda_N = \overline{\lambda}^*_N \, , \qquad \lambda_{N + 1} = \overline{\lambda}^*_{N + 1} \, ,
\eeq
rather than \eq{free}. In \eq{improved}, the barred couplings $\overline{\lambda}^*_N$ denote accurate estimates for the fixed point couplings deduced independently from a sufficiently higher order in the expansion, typically from an expansion up to $N_{\rm max}=144$ with free boundary conditions \eq{free}. With this choice of boundary condition all fixed point couplings ${\lambda}^*_{n\le N}$ at approximation order $N$ automatically take the more accurate values of the fixed point couplings $\overline{\lambda}^*_n$ as determined from the highest approximation order $N_{\rm max}$. Besides  stabilising the fixed point coordinates from the outset, the virtue of the {\it a posteriori} choice \eq{improved} is that it also leads to improved results for scaling exponents, most notably at low orders \cite{Falls:2014tra}. For later reference, and opposed to \eq{free}, we refer to choices of the type \eq{improved} as improved boundary conditions.

\bfi
		\includegraphics[width=.6\linewidth]{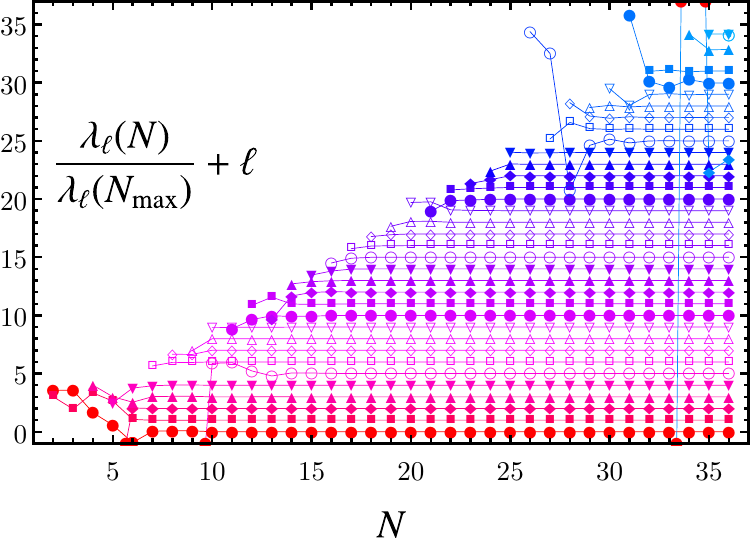}
		\caption{Convergence of couplings $\lambda_\ell (N)$ 
		with approximation order $N$ at the Riemann fixed point, and for $\ell=0,\cdots,36$ (bottom to top). Couplings are  normalised to their values at order $N_{\text{max}} = 144$. The convergence pattern persists all the way up to $N=N_{\text{max}}$ (not shown).}
				\label{fig:fp05522couplingsstandard}
\efi

\section{\bf Riemann  Fixed Point}
\label{sec:mainfp}

We report in this section the evidence for the primary  fixed point with Riemann tensor interactions
and provide an overview of 
 its main features such as coupling coordinates, convergence of effective action, de Sitter solutions and residuals. 
 
\subsection{Fixed Point Coordinates}
\label{sec:fixedpoint}

For the Riemann fixed point FP$_{4s}$
we have calculated its fixed point coordinates  up to approximation order $N_{\rm max} = 144$, involving up to $N_{\rm max} $ powers of the Riemann tensor. 
We find that the fixed point coordinates  fluctuate initially at low approximation orders, but then converge nicely starting with approximation orders $N \approx 6$. This is illustrated  in \fig{fig:fp05522couplingsstandard} where we show the convergence of couplings, derived with standard boundary conditions and normalised to their asymptotic values at $N_{\rm max}$, up to approximation order $N = 36$. The pattern of convergence continues up to the maximal order $N_{\rm max}$ (not shown). 

Quantitative results are summarised in \tab{tab:fp05522couplings} for the first 14 fixed point couplings. The red-shaded areas indicate couplings which arise with the wrong sign at the leading order. Most of these adjust to the correct sign after a single step $N\to N+1$, except for $\lambda_2$ or $\lambda_5$ where the convergence requires 3 or 4 steps, respectively.  Moreover, a fast convergence of couplings towards large $N$ is observed starting with approximation orders $N \approx 6$. Qualitatively, this is associated  to the 
 deviation of $\lambda_2(N)$  from its large $N$  limit in the first few approximation orders $N=3,4,5$. Once $\lambda_2(N)$ has  settled, all other couplings  follow suit. The coupling $\lambda_2$  proportional to the square of the Riemann tensor is expected to play a key role because it is canonically dimensionless. On the other hand, the initially 
  slow convergence of $\lambda_5(N) $ at $N=6,7,8,9$ has no noticeable impact on the convergence of other couplings due to its  smallness.   Overall, we conclude that a viable Riemann fixed point has been identified which is very stable in the polynomial approximation.
  
Note, that the fixed point value $g_*\approx 2.8$ for Newton's coupling (Tab.~\ref{tab:fp05522couplings})  comes out nearly thrice as large as the result $g_*\approx 0.9 - 1.2$  in the Einstein Hilbert approximation, in $f(R)$ models, and in $f(R,{\rm Ric}^2)$ models, with all other technical parameters the same. 
  Hence, this is an example where the higher order interactions impact noticeably onto the lower order fixed point couplings. A more detailed comparison  is postponed until Sec.~\ref{sec:discus}.  
\begin{center}
	\begin{table*}
		\addtolength{\tabcolsep}{3pt}
		\setlength{\extrarowheight}{3pt}
		\centering
		\scalebox{0.8}{
\begin{tabular}{`G?cGcGcGc`}
			\toprule 
			\rowcolor{Yellow} $\bm{N}$ & $\bm{10 \times \lambda_{0}} $ & $\bm{\lambda_{1}} $ & $\bm{10 \times \lambda_{2}} $ & $\bm{10 \times \lambda_{3}} $ & $\bm{\lambda_{4}} $ & $\bm{1000 \times \lambda_{5}} $ & $\bm{\lambda_{6}} $ \\[0.5ex]
			\midrule
			$2$ & $2.51678$ & $-1.06039$ &&\cellcolor{white}  & & \cellcolor{white} &\\ 
			$3$ & $2.51891$ & $-0.686794$ & \cellcolor{LightRed}$2.26290$ & \cellcolor{white} & &\cellcolor{white}  & \\
			$4$ & $1.45783$ & $-1.17577$ & \cellcolor{LightRed}$4.54396$ & $1.31729$ & &\cellcolor{white}  & \\
			$5$ & $0.867111$ & $-0.926808$ & \cellcolor{LightRed}$3.83082$ & $0.662210$ & \cellcolor{LightRed}$0.140972$ & \cellcolor{white} & \\
			$6$ & $0.0419731$ & $-0.375446$ & $-0.518439$ & $0.354710$ & $-0.208929$ & \cellcolor{LightRed}$69.8789$ & \\
			$7$ & $0.576504$ & $-0.346204$ & $-0.487432$ & $0.668691$ & $-0.263364$ & \cellcolor{LightRed}$25.7418$ & $-0.190734$ \\
			$8$ & $0.597978$ & $-0.348344$ & $-0.491396$ & $0.687706$ & $-0.280608$ & \cellcolor{LightRed}$5.18091$ & $-0.278698$ \\
			$9$ & $0.599183$ & $-0.348194$ & $-0.491317$ & $0.688467$ & $-0.280511$ & \cellcolor{LightRed}$5.60590$ & $-0.277357$ \\
			$10$ & $0.552336$ & $-0.352094$ & $-0.492529$ & $0.656860$ & $-0.276576$ & $-0.566918$ & $-0.290361$ \\
			$11$ & $0.552505$ & $-0.352087$ & $-0.492532$ & $0.656981$ & $-0.276618$ & $-0.588649$ & $-0.290484$ \\
			$12$ & $0.552485$ & $-0.352054$ & $-0.492494$ & $0.656932$ & $-0.276487$ & $-0.372249$ & $-0.289603$ \\
			$13$ & $0.552835$ & $-0.352013$ & $-0.492468$ & $0.657154$ & $-0.276470$ & $-0.241403$ & $-0.289146$ \\
			$14$ & $0.552300$ & $-0.352057$ & $-0.492485$ & $0.656796$ & $-0.276424$ & $-0.319270$ & $-0.289350$ \\
			$15$ & $0.552219$ & $-0.352061$ & $-0.492485$ & $0.656739$ & $-0.276409$ & $-0.317244$ & $-0.289325$ \\
			$16$ & $0.552202$ & $-0.352061$ & $-0.492484$ & $0.656726$ & $-0.276402$ & $-0.309584$ & $-0.289290$ \\
			$17$ & $0.552207$ & $-0.352060$ & $-0.492483$ & $0.656728$ & $-0.276399$ & $-0.303941$ & $-0.289269$ \\
			$18$ & $0.552210$ & $-0.352060$ & $-0.492483$ & $0.656730$ & $-0.276399$ & $-0.302915$ & $-0.289265$ \\
			$19$ & $0.552210$ & $-0.352060$ & $-0.492482$ & $0.656731$ & $-0.276399$ & $-0.302833$ & $-0.289265$ \\
			$144$ & $0.552211$ & $-0.352060$ & $-0.492482$ & $0.656731$ & $-0.276399$ & $-0.302828$ & $-0.289265$ \\
			\bottomrule
			\rowcolor{Yellow} $\bm{N}$ & $\bm{10 \times \lambda_{7}} $ & $\bm{\lambda_{8}} $ & $\bm{10 \times \lambda_{9}} $ & $\bm{10 \times \lambda_{10}} $ & $\bm{10 \times \lambda_{11}} $ & $\bm{10 \times \lambda_{12}} $ & $\bm{10 \times \lambda_{13}} $ \\
			\midrule
			$8$ & $-0.517518$ &\cellcolor{white}  & &\cellcolor{white}  & &\cellcolor{white}  & \\
			$9$ & $-0.504042$ & \cellcolor{LightRed}$0.00394341$ & & \cellcolor{white} & & \cellcolor{white} & \\
			$10$ & $-0.788545$ & $-0.148690$ & $-0.967082$ & \cellcolor{white} & & \cellcolor{white} & \\
			$11$ & $-0.788653$ & $-0.148214$ & $-0.959884$ &\cellcolor{LightRed} $0.0297045$ & & \cellcolor{white} & \\
			$12$ & $-0.782596$ & $-0.147738$ & $-0.979088$ & $-0.130361$ & $-0.131648$ & \cellcolor{white} & \\
			$13$ & $-0.777856$ & $-0.146315$ & $-0.979261$ & $-0.197352$ & $-0.239536$ &\cellcolor{LightRed} $-0.475794$ & \\
			$14$ & $-0.781649$ & $-0.148185$ & $-0.989428$ & $-0.184008$ & $-0.149578$ & $0.235337$ & $0.597885$ \\
			$15$ & $-0.781833$ & $-0.148433$ & $-0.992146$ & $-0.192045$ & $-0.144420$ & $0.341166$ & $0.755831$ \\
			$16$ & $-0.781670$ & $-0.148470$ & $-0.993356$ & $-0.199061$ & $-0.147722$ & $0.363293$ & $0.825565$ \\
			$17$ & $-0.781495$ & $-0.148440$ & $-0.993679$ & $-0.202765$ & $-0.151586$ & $0.355885$ & $0.843862$ \\
			$18$ & $-0.781457$ & $-0.148427$ & $-0.993668$ & $-0.203258$ & $-0.152464$ & $0.351623$ & $0.843120$ \\
			$19$ & $-0.781453$ & $-0.148425$ & $-0.993658$ & $-0.203274$ & $-0.152558$ & $0.350902$ & $0.842530$ \\
			$144$ & $-0.781452$ & $-0.148425$ & $-0.993652$ & $-0.203262$ & $-0.152576$ & $0.350644$ & $0.842196$ \\
			\bottomrule
		\end{tabular}
}
		\caption{Fixed point coordinates of the first $14$ couplings   up to approximation order $N = 144$. A fast convergence is observed starting with approximation order $N\approx 6$. Notice that some couplings arise with the wrong sign at the leading orders (red). Most of these adjust to the correct sign after a single step $N\to N+1$, except for $\lambda_2$ ($\lambda_5$) where the convergence requires 3 (4) steps.}
		\label{tab:fp05522couplings}
	\end{table*}
\end{center}

\bfi
	\includegraphics[width=.6\linewidth]{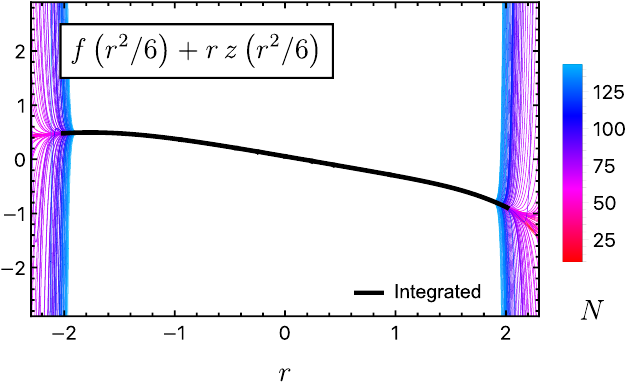}
	\caption{Effective action of the Riemann fixed point from polynomial solution
	at approximation orders $10 \leq N \leq 144$ (colour-coded in the legend) 
	and numerical solution (black line). Results from Pad\'e\ approximants coincide with those from numerical integration (not shown).}
	\label{fig:fp4smallfixpointfunc}
\efi

\subsection{Quantum Effective Action}
\label{sec:effac}
In \fig{fig:fp4smallfixpointfunc} we show the fixed point effective action $f(r^2/6) + r z(r^2/6)$ at the Riemann fixed point  at different approximation orders $N$. A converging behaviour within $r \approx \pm 1.9$ can be observed and it can be seen that the polynomial solution diverges alternatingly at the boundary of the converging regime. This suggests that the poles which limit the radius of the convergence of the polynomial solution are not on the real axis, but somewhere in the complex plane. 

This behaviour and the fact that the fixed singularities of the differential equation \eq{floweqnabc} are at $r \approx 2.00648$ and $r \approx -9.99855$ suggest that the radius of convergence of the polynomial solution is not maximal and can be extended. It is then natural to numerically integrate the differential equation and enlarge the converging region. The result of the numerical integration is shown by the thick black line in \fig{fig:fp4smallfixpointfunc}. As expected, the radius of convergence is enlarged comparing to the polynomial case. 
The new radius of convergence reaches up to $r \approx 2$. More work is required in order to go beyond the point $r \approx 2.00648$ which necessitates the (numerical) lifting of an apparent pole in the flow; see \cite{Demmel:2015oqa,Gonzalez-Martin:2017gza} for examples.

We close with a general remark on global solutions. From the viewpoint of the asymptotic safety conjecture, it will be important to understand how 
the Reuter fixed point extends into a global fixed point, including the regime of asymptotically large fields. Using the method of functional renormalisation in simpler theories -- such as $O(N)$ or $U(N)$ symmetric scalar $(\phi_a\phi_a)^2_{\rm 3d}$ theories or Gross-Neveu $(\bar\psi_a\psi_a)^2_{\rm 3d}$ theories --  global  fixed point  have  indeed been found rigorously based on the  invariants~$\phi^{2}$ or $\bar\psi\psi$, respectively. Identifying global solutions in gravity is   more demanding,  because many more independent invariants  can be constructed out of the Riemann tensor.
Further, it is not known  which of these dominate the large-field asymptotics, nor whether  the very same set will also dominate the small field region \cite{Falls:2018ylp}. Therefore, it is not warranted to  presuppose that global solutions must exist for any given subset of curvature invariants retained in the action.

\bfi
	\includegraphics[width=.6\linewidth]{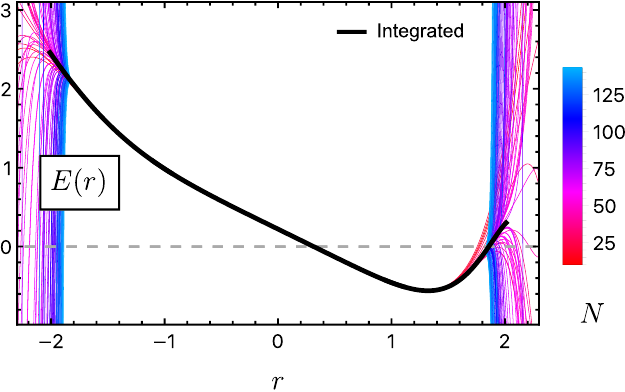}
	\caption{Quantum equation of motion \eq{eqn:eom} at the Riemann fixed point 
	using various polynomial approximation up to order $N\le 144$  (colour-coded in the legend) and numerical  integration (black line). Results from Pad\'e\ approximants coincide with those from numerical integration (not shown).}
	\label{fig:EOM_FP4}
\efi

\subsection{Equations of Motion and de Sitter Solutions}
\label{sec:eom}
Solutions to the  equations of motion for quantum effective actions   \eq{model}   evaluated on the sphere are of physical interest for inflationary cosmology as they correspond to Euclidean de-Sitter and anti de-Sitter solutions \cite{Baumann:2009ds, Hindmarsh:2011hx, Falls:2016wsa}.\footnote{Note that this  assumes the validity of analytic continuation to Lorentzian signature \cite{Falls:2016wsa}. For recent  discussions of subtleties related to the Wick rotation from Euclidean to Lorentzian quantum gravity, see  \cite{Baldazzi:2018mtl,Bonanno:2021squ,Fehre:2021eob}.} 
Previously, de Sitter solutions have been found in the UV scaling regime  of asymptotically safe $f(R)$ and $f(R,{\rm Ric}^2)$ models of quantum gravity  \cite{Bonanno:2010bt,Falls:2014tra,Falls:2017lst,Falls:2018ylp}. 

Denoting the equation of motion as $E(r)$, we find for our model
\beq
	\frac{\delta \Gamma_k}{\delta g_{\mu \nu}} \propto \left[ 4 f \left( \frac{r^2}{6} \right) - \frac{2}{3} r^2 f' \left( \frac{r^2}{6} \right) + r \left( 2 z \left( \frac{r^2}{6} \right) - \frac{2}{3} r^2 z' \left( \frac{r^2}{6} \right) \right) \right] = E(r) \, .
	\label{eqn:eom}
\eeq
Plugging in the  couplings of the Riemann fixed point we can see how this equation behaves at different values for the dimensionless Ricci scalar. Using the polynomial approximation, this is shown at different orders $N$ in \fig{fig:EOM_FP4}, where the first ten approximations have been neglected. Solutions to the quantum equations of motion obey $E(r_{\rm dS})=0$. A first Euclidean de-Sitter solution is found at 
\beq\label{dS1}
r_{{\rm dS}_1} \approx 0.31318\,.
\eeq 
Moreover, had we limited ourselves to just  the Einstein Hilbert approximation, a de Sitter solution would be located at
\beq\label{deSitter-EH}
r_{\rm dS}|_{\rm EH} \approx 0.475\,.
\eeq
Comparing \eq{deSitter-EH} with \eq{dS1} we observe a strong downward shift of the de Sitter solution due to the Riemann fixed point being more strongly interacting than the Einstein Hilbert one. Overall, the shift goes into the opposite direction as the shift induced by Ricci scalar or Ricci tensor interactions \cite{Falls:2016wsa,Falls:2017lst,Falls:2018ylp}.
Furthermore, if we take the Riemann fixed point and disregard all invariants beyond Einstein Hilbert (i.e.~all Riemann interactions), the de Sitter solution \eq{dS1} reduces to
\beq\label{deSitter-Riem-EH}
r_{{\rm dS}_1}|_{\rm EH} \approx 0.31370\,.
\eeq
Comparing \eq{dS1} with \eq{deSitter-Riem-EH} we conclude that the vacuum solution \eq{dS1} arises primarily due to the fixed point  value of the cosmological constant, while the set of higher order Riemann interaction terms has only a  minor effect.

Further, $E(r)$ approaches zero again just before the radius of convergence ends (\fig{fig:EOM_FP4}). However, within the polynomial solution it is uncertain whether a second Euclidean de-Sitter solution is apparent at the fixed point or not.  This question can be answered using numerical integration, which enhances the radius of convergence of the polynomial expansion and renders it maximal.
In \fig{fig:EOM_FP4} we show the corresponding plot for the equation of motion. Indeed, the enlarged radius of convergence gives rise to a second Euclidean de-Sitter solution at 
\beq\label{dS2}
r_{{\rm dS}_2} \approx 1.8567\,.
\eeq This vacuum solution is entirely due to higher order Riemann interactions and cannot be understood from within an Einstein Hilbert approximation. We conclude that the Riemann fixed point displays two dS solutions in the UV scaling regime. Further dS or AdS solutions may exist for larger positive or negative curvature, beyond the range studied here.

\subsection{Residuals and Absence  of Poles}

\label{sec:polyconv}

So far, we have considered the stability of couplings as well as the resulting effective action and de Sitter solutions of the Riemann fixed point FP$_{4s}$. Now, we want to test the convergence of the fixed point as a polynomial solution of the flow equation.

\bfi
	\includegraphics[width=.6\linewidth]{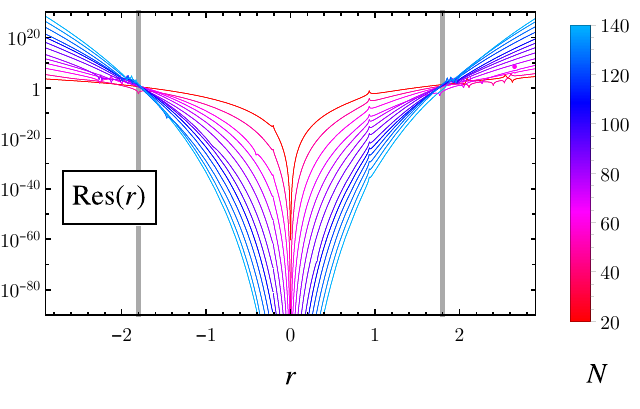}
	\caption{Shown are the residuals \eq{eqn:ployres} of the flow equation in \eq{floweqnabc} with \eq{Riemann} for the solution FP$_{4s}$. Different approximation orders $N$ are distinguished through different colours as shown in the legend. For illustrative reasons, only every tenth order is shown. The grey lines indicate the point beyond which the polynomial solution stops converging. We can read off a radius of convergence of $r_c \approx 1.8$.}
	\label{fig:fp4smallresi}
\efi

Firstly, in \fig{fig:fp4smallresi} we show the residuals of the flow equation \eq{floweqnabc} with \eq{Riemann}, see also \eq{eqn:floweqnriemfp}, at the Riemann fixed point,
\beq
	\text{Res} (r) = 4 f + 2 z r - \frac{2}{3} r^2 \left( f' + r z \right) - \frac{1}{c} I_0 [f, z] \, .
	\label{eqn:ployres}
\eeq
It can be seen that the residuals shrink when more operators are included in the truncation, confirming the convergence of the fixed point solution. Further, the radius of convergence can be estimated to be $r_c \approx 1.8$. Note that the flow equation in \eq{floweqnabc} has fixed singularities at $r_{\rm f}^+ \approx 2.00648$ and $r_{\rm f}^- \approx -9.99855$ coming from the zeros of $P^{S 3}_0$. As we have discussed already, the fact that the radius of convergence of the polynomial solution is smaller than these fixed singularities suggests that it is not maximal.

\bfi
	\includegraphics[width=.6\linewidth]{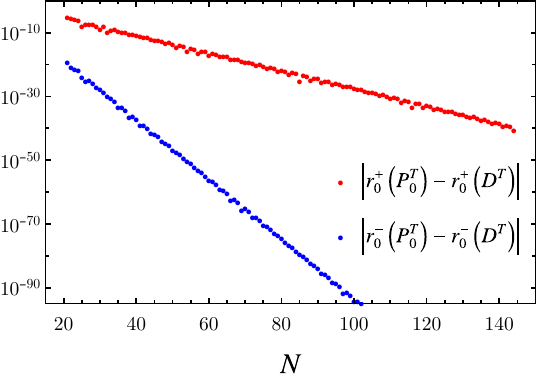}
	\caption{We show the absolute value of the differences of the zeros of $P_0^T (r)$ and $D^T (r)$ at the poles $r_0^+$ (red) and $r_0^-$ (blue) at different orders $N$. Even though the difference decreases much faster at $r_0^-$ with increasing $N$, we observe a strong decrease for this quantity at both poles. This suggests that the poles coincide in an exact solution.}
	\label{fig:fp05522_removdiv_differences}
\efi

Apart from that, we observe small dips in the residuals of \fig{fig:fp4smallresi} at $r_0^+ \approx 0.932$ and $r_0^- \approx -0.204$. These features originate from removable singularities of the flow  at the Riemann fixed point: At these points the denominator $D^T (r)$ defined in \eqn{floweqnI0} changes sign and has a zero. However, at the same time the corresponding numerator $P_0^T (r)$ has a sign change and goes to zero as well, leaving the flow term $I_0[f,z]$ finite, \eq{In}. In numerical approximations, however, these zeros are not at the exact same position. This leads to a spurious singularity responsible for the  dips in \fig{fig:fp4smallresi}. However, \fig{fig:fp05522_removdiv_differences} shows that the differences between the zeros of the numerator $P_0^T (r)$ and the denominator $D^T (r)$ decrease strongly order by order at both points, $r_0^+$ and $r_0^-$. This supports the view that the zeros of $P_0^T$ and $D^T$ are at the same position in the exact solution of the Riemann fixed point. This alone does not yet tell us that the singularity at those points is removable. However, we note that the derivative of the denominator converges to ${D^T}' \left( r_0^+ \right) \approx -42.6$ and ${D^T}' \left( r_0^- \right) \approx -19.5$ at higher orders. Thus, we expect the derivative of the denominator to be non-vanishing at both $r_0^+$ and $r_0^-$ from which we conclude that the singularities are removable. Hence, we expect the full non-polynomial solution of the Riemann fixed point  to be free of dips such as those in \fig{fig:fp4smallresi} and to be well-defined in the entire regime $|r|\le r_{\rm f}^+$.

Finally, we use numerical integration to see whether the radius of convergence of the polynomial solution can be enhanced. To perform the numerical integration, we seek two differential equations for the highest derivatives of $f(r^2/6)$ and $z(r^2/6)$. Starting from the flow equation, these can be obtained by splitting the flow into even and odd parts (in $r$) \cite{Falls:2017lst}. In this way, we deduce a system of two coupled third order differential equations,
\begin{equation}
	\begin{split}
		f''' (x) = \mathcal{I}_f [f, z](x) \, , \\
		z''' (x) = \mathcal{I}_z (f, z](x) \, ,
	\end{split}
	\label{eqn:thirdderivs}
\end{equation}
with $x = r^2/6$. The functions $\mathcal{I}_f$ and $\mathcal{I}_z$ are rational functions depending only on $f$ and $z$ as well as their derivatives up to second order and explicit dependencies on $x$.

In the form \eq{eqn:thirdderivs}, we can integrate the fixed point equation and find the fixed point solution of $f(x)$ and $z(x)$.
As a consequence of the presence of spurious singularities for FP$_{4s}$, however, this must be performed in patches.
Otherwise, numerical uncertainties might lead to the numerical solution developing an artificial pole coming from a removable singularity. In practice, it is easiest to use the polynomial solution until we are beyond the removable singularity. This can also be motivated by the very small residuals of the polynomial solution for small $r$, which are hard to achieve relying solely on a numerical integration. All in all, we use the polynomial expansion to provide $(i)$ the fixed point solution in the range $|r| < r_i$, and further use these as initial conditions for the numerical integration starting at $|r| = r_i$, with $r_i$ within $(r_0^+,r_c)$. In this manner, we   obtain a reliable fixed point solution over the maximal domain $|r| < r_{\rm f}^+$.

As an alternative method, we have also used Pad\'e resummation to extend the polynomial solution. 
However, similar to \fig{fig:fp4smallresi} the residuals of Pad\'e approximants can give rise to spikes. Such spikes originate from the removable singularities found in the polynomial approximation. As discussed above, we expect these singularities to be removable in the full solution. Within numerical precision, the result obtained from Pad\'e resummation is in full agreement with the numerical integration and gives the maximal radius of convergence as seen in \fig{fig:fp4smallfixpointfunc} and \fig{fig:EOM_FP4}.

\section{\bf Universality and Critical Surface}\label{universality}

\bfi
	\includegraphics[width=.6\linewidth]{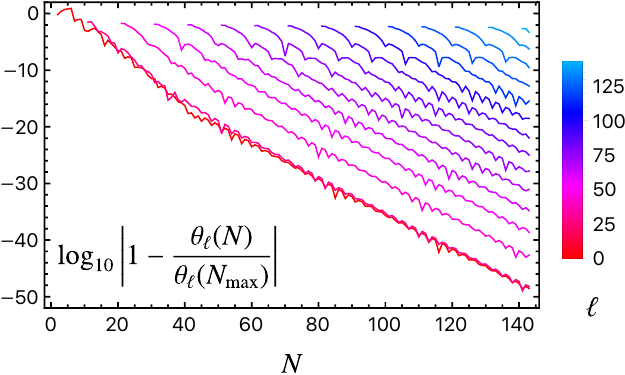}
	\caption{Convergence of eigenvalues at the Riemann fixed point with increasing approximation order $N$.
	A decimal place of accuracy is gained for any $\Delta N \approx 2.8$  in the approach towards $N_{\text{max}} = 144$. For better visualisation, we only show the convergence for every $10$th eigenvalue.}
	\label{fig:fp05522eigenvalueconvergence}
\efi

In  this section, we report our results for  universal scaling exponents, the dimension of the UV critical surface, the set of fundamentally free parameters, and the bootstrap test for the Riemann fixed point. We also discuss the large order behaviour of scaling dimensions and signatures for weak coupling.

\subsection{Scaling Exponents and Infinite $N$ Limit}

Universal scaling exponents describe how renormalisation group trajectories scale towards or away from interacting fixed points. In practice, they are deduced as the eigenvalues of the stability matrix $M_{ij}=\partial\beta_i/\partial g_j|_*$. For each and every order of the approximation, we have computed the set of universal scaling exponents 
\beq\label{thetas}
\{\theta_\ell(N),\,\ell=0,\cdots, N-1\}
\eeq
with eigenvalues sorted according to magnitude, and up to the maximal order $N_{\rm max}=144$. With increasing $N$, we find that eigenvalues converge rapidly to their asymptotic values. This is illustrated in \fig{fig:fp05522eigenvalueconvergence} which displays (minus) the number of relevant digits of scaling exponents which agree with the scaling exponents at the highest approximation order $N_{\rm max}=144$. 
By and large, the convergence is exponential with increasing $N$. Roughly, enhancing approximation orders by $\Delta N\approx 2.8$ leads to an additional significant digit in any of the scaling exponents.
\bfi
		\includegraphics[width=.45\linewidth]{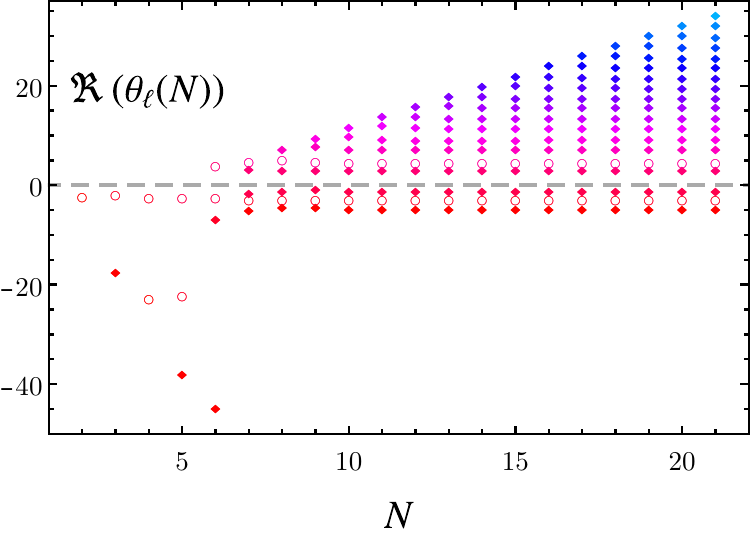}\quad
	\includegraphics[width=.45\linewidth]{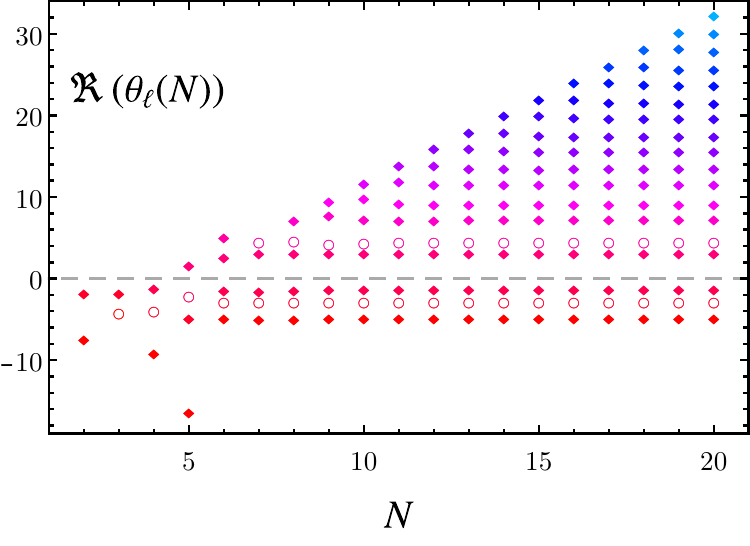}
		\caption{Scaling exponents at the Riemann fixed point from  standard (left panel) and improved boundary conditions (right panel).
		Real (complex) eigenvalues are visualised as filled diamonds (empty circles).} 		\label{fig:fp05522eigvstandard}
\efi

Using our results at the highest order $N_{\rm max} = 144$ and taking an average over the preceeding ten approximation orders we find an infinite $N$ estimate for the first seven scaling exponents 
\beq
	\begin{split}
		\langle \theta_0 \rangle \approx& -4.96188 \, , \\
		\langle \theta_1 \rangle \approx& -3.00645 - 1.45800 i \, , \\
		\langle \theta_2 \rangle \approx& -3.00645 + 1.45800 i \, , \\
		\langle \theta_3 \rangle \approx& -1.32993 \, , \\
		\langle \theta_4 \rangle \approx& \,\, 2.98723 \, , \\
		\langle \theta_5 \rangle \approx& \,\, 4.41530 - 2.85549 i \, , \\
		\langle \theta_6 \rangle \approx& \,\, 4.41530 + 2.85549 i \, .
	\end{split}
	\label{eqn:fp05522eigvfinal}
\eeq
We have determined the leading $46$ relevant digits of these eigenvalues which have become independent of the approximation order, though for practical purposes, we only  show the first six relevant digits.  Hence, the errors of the calculated values are of order $10^{-46}$. Note that the error reflects the accuracy of the numerically determined fixed point solution. It does neither reflect  truncation errors nor systematic errors. For strategies to access truncation and systematic errors within the functional RG framework, we refer to \cite{Litim:2010tt,Litim:2000ci, Litim:2001up, Litim:2001fd}.
As such, however, the values \eq{eqn:fp05522eigvfinal} can be viewed as the infinite-$N$ limits for the various fixed point couplings. 
Note also that the canonical values of the four relevant eigenvalues are $\vartheta_0 = -4$, $\vartheta_1 = -2$, $\vartheta_2 = 0$, and $\vartheta_3 = 2$, see \eq{classicalscaling}. Hence, the quantum corrections shift the real parts of the relevant eigenvalues by an amount of $\approx -2.1$ on average, pointing into the relevant direction for all of these eigenvalues (see also Sec.~\ref{EVs}).

We should also compare findings at the Riemann fixed point with those in the Einstein Hilbert approximation, where the scaling exponents are
\beq\label{EH-exponents}
\begin{array}{rcl}
\theta_0&=&-2.41i-1.95i\\
\theta_1&=&-2.41i+1.95i\,.
\end{array}
\eeq
For the first two scaling exponents, the real part of eigenvalues differ from classical values by $+1.6$ and $-0.4$, respectively. Instead, the leading two exponents at the Riemann fixed point \eq{eqn:fp05522eigvfinal} differ from classical values by roughly one unit,  $-.96$ and $-1.0$, respectively. Most notably, the exponent $\theta_0$ receives shifts into opposite directions. On the other hand, the exponent $\theta_1$ becomes more relevant due to quantum corrections in both cases. We conclude that quantum effects due to Riemann curvature invariants have a noticeable effect on the scaling exponents, both in view of classical or Einstein Hilbert exponents.

It is  worth pointing out that the vast majority of eigenvalues are real rather than complex conjugate pairs. This is in  contrast to $f(R)$ approximations 
where about half of the eigenvalues are complex \cite{Falls:2014tra,Falls:2016wsa}, or $f(R,{\rm Ric}^2)$-type approximations 
where most eigenvalues come in complex conjugate pairs \cite{Falls:2017lst}. In the present case, only the second and third, and fifth and sixth most relevant eigenvalues turn out to be complex conjugate pairs while all other eigenvalues are real.

\label{sec:impbc}

\begin{center}
	\begin{table*}
		\centering
		\addtolength{\tabcolsep}{3pt}
		\setlength{\extrarowheight}{3pt}
		\scalebox{0.8}{
		\begin{tabular}{`G?cGcGcG`}
			\toprule
			\rowcolor{Yellow} $\bm{N}$ & $\bm{\theta_{0}} $ & $\bm{\theta_{1}} $ & $\bm{\theta_{2}} $ & $\bm{\theta_{3}} $ & $\bm{\theta_{4}} $ & $\bm{\theta_{5}} $ \\[0.5ex]
			\midrule
				$2$ & \cellcolor{LightRed} $-2.41 - 1.95 I$ & $-2.41 + 1.95 I$ & & \cellcolor{white} & \multicolumn{2}{c`}{\underline{\bf Standard\ Boundary}} \\
				$3$ &\cellcolor{LightRed} $-17.6$ & $-2.00 - 2.09 I$ & $-2.00 + 2.09 I$ & \cellcolor{white} &  \multicolumn{2}{c`}{\bf \underline{Conditions${}_{}$}} \\
				$4$ &\cellcolor{LightRed} $-23.0 - 16.4 I$ &\cellcolor{LightRed} $-23.0 + 16.4 I$ & $-2.78 - 0.750 I$ &\cellcolor{LightRed}  $-2.78 + 0.750 I$ & &\cellcolor{white}  \\
				$5$ & \cellcolor{LightRed}$-38.1$ &\cellcolor{LightRed} $-22.3 - 5.26 I$ & \cellcolor{LightRed}$-22.3 + 5.26 I$ & \cellcolor{LightRed} $-2.77 - 0.705 I$ & \cellcolor{LightRed} $-2.77 + 0.705 I$ &\cellcolor{white}  \\
				$6$ & \cellcolor{LightRed}$-45.0$ & \cellcolor{LightRed}$-6.95$ & $-2.74 - 1.15 I$ & \cellcolor{LightRed}  $-2.74 + 1.15 I$ & $3 .74 - 1.36 I$ & $3 .74 + 1.36 I$ \\
				$7$ & $-5.19$ & $-3.05 - 1.07 I$ & $-3.05 + 1.07 I$ & $-1.78$ & $3 .04$ & $4 .62 - 2.29 I$ \\
				$8$ & $-4.52$ & $-3.07 - 1.44 I$ & $-3.07 + 1.44 I$ & $-1.24$ & $2 .97$ & $4 .95 - 2.90 I$ \\
				$9$ & $-4.50$ & $-3.04 - 1.43 I$ & $-3.04 + 1.43 I$ & $-0.990$ & $2 .97$ & $4 .57 - 2.90 I$ \\
				$10$ & $-4.96$ & $-3.00 - 1.45 I$ & $-3.00 + 1.45 I$ & $-1.34$ & $2 .98$ & $4 .29 - 2.79 I$ \\
				$11$ & $-4.96$ & $-3.01 - 1.46 I$ & $-3.01 + 1.46 I$ & $-1.33$ & $2 .99$ & $4 .37 - 2.82 I$ \\
				$17$ & $-4.96$ & $-3.01 - 1.46 I$ & $-3.01 + 1.46 I$ & $-1.33$ & $2 .99$ & $4 .42 - 2.86 I$ \\[0.5ex]
			\bottomrule 
			\toprule 
			\rowcolor{Yellow} $\bm{N}$ & $\bm{\theta_{0}} $ & $\bm{\theta_{1}} $ & $\bm{\theta_{2}} $ & $\bm{\theta_{3}} $ & $\bm{\theta_{4}} $ & $\bm{\theta_{5}} $ \\[0.5ex]
			\midrule
				$2$ & $-7.51$ & $-1.88$ & &\cellcolor{white}  & \multicolumn{2}{c`}{\underline{\bf Improved\ Boundary}} \\
				$3$ & $-4.28 - 0.217 I$ & $-4.28 + 0.217 I$ & $-1.86$ &\cellcolor{white}  &  \multicolumn{2}{c`}{\bf \underline{Conditions${}_{}$}}\\
				$4$ & \cellcolor{LightRed}$-9.19$ & $-4.13 - 1.35 I$ & $-4.13 + 1.35 I$ & $-1.21$ & &\cellcolor{white}  \\
				$5$ & \cellcolor{LightRed}$-16.5$ & $-5.00$ & $-2.29 - 1.17 I$ & $-2.29 + 1.17 I$ & $1 .59$ &\cellcolor{white}  \\
				$6$ & $-4.97$ & $-3.02 - 1.48 I$ & $-3.02 + 1.48 I$ & $-1.55$ & $2 .55$ & $4 .96$ \\
				$7$ & $-5.12$ & $-3.01 - 1.33 I$ & $-3.01 + 1.33 I$ & $-1.66$ & $3 .04$ & $4 .41 - 1.79 I$ \\
				$8$ & $-5.01$ & $-3.03 - 1.48 I$ & $-3.03 + 1.48 I$ & $-1.52$ & $3 .03$ & $4 .55 - 2.83 I$ \\
				$9$ & $-4.98$ & $-2.99 - 1.48 I$ & $-2.99 + 1.48 I$ & $-1.37$ & $3 .02$ & $4 .18 - 2.77 I$ \\
				$10$ & $-4.97$ & $-3.00 - 1.45 I$ & $-3.00 + 1.45 I$ & $-1.34$ & $2 .98$ & $4 .29 - 2.80 I$ \\
				$11$ & $-4.97$ & $-3.01 - 1.46 I$ & $-3.01 + 1.46 I$ & $-1.34$ & $2 .99$ & $4 .36 - 2.83 I$ \\
				$17$ & $-4.96$ & $-3.01 - 1.46 I$ & $-3.01 + 1.46 I$ & $-1.33$ & $2 .99$ & $4 .42 - 2.86 I$ \\
			\bottomrule
		\end{tabular}
		}
		\caption{The first six eigenvalues of the Riemann fixed point based on  standard  (upper panel) and improved boundary conditions (lower panel).  
		Red-shaded areas indicate that scaling exponents deviate by more than a factor of two from  asymptotic values. We observe four relevant eigendirections at the UV critical point.}
		\label{tab:fp05522eigvlow}
	\end{table*}
\end{center}

Next, we discuss the improved convergence  of scaling exponents due to improved boundary conditions (see Sect.~\ref{sec:boundarycond}). 
It has been shown previously that improved boundary conditions lead to an enhanced convergence of scaling exponents, in particular at low orders \cite{Falls:2014tra}. Setting the boundary conditions according to the highest order fixed point couplings, fixed points will take the coordinates dictated by the highest order approximation, for all approximation orders, and the initial fluctuations as seen in Tab.~\ref{tab:fp05522couplings} are avoided. Eigenvalues, however, change non-trivially under this choice of boundary conditions. 

Results for  eigenvalues are shown in \fig{fig:fp05522eigvstandard} where we also compare standard and improved boundary conditions. In either case   eigenvalues converge rapidly  after $N \approx 6$.
With standard boundary conditions   (left panel)  and smaller $N$, couplings  fluctuate strongly and large relevant eigenvalues arise.
The appearance of spurious large negative eigenvalues at low orders has already been seen in earlier works \cite{Lauscher:2001ya,
Falls:2017lst, Falls:2018ylp} at $N = 3$.  
In contrast, with improved boundary conditions (right panel) the magnitudes of the large negative eigenvalues are substantially smaller. 
A quantitative comparison of eigenvalues is given in  \tabs{tab:fp05522eigvlow} and \ref{tab:fp05522eigvhigh}. In  \tab{tab:fp05522eigvlow}
the first six eigenvalues of the Riemann fixed point are shown based on  standard  (upper) and improved (lower panel)  boundary conditions.  
Red-shaded areas indicate if scaling exponents deviate by more than a factor of two from their asymptotic values. This affects many of the leading 
critical exponents from standard boundary condition including $\theta_0(N=2,3,4,5,6)$, $\theta_1(N=4,5,6)$, $\theta_2(N=5)$, $\theta_3(N=4,5,6)$ and $\theta_4(N=5)$. In contrast, re-adjusting the underlying fixed point couplings to their asymptotic values via improved boundary 
conditions, we find that only the exponents  $\theta_0(N=4,5)$ are off by a factor $2-3$. For example, the largest relevant eigenvalue at $N = 6$ is reduced in magnitude from the unreliably large value $-45.0$ down to $-4.7$ by using improved boundary conditions. Most notably, the improved result at $N=6$ comes out very close to the exact result in the infinite-$N$ limit. Another feature  of improved boundary conditions is that we now  find a four-dimensional UV critical surface at $N = 5$, in accord with the result \eq{eqn:fp05522eigvfinal}. In contrast, standard boundary conditions spuriously lead to a five dimensional UV critical surface. 

\begin{center}
	\begin{table}
		\addtolength{\tabcolsep}{3pt}
		\setlength{\extrarowheight}{3pt}
		\scalebox{0.8}{
		\begin{tabular}{`G?cGcGc`}
			\toprule 
			\rowcolor{Yellow} $\bm{N}$ & $\bm{\theta_{0}} $ & $\bm{\theta_{1}} $, $\bm{\theta_{2}^*} $ & $\bm{\theta_{3}} $ & $\bm{\theta_{4}} $ & $\bm{\theta_{5}} $, $\bm{\theta_{6}^*} $ \\[0.5ex]
			\midrule
			$7$ & $-5.19405$ & $-3.05354 - 1.06701 I$ & $-1.77830$ & $3.03790$ & $4.62121 - 2.28727 I$ \\
			$8$ & $-4.52166$ & $-3.07089 - 1.43645 I$ & $-1.23942$ & $2.97158$ & $4.95260 - 2.89518 I$ \\
			$9$ & $-4.49733$ & $-3.04415 - 1.42823 I$ & $-0.989626$ & $2.97152$ & $4.57428 - 2.89885 I$ \\
			$10$ & $-4.95655$ & $-3.00452 - 1.45296 I$ & $-1.33724$ & $2.98371$ & $4.29308 - 2.79177 I$ \\
			$11$ & $-4.95839$ & $-3.00709 - 1.45810 I$ & $-1.33429$ & $2.98873$ & $4.37013 - 2.81807 I$ \\
			$12$ & $-4.95392$ & $-3.00832 - 1.45856 I$ & $-1.33398$ & $2.98684$ & $4.41313 - 2.85557 I$ \\
			$13$ & $-4.94871$ & $-3.00794 - 1.45668 I$ & $-1.32518$ & $2.98684$ & $4.41988 - 2.85881 I$ \\
			$14$ & $-4.95832$ & $-3.00670 - 1.45791 I$ & $-1.32851$ & $2.98704$ & $4.41527 - 2.85517 I$ \\
			$15$ & $-4.96096$ & $-3.00644 - 1.45814 I$ & $-1.32929$ & $2.98722$ & $4.41468 - 2.85469 I$ \\
			$16$ & $-4.96189$ & $-3.00641 - 1.45813 I$ & $-1.32984$ & $2.98723$ & $4.41496 - 2.85522 I$ \\
			$17$ & $-4.96197$ & $-3.00643 - 1.45803 I$ & $-1.32996$ & $2.98723$ & $4.41520 - 2.85547 I$ \\
			$18$ & $-4.96190$ & $-3.00645 - 1.45800 I$ & $-1.32994$ & $2.98723$ & $4.41529 - 2.85550 I$ \\
			$19$ & $-4.96188$ & $-3.00645 - 1.45800 I$ & $-1.32994$ & $2.98723$ & $4.41530 - 2.85549 I$ \\
			$144$ & $-4.96188$ & $-3.00645 - 1.45800 I$ & $-1.32993$ & $2.98723$ & $4.41530 - 2.85549 I$ \\
			\bottomrule
		\end{tabular}
}
		\caption{Convergence of eigenvalues at the Riemann fixed point  
		with approximation order $N$.}
		\label{tab:fp05522eigvhigh}
	\end{table}
\end{center}

We conclude that the large fluctuations in the eigenvalue spectrum found with standard boundary conditions at low orders are significantly reduced using improved boundary conditions. At higher orders, starting at  $N = 6$, eigenvalues from improved boundary conditions are already close to their asymptotic values  \eq{eqn:fp05522eigvfinal}, and only differ mildly  from standard boundary conditions. Starting at $N = 17$,  the significant digits in \tab{tab:fp05522eigvlow} no longer change with increasing $N$ and the differences in the boundary condition become irrelevant  (\tab{tab:fp05522eigvhigh}).

\subsection{Dimensionality of UV Critical Surface}
\label{sec:gap}

Another novel feature of the Riemann fixed point is its four dimensional UV-critical surface. As can be seen in \fig{fig:fp05522eigvstandard}, the fixed point has four negative and therefore relevant eigenvalues, one of which is a complex conjugate pair. The gravitational action \eq{eqn:effansatz} with \eq{X} and \eq{Riemann} only includes two canonically relevant and one canonically marginal operator. Thus, we are led to the conclusion that quantum effects have turned a classically marginal and a  classically irrelevant operator into relevant directions at the UV fixed point. 
\bfi
	\includegraphics[width=.6\linewidth]{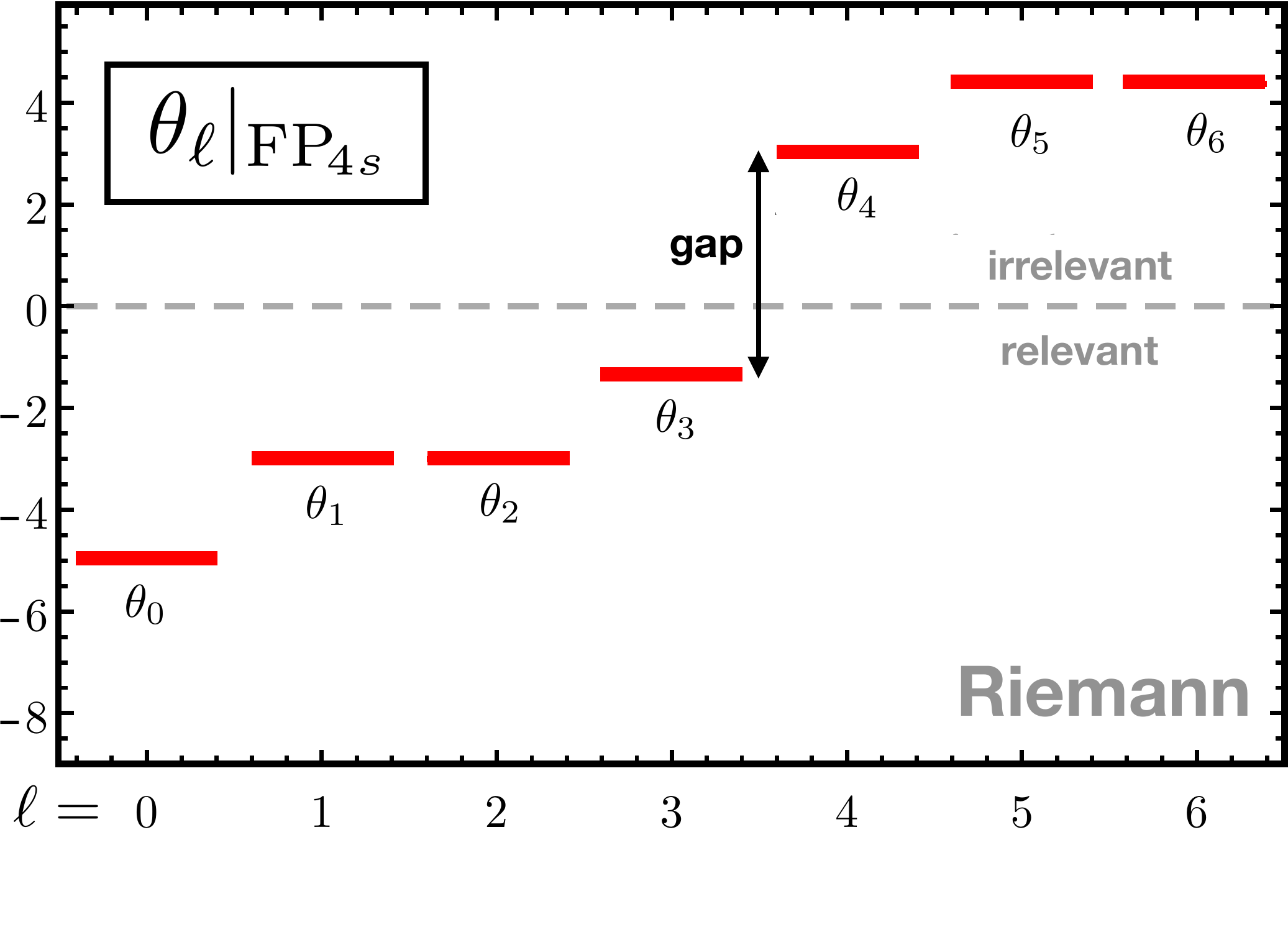}
	\caption{UV critical surface and gap. Shown are the converged results for the seven most relevant eigenvalues $\theta_\ell$ at the Riemann fixed point at $N=N_{\rm max}$. We observe a four-dimensional UV critical surface. The gap in the eigenvalue spectrum between the largest relevant  and the smallest irrelevant eigenvalue is indicated by an arrow.}
	\label{fig:fp05522Gap}
\efi

The UV critical surface is further illustrated in \fig{fig:fp05522Gap} where we show the seven most relevant eigenvalues for different approximation orders. We observe that eigenvalues (thin black lines) convergence rapidly to asymptotic values (red thick line), four of which settle at negative values. Of particular interest is the gap in the eigenvalue spectrum, corresponding to the difference between the largest relevant and the smallest irrelevant scaling exponents. In the free theory, this gap is $\Delta_{\rm free} = 2$. At the interacting fixed point,  the most irrelevant of the relevant eigenvalues is  $\theta_3$ and the most relevant of the irrelevant eigenvalues is $\theta_4$. Their gap $\Delta_{\rm Riem} $ is given by
\beq\label{gap-Riem}
	\Delta_{\rm Riem} = \theta_4 - \theta_3 \approx 4.32 \, .
\eeq
We observe that the gap is larger than the classical gap. However, it is smaller than the gap $\Delta_R \approx 5.5$  in $f(R)$ theories  \cite{Falls:2014tra,Falls:2018ylp}  and smaller than the gap $\Delta_{\rm Ric} \approx 5.98$  in theories with Ricci tensor interactions  \cite{Falls:2017lst}.  
The difference originates from the fact that the quantum effects make  the first classically irrelevant eigenvalue more  relevant quantum-mechanically in the present case, leading to a smaller gap and to the ordering $0<\Delta_{\rm free}<\Delta_{\rm Riem} < \Delta_{\rm Ric} < \Delta_{R}$.

\subsection{Fundamentally Free Parameters}
It is interesting to discuss the set of  fundamentally free parameters $N_{\rm free}$ from the viewpoint of the full quantum theory of gravity  \cite{Falls:2018ylp}.
The family of quantum gravity models \eq{eqn:effansatz}, \eq{X} always contains the Einstein-Hilbert interaction terms
$\sim \sqrt{g}$ and $\sim\sqrt{g}R$. For both of these classically relevant interactions, the corresponding couplings are found to be relevant in the quantum theory, giving the Reuter fixed point \cite{Reuter:1996cp,Souma:1999at,Litim:2003vp}. On the level of fourth order interactions, our models retain any one of the invariants 
\beq\label{4th}
\sqrt{g}R^2\,,\quad  \sqrt{g}R_{\mu \nu} R^{\mu \nu}\,,\quad  \sqrt{g}R_{\rho \sigma \mu \nu} R^{\rho \sigma \mu \nu}\,,
\eeq
or a linear combination thereof. We recall that the invariants \eq{4th} cannot be distinguished within projections on maximally symmetric spaces (adopted here). Nevertheless, for either of these classically marginal interactions, it is found that the corresponding coupling becomes relevant in the quantum theory \cite{Falls:2013bv,Falls:2014tra,Falls:2017lst}.  This is in accord with the lower bound 
\beq\label{free}
  N_{\rm free}\ge 3
 \eeq
on the number of fundamentally free parameters \cite{Falls:2018ylp}.   Based on  fourth order extensions of gravity, and neglecting  the potential impact of higher order curvature invariants, similar conclusion have been reached in \cite{Benedetti:2009rx,Niedermaier:2009zz,Falls:2020qhj}. 
An important new observation of this work arises for dim-6 invariants. Of these, our models  \eq{eqn:effansatz}, \eq{X}  can retain any one of the invariants
\beq\label{6th}
\sqrt{g}R^3\,,\quad  \sqrt{g}R\,R_{\mu \nu} R^{\mu \nu}\,,\quad \sqrt{g}R\,R_{\rho \sigma \mu \nu} R^{\rho \sigma \mu \nu}\,,
\eeq
or any linear combination thereof.  In models which retain the term with Ricci scalar or Ricci tensor interactions, no new relevant direction is created and we are left with \eq{free}. In contrast, for models with Riemann interactions, \eq{XRiem}, \eq{model}, a new relevant eigendirection can be found.
In other words, the dynamics of Riemann tensor interactions displays qualitative differences from Ricci scalar or Ricci tensor interactions starting at dim-6. A discussion of the corresponding eigenperturbations is relegated to Sec.~\ref{EV-Riemann} below. 

 A conservative interpretation of this result for the full theory of quantum gravity is that at least one of the canonically irrelevant dim-6 invariants (see Tab.~\ref{tab:dim246operators} for an overview) becomes a relevant operator due to quantum fluctuations. This does require quantum corrections to scaling dimensions to be of order unity, which has previously been observed in many asymptotically safe models. What is new here is that the quantum corrections make the dim-6 interaction more relevant rather than more irrelevant. In consequence, it indicates that the UV theory of pure quantum gravity has at least
\beq\label{nonfree}
  N_{\rm free}\ge 4
 \eeq
 fundamentally free parameters. In order to further clarify  the number of free parameters, it will be important to investigate other dim-6 invariants in future works (Tab.~\ref{tab:dim246operators}) including the Goroff-Sagnotti term \cite{Goroff:1985sz,Goroff:1985th,Gies:2016con}, while also retaining sufficiently many higher dimensional interaction terms for eigenvalue spectra to become reliable.

\begin{figure*}
	\centering
	\includegraphics[width=0.6\textwidth]{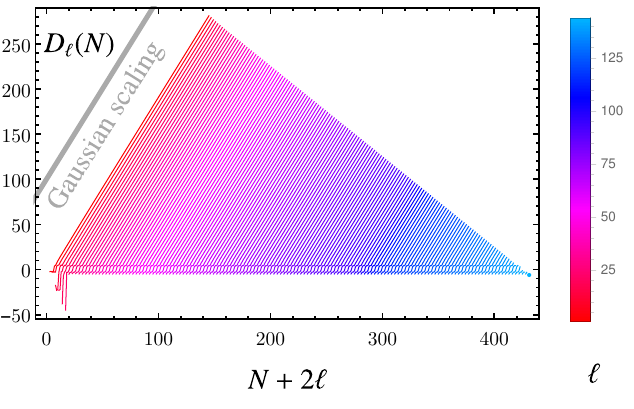}
	\caption{Bootstrap test for asymptotic safety and canonical power counting, with  lines of scaling exponents $D_\ell (N)$ defined in \eq{DellN}.
	From the left to the right, each line $D_\ell$, as a function of the approximation order $N$, shows the $\ell^{\text{th}}$ largest eigenvalue, connected by a line, against  $N + 2\ell$  ($\ell$ fixed). 
	The linear growth of all curves establishes that canonical mass dimension (grey line) is a viable ordering principle.}
	\label{fig:fp05522BootStrap}
\end{figure*}

\subsection{Bootstrap and Large Order Behaviour}
\label{sec:boot}

The bootstrap approach to quantum gravity is a systematic way of testing whether a fixed point candidate is viable or otherwise \cite{Falls:2013bv}. The fundamental assumption is that canonical power counting remains to be a good ordering principle for operators even at an interacting fixed point, which is known to hold true for fixed points observed in nature. A virtue of the bootstrap hypothesis is that it can be confirmed a posteriori. Specifically, starting at some approximation with $N$ different operators, the eigenvalue introduced by a new operator of higher mass dimension should turn out to be more irrelevant than other eigenvalues. Further, the eigenvalues already included in the truncation before should only change by small fluctuations. In particular, from a sufficiently high order onwards, new relevant eigenvalues should not be created.
To test this conjecture for the Riemann fixed point, we take a look at the quantity
\beq\label{DellN}
	D_\ell (N) = \theta_{N - \ell} (N) \, ,
\eeq
This quantity gives the $\ell^{\text{th}}$ largest eigenvalue in the eigenvalue spectrum at approximation order $N$. For  the bootstrap to be viable, we expect that the functions $D_\ell (N)$ are growing with increasing $N$, which manifests that the addition of a new high-order invariant  in the step from $N\to N+1$ leads to a new eigenvalue which is less relevant than those identified in previous approximations. This pattern also ensures that no new relevant eigenvalues are created when higher operators are included. In \fig{fig:fp05522BootStrap} we see that this is clearly the case for the Riemann fixed point as soon as $N\approx 6$ or larger,  in agreement with the bootstrap conjecture, which verifies our working hypothesis.

\bfi
	\includegraphics[width=.55\linewidth]{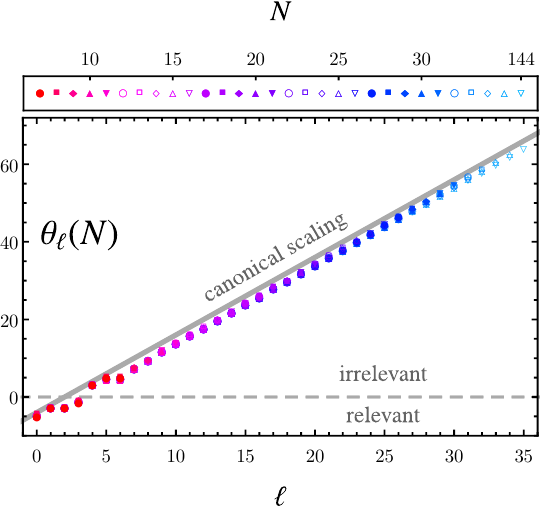}
	\caption{Shown are the real parts of the eigenvalues $\theta_\ell (N)$ at different approximation orders $N$ up to $N_{\text{max}} = 35$ using standard boundary conditions. Different orders can be distinguished using different symbols and colour coding as explained in the legend. The first orders up to $N = 6$ have been excluded due to their previously found inconsistencies with standard boundary conditions. Hence, the plot shown starts at $N = 7$. The grey line indicates the eigenvalues of a Gaussian fixed point following from the mass dimension of included operators, see \eq{eqn:Gausseigv}.}
	\label{fig:fp4Gaussian}
\efi

Having seen that the bootstrap approach is fulfilled for the Riemann fixed point we want to further investigate the eigenvalue spectrum. In particular we are interested to ask how far away the observed eigenvalues are from Gaussian scaling. For that purpose, \fig{fig:fp4Gaussian} shows the eigenvalues $\theta_\ell (N)$ obtained at different orders $N$ in comparison with their Gaussian values
\eq{classicalscaling}, 
\beq
	\vartheta_\ell = 2 \ell - 4 \, , 
	\label{eqn:Gausseigv}
\eeq
indicated by a grey line. As can be seen, eigenvalues are close to their Gaussian values with small deviations due to quantum corrections. This is in accordance with other works \cite{Falls:2018ylp, Falls:2017lst,Falls:2014tra,Codello:2008vh} where eigenvalues are observed to be ``as Gaussian as it gets''. A difference to these works is that quantum corrections are strictly shifting eigenvalues towards relevant directions, leading to, among other, the observed four dimensional UV critical surface. It is also worth noting that the largest corrections towards relevant directions are received by the eigenvalues with $\ell = 2, 3, $ and $6$ as seen by the dips in \fig{fig:fp4Gaussian}. The largest correction, at $\ell = 6$, is given by
\beq
	\theta_6 (N_\text{max}) - \vartheta_6 \approx - 3.6 \, .
\eeq

\begin{figure*}
	\centering
	\includegraphics[width=0.6\textwidth]{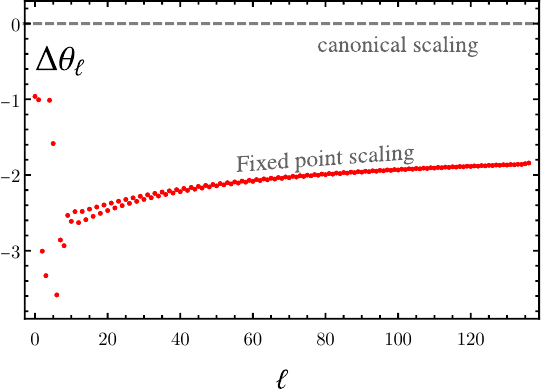}
	\caption{Shown are the differences $\Delta\theta_\ell$ between the real part of the converged universal eigenvalues $\theta_\ell (N)$ and their canonical counterpart $\vartheta_\ell$ at the highest approximation order $(N = 144)$. 
	}
	\label{fig:fp4smallgaussappro}
\end{figure*}

We further investigate the deviations from Gaussian scaling in \fig{fig:fp4smallgaussappro} where the difference between the real part of the eigenvalues  $\theta_\ell (N)$ and their canonical counterparts in \eq{eqn:Gausseigv}
at different orders $N$ is shown. As we see, the quantum corrections to scaling are of the order of $1 -3$ for all eigenvalues, which substantiates the observation that, except for the leading few, most of the quantum gravitational scaling dimensions  are near-Gaussian with only small relative deviations.

More precisely, the first ten eigenvalues follow rather stochastic fluctuations around Gaussian scaling while higher eigenvalues show very systematic deviations: Eigenvalues $\theta_l (N)$ with $l > 10$ enter relatively close to Gaussian scaling when they are included for the first time and then drift away to their final value when the truncation order is increased and more operators are included. Furthermore, the line of final values for the deviation is bended in \fig{fig:fp4smallgaussappro}, showing that higher eigenvalues are getting closer to Gaussian scaling.

The mean deviation of the first ten eigenvalues from Gaussian scaling at order $N = 144$, $\sigma_{10}^{N = 144}$, is given by
\beq\label{shift}
	\left| \sigma_{10}^{N = 144} \right| = 2.3 \pm 1.0 \,,
\eeq
and points into the relevant direction. This deviation is of the order of three within one standard deviation. Returning to our assumption that only eigenvalues corresponding to operators from \tab{tab:dim246operators} should become relevant in the UV, we see that the observed shift can indeed be big enough to do so. The Riemann fixed point FP$_{4s}$ is a particular example for this to happen. 

Moreover, we note that within our setup a shift of $4$ in the relevant direction for the low eigenvalues is only $2 \sigma$ away. Hence, our observations in the eigenvalue spectrum do not rule out the possibility of rendering an eigenvalue corresponding to a dim-8 operator relevant. How likely it is for an eigenvalue of a dim-8 operator to become relevant is hard to estimate from our analysis, since the offset from Gaussian scaling and its standard deviation changes between settings and approximations. However, it is clear that we should not rule out the possibility of eigenvalues corresponding to dim-8 operator becoming relevant at an interacting fixed point for gravity.

\bfi
	\includegraphics[width=.4\linewidth]{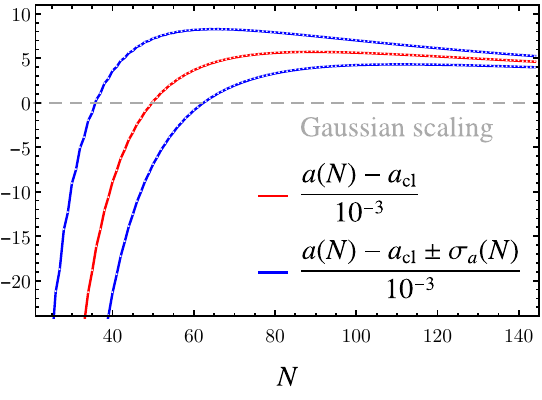}\quad
	\includegraphics[width=.4\linewidth]{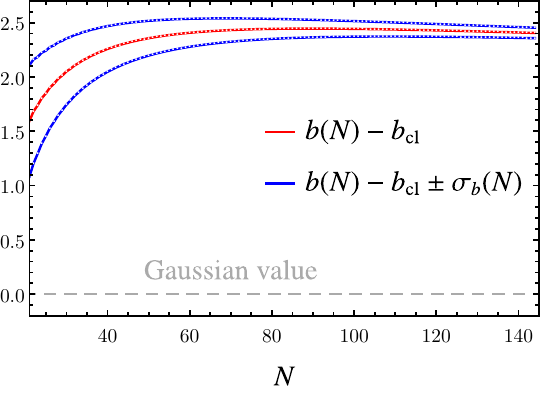}
	\caption{Shown are the quantum-induced shifts in the scaling exponents, expressed  in terms of the slope parameter $a(N)$ (left panel) and the  offset parameter $b(N)$ (right panel) using the fit   \eq{eqn:slopefit} and \eq{abfit}, and  in comparison to classical values \eq{ab}. The blue lines show the $1\sigma$ band.}
	\label{fig:fp05522slope}
\efi
The pattern in the deviations of higher eigenvalues from Gaussian scaling in \fig{fig:fp4smallgaussappro} suggests that there might be a simple linear relation between these higher eigenvalues. Therefore, we use the results obtained at different orders and fit a linear function to the eigenvalue spectrum. Since the last eigenvalues of every approximation need a few more orders to settle to their final values, we exclude the last ten eigenvalues of each order from the fit. Then, fitting the linear function
\beq
	\theta_\ell^{\text{fit}} = a \cdot \ell - b \, ,
	\label{eqn:slopefit}
\eeq
we obtain the results shown in \fig{fig:fp05522slope} 
for different approximations $N$ starting from $N = 20$ up to $N = 144$. At the highest approximation, we find
\beq\label{abfit}
	\begin{split}
		a|_{\rm Riemann}  &\approx 2.0046 \pm 0.0006 \, , \\
		b |_{\rm Riemann} &\approx  6.41 \pm 0.05 \, .
	\end{split}
\eeq
According to \eq{eqn:Gausseigv} the Gaussian values for these parameters are given by
\beq\label{ab}
	a_{\text{cl}} = 2 \, , \qquad b_{\text{cl}} = 4 \, .
\eeq
Hence, the fitted slope of the Riemann fixed point  is very close to Gaussian scaling. The y-intercept ($\ell=0)$, on the other hand, is shifted away from its Gaussian value by an amount of $b - b_{\text{cl}} = 2.41$. This is roughly the same amount as the observed average shift in the first ten eigenvalues of the spectrum \eq{shift}.

To summarise, we find that the eigenvalues of the Riemann fixed point  follow near-Gaussian scaling with quantum induced shifts. These shifts in effective mass dimension are such that operators invariably become more relevant. The magnitude  of shifts is of the order of $1 - 3$. Note that this is different from what has been found in $f(R)$ models of gravity   where quantum corrections mostly shift eigenvalues into irrelevant directions. Hence, we come to the conclusion that the dynamics induced by the Riemann tensor interactions included in settings of the form \eq{model} are distinct from those found for a $f(R)$-type models, or of models involving Ricci tensor interactions such as  \eq{eqn:effansatz} with \eq{X} and $(a, b, c) = (0, 1, 0)$. In particular, this manifests itself in generating a four dimensional UV critical surface for the Riemann fixed point.

\subsection{Signatures of Weak Coupling}
\label{sec:weakcoup}
\begin{figure}
	\includegraphics[width=.6\linewidth]{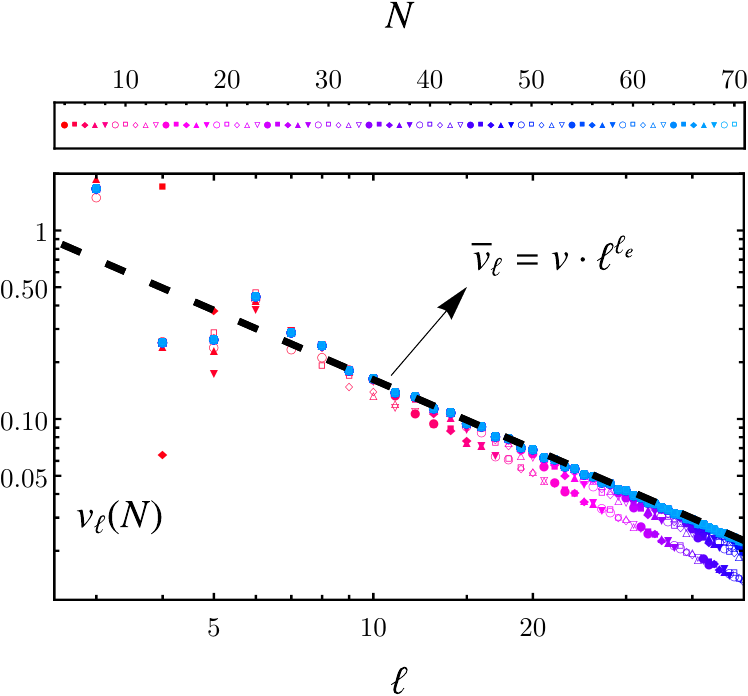}
	\caption{Signatures of weak coupling  at the Riemann fixed point. Shown are the parameters ${\rm v}_\ell$ as defined in \eq{eqn:vldefinition} which measure the differences between eigenvalues $\theta_\ell (N)$ and their Gaussian values $\vartheta_\ell$. Different approximation orders $N$ can be distinguished using different symbols and colour codings as explained in the legend. The dashed black line corresponds to a fit of $\bar{\rm v}_\ell = v \cdot \ell^{\ell_e}$ with $v \approx 2.690$ and $\ell_e \approx -1.221$. For visualisation purposes only data up to $\ell = 50$ and $N = 70$ is shown. However, we note that this picture continues up to highest order $N = 144$ and corresponding $\ell = 143$. The fit was done with data from $N = 144$ with the ten highest eigenvalues of this order being excluded.
	}
	\label{pDelta17_FP4}
\end{figure}

The near-Gaussian behaviour as observed above can be further discussed. For that purpose, we define the quantity $v_\ell$ which measures the relative deviation of scaling dimensions from classical values, 
\beq
	v_\ell = \left| 1 - \frac{\Re \left( \theta_\ell (N) \right)}{\vartheta_\ell} \right| \,.
	\label{eqn:vldefinition}
\eeq
Here $\vartheta_\ell$ denotes classical eigenvalues and $\theta_\ell (N)$ are observed eigenvalues at different approximation orders $N$. \fig{pDelta17_FP4} shows this quantity in a double logarithmic plot for orders up to $N = 70$ and eigenvalues $\theta_\ell (N)$ up to $\ell = 50$. As we see, the plot shows a linear relation between the quantities $\log (v_\ell)$ and $\log(\ell)$. This suggests a functional relation of the form
\beq
	\bar{v}_\ell = v \cdot \ell^{\ell_e} \, .
\eeq
Fitting this function to the observed eigenvalues at highest order $N = 144$ with the last ten eigenvalues being excluded gives $v \approx 2.690$ and $k_e \approx -1.221$. \fig{pDelta17_FP4} shows that this relationship with these values is fulfilled to a good approximation. This can also be confirmed to higher orders which are not shown here for illustrative reasons. If this functional dependence remains to be true at higher orders the quantity $v_k$ will approach zero for high eigenvalues. This means that higher eigenvalues are getting closer and closer to their canonical values and eventually end up there. Thus, the induced quantum corrections to scaling exponents become smaller for higher eigenvalues which gives another notion of the term ``as Gaussian as it gets''. Moreover, it can be interpreted as a signature of a hidden weak coupling controlling Gaussian scaling of higher operators.

It is worth pointing out that the approach to Gaussian values is algebraic, and much slower than the  approach observed in  truncations involving Ricci scalar or Ricci tensor interactions. Specifically, for $f(R)$ actions or actions of the form $X = R_{\mu \nu} R^{\mu \nu}$ the Gaussian approach have been found out to be exponential and therefore faster than the observed power law behaviour in the $X = R_{\rho \sigma \mu \nu} R^{\rho \sigma \mu \nu}$ truncation. However, in the case of $X = R_{\mu \nu} R^{\mu \nu}$ only $20$ different approximation orders have been taken into account. Also, $f(R)$ approximations tend to have stronger fluctuations in couplings and eigenvalues. Thus, our results are somewhat more precise in that matter.

\section{\bf Weakly Interacting Riemann 
Fixed Points}
\label{sec:2ndphyfp}
In Fig.~\ref{scatter}, 
we have pointed out that our fixed point search gives rise to three candidates which pass the viability tests and which do not suffer from unnaturally large relevant/irrelevant eigenvalues or strong instabilities in eigenvalues. In this section, we discuss our results for  the two secondary, weakly interacting Riemann fixed points  FP$_4$ and FP$_3$ (magenta dots in Fig.~\ref{scatter}).
$\text{FP}_4$ should not be confused with the primary Riemann fixed point (FP$_{4s}
$) of the previous sections.

\subsection{Fixed Point Coordinates}
Within the polynomial expansion,  the coordinates of the secondary fixed points $\text{FP}_4$ and $\text{FP}_3$ fluctuate at low orders but stabilise thereafter. This is not unexpected, and very similar to what has been observed for the primary  Riemann fixed point. The fixed point coordinates of $\text{FP}_3$ stabilise around the order $N \approx 9$. We notice that its couplings are very close to those of $\text{FP}_4$. 
In this light, we may interpret this as the fixed point $\text{FP}_4$ splitting up into two fixed points $\text{FP}_4$ and $\text{FP}_3$ starting from order $N\approx9$ onwards. After these orders both fixed points are  stable and arise consistently at each and every order, up  to the highest orders. 

Quantitatively, the first  couplings of the fixed point effective action \eq{model} at $\text{FP}_4$ are
\beq\label{FP4}
	\begin{aligned}
		\lambda_0 =&  \ \ \ \,\, 0.26914  \, ,\\
				\lambda_1 =& \, -0.90665  \, , \\
		{\text{FP}_4}:\quad\quad\lambda_2 =& \ \ \ \,\, 0.24452  \, , \\
		\lambda_3 =& \ \ \ \,\,  0.061344  \, , \\
		\lambda_4 =& \ \ \ \,\, 1.3278  \,. \\
	\end{aligned}
\eeq
To achieve these, we have used a polynomial expansion of the action \eq{eqn:effansatz} with \eq{XRiem} and using \eq{series} up to the order $N = 144$. 
In terms of the couplings \eq{gla} we have
\beq\label{FP4-data}
g_*|_{\text{FP}_4}=1.103\,,\quad
 \lambda_*|_{\text{FP}_4}=0.148\,,\quad 
  \lambda_*g_*|_{\text{FP}_4}=0.164\,.
\eeq
Clearly, the values \eq{FP4-data} are close to those of the Einstein Hilbert approximation \eq{EH-data}, see Fig.~\ref{scatter}.
This data should be compared with the first few couplings of $\text{FP}_3$, for which we find 
\beq\label{FP3}
	\begin{aligned}
		\lambda_0 =& \ \ \ \,\, 0.27059  \, ,\\
		\lambda_1 =&  \, -0.97217  \, , \\
		{\text{FP}_3}:\quad\quad\lambda_2 =&  \ \ \ \,\, 0.23082  \, ,\\
		\lambda_3 =&  \ \ \ \,\,  0.15913  \, , \\
		\lambda_4 =&  \ \ \ \,\, 1.8557  \,. \\
	\end{aligned}
\eeq
To achieve these, we have used a polynomial expansion of the action \eq{eqn:effansatz} with \eq{XRiem} and using \eq{series} up to the order $N = 72$. 
We notice that the first three couplings at FP${}_3$  \eq{FP3} are close to those at FP${}_4$ \eq{FP4}. The first coupling which differs substantially is $\lambda_3$.  In terms of the couplings \eq{gla} we find
\beq\label{FP3-data}
g_*|_{\text{FP}_3}=1.029\,,\quad
\lambda_*|_{\text{FP}_3}=0.139\,,\quad 
\lambda_*g_*|_{\text{FP}_3}=0.143\,.
\eeq
which, again, comes out quite close to the Einstein-Hilbert result \eq{EH-data}. 
The vicinity of \eq{FP4-data} and \eq{FP3-data} to \eq{EH-data} allows to interpret
these fixed points as higher order extensions of the Reuter fixed point.

\bfi
	\includegraphics[width=.55\linewidth]{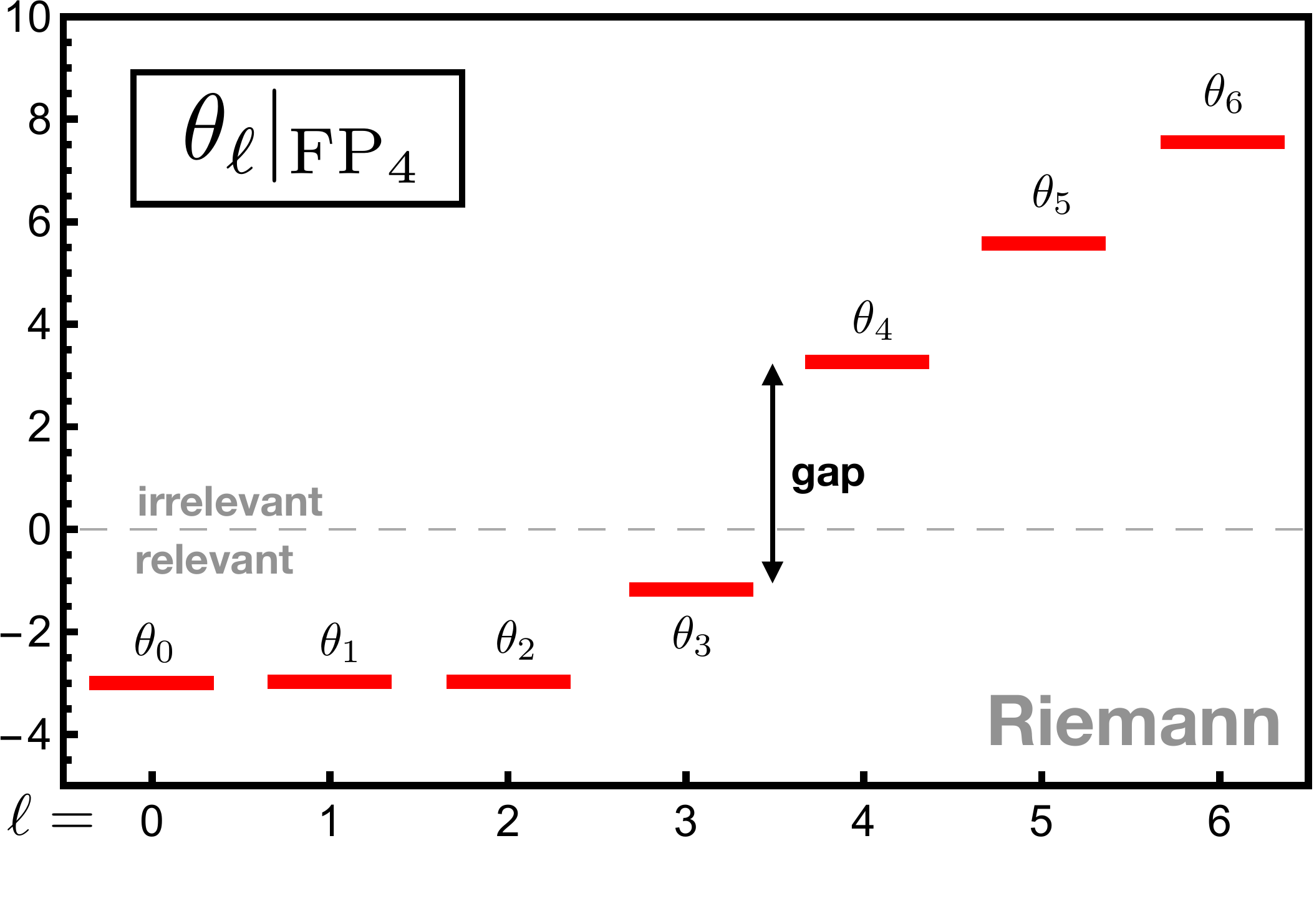}
	\caption{Shown are the converged results for the seven most relevant eigenvalues $\theta_\ell$ at the $\text{FP}_{4s}$ fixed point at $N = 144$. 	
	We observe a four-dimensional UV critical surface. The gap in the eigenvalue spectrum between the largest relevant  and the smallest irrelevant eigenvalue is indicated by an arrow.}
	\label{fig:FP4-UV-Surface}
\efi

\subsection{Scaling Exponents and Critical Surface}
 For both $\text{FP}_4$ and $\text{FP}_3$ we have computed scaling exponents up to very high order in the polynomial expansion.
 In either case, we found that the scaling exponents satisfy the bootstrap test as performed for the Riemann fixed point in Sec.~\ref{sec:boot}.
 Moreover, we have also confirmed that large order scaling exponents for $\text{FP}_4$ and $\text{FP}_3$ become near-Gaussian, in full analogy with
 the Riemann fixed point and earlier findings for $f(R)$ and $f(R,{\rm Ric}^2)$ type theories. Quantitatively, the first few scaling exponents
 at $\text{FP}_4$ read
\beq\label{FP4-exponents}
	\begin{aligned}
		 \theta_0 =& -3.0191  \, , \\
		 \theta_1 =& -2.9332 - 2.0659 i  \, , \\
		 {\text{FP}_4}:\quad\quad\theta_2 =& -2.9332 + 2.0659 i  \, , \\
		 \theta_3 =& -1.1590  \, , \\
		 \theta_4 =& \ \ \ \,3.2157  \,.
	\end{aligned}
\eeq
Several comments are in order. Firstly, we notice that the critical surface is four-dimensional, similar to what has been found at the  Riemann fixed point \eq{eqn:fp05522eigvfinal}. Quantitatively, three of the four relevant exponents 
are close to those at the Riemann fixed point, with $\theta_i|_{\rm Riemann}-\theta_i|_{\text{FP}_4}\approx -0.1$ $(i=1,2,3)$, except the leading one which differs more substantially, $\theta_0|_{\rm Riemann}-\theta_0|_{\text{FP}_4}\approx -1.9$. On the other hand, the gap in the eigenvalue spectrum \eq{FP4-exponents} is given by
\beq\label{gap-FP4}
\Delta_{\text{FP}_4} = \theta_4-\theta_3\approx 4.37
\eeq
and is very close to, if mildly larger than, the Riemann gap \eq{gap-Riem}.
In comparison with the Einstein Hilbert approximation \eq{EH-exponents}, the two leading exponents in \eq{FP4-exponents} are more relevant, with $\theta_i|_{\text{FP}_4}-\theta_i|_{\rm EH}\approx -0.5$ $(i=0,1)$. We conclude that quantum effects due to Riemann curvature invariants make the $\text{FP}_4$ fixed point more relevant compared to the Einstein Hilbert approximation, and despite of the fact that the two leading couplings remain essentially unchanged.

\bfi
	\includegraphics[width=.55\linewidth]{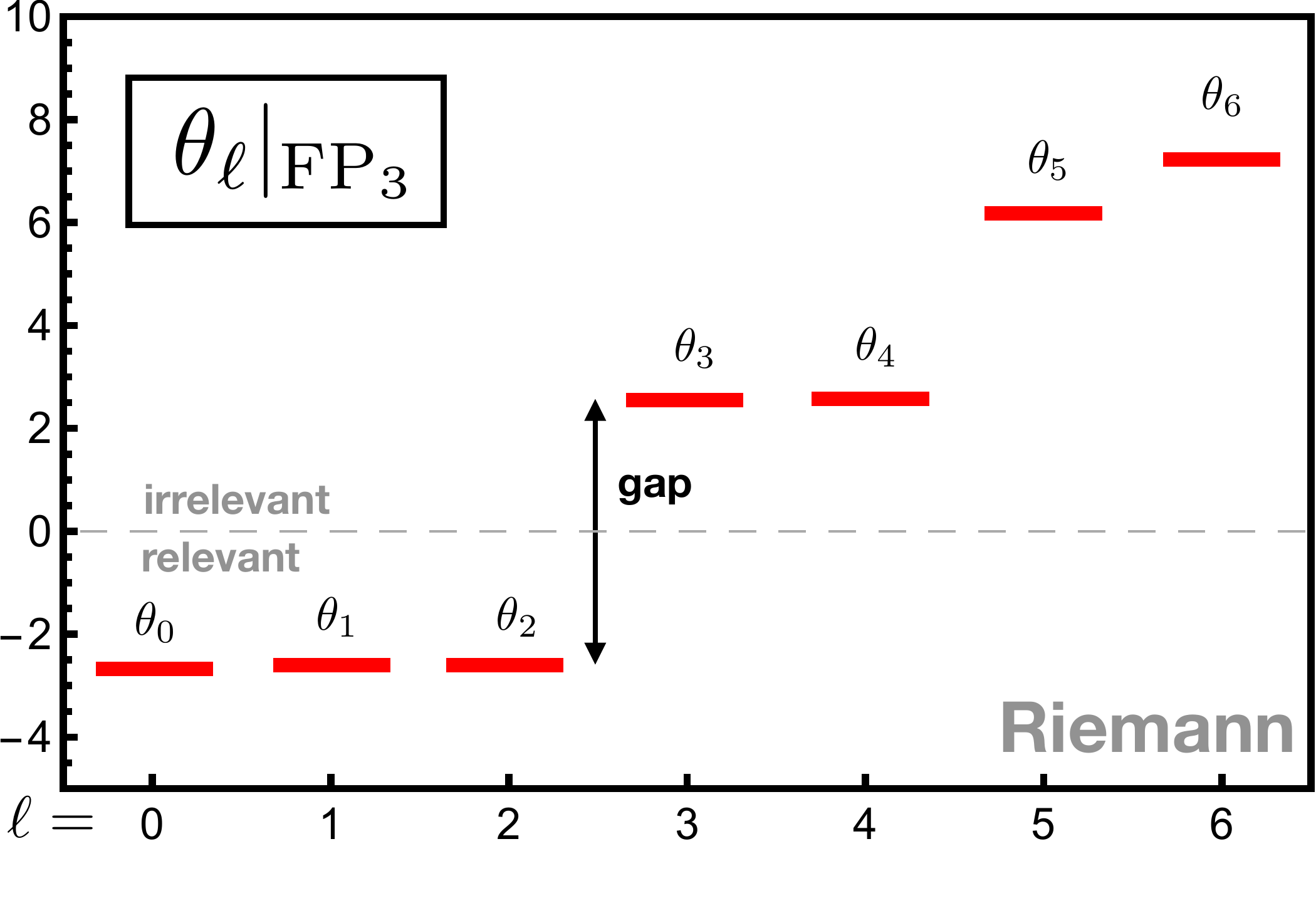}
	\caption{Shown are the converged results for the seven most relevant eigenvalues $\theta_\ell$ at the $\text{FP}_3$ fixed point at $N = 144$. 	
	We observe a three-dimensional UV critical surface. The gap in the eigenvalue spectrum between the largest relevant  and the smallest irrelevant eigenvalue is indicated by an arrow.}
	\label{fig:FP3-UV-Surface}
\efi

Comparing the scaling exponents of $\text{FP}_4$ with classical values, we find that the real part of the first two scaling exponents differ from classical ones by $+1.0$ and $-0.9$, respectively. This can be compared with the shifts of $+1.6$ and $-0.4$ induced by the Einstein Hilbert approximation alone.  In both cases, the leading exponent receives shifts towards positive values, while the second most relevant exponent receives negative quantum corrections. Moreover, all higher order Riemann tensor interactions together make the leading scaling exponents $\theta_0$ and $\theta_1$ more relevant  by $-0.6$ and $-0.5$ units compared to the Einstein-Hilbert ones. 

Let us now discuss the large-order behaviour of universal eigenvalues $\{\theta_n\}$ in more detail by performing a linear fit as in \eq{eqn:slopefit}.
Overall, we find a pattern similar to the one observed in Fig.~\ref{fig:fp05522slope} for the Riemann fixed point. Due to a slowed-down convergence of eigenvalues (to be discussed in \sct{sct:weakFpconvergence}) we take the average of values for the slope and the offset parameter over the ten highest approximation 
orders  $(N=134 - 143)$ to find
\beq\label{fit-FP4}
	\begin{split}
		 a  |_{{\rm FP}_4}\approx & \, 2.0055 \pm 0.0007 \, , \\
		 b |_{{\rm FP}_4} \approx & \, 4.78 \pm 0.05 \, .
	\end{split}
\eeq
Similar to what is found at the Riemann fixed point \eq{abfit}, the slope is very close to its Gaussian value \eq{ab}. The shift $b-b_{\rm cl}\approx 0.78$ in the offset is pushing the eigenvalues into the relevant direction and comes out much smaller than at the Riemann fixed point (\fig{fig:fp05522slope}).

It is well-known that the Einstein-Hilbert fixed point and $f(R)$ fixed points are connected with the Gaussian fixed point of classical GR by well-defined RG trajectories. Given that $\text{FP}_4$ is numerically close to the Einstein-Hilbert and $f(R)$ values of  the leading two fixed point couplings, 
it is conceivable that RG trajectories should exist which flow from $\text{FP}_4$ in the UV to classical GR in the infrared  (see Sec.~\ref{GR}). Indeed, we have checked by numerical integration that  a weakly coupled low energy regime with \eq{Gauss0} and the emergence of General Relativity  is realised from the  UV fixed point.

For the $\text{FP}_3$ fixed point, we find the scaling exponents
\beq\label{FP3-exponents}
	\begin{aligned}
		 \theta_0 =& -2.7283  \, , \\
		 \theta_1 =& -2.6562 - 1.9498 i  \, , \\
		 {\text{FP}_3}:\quad\quad\theta_2 =& -2.6562 + 1.9498 i  \, , \\
		 \theta_3 =& \ \ \ \,2.5133 - 0.9150 i  \, , \\
		 \theta_4 =& \ \ \ \,2.5133 + 0.9150 i  \, .
	\end{aligned}
\eeq
The critical surface is three-dimensional, similar to what has previously been found at the  $f(R)$ and $f(R,{\rm Ric}^2)$ fixed points. 
Quantitatively, the three relevant exponents are close to the leading three exponents at the $\text{FP}_4$ fixed point, and less relevant with $\theta_i|_{\text{FP}_4}-\theta_i|_{\text{FP}_3}\approx -0.3$ $(i=0,1,2)$. On the other hand, the gap in the eigenvalue spectrum \eq{FP3-exponents} is given by
\beq\label{gap-FP3}
\Delta_{\text{FP}_3} = \theta_3-\theta_2\approx 5.17
\eeq
and is larger than the Riemann gap  \eq{gap-Riem}, and larger than the gap at the $\text{FP}_4$ fixed point   \eq{gap-FP4}. We also notice that the gap is much closer to the gap $\Delta_{f(R)}\approx 5.48$ observed in $f(R)$ theories, yet still much  smaller than $\Delta_{\rm Ric}\approx 5.98$ found for $f(R,{\rm Ric}^2)$ theories with three relevant couplings.

Let us now compare exponents with classical and Einstein Hilbert values. For the first two scaling exponents $\theta_0$ and $\theta_1$, the real part of eigenvalues differ from classical values by $+1.3$ and $-0.6$, respectively. This should be compared with the shifts by $+1.6$ and $-0.4$ in the EH approximation, \eq{EH-exponents}.  Hence, not only are the fixed point couplings at $\text{FP}_3$ very close to those of the EH approximation, but the quantum corrections of the two leading exponents are similarly close to those already induced when only the two leading operators are retained. In other words, at $\text{FP}_3$ the quantum effects due to all  curvature invariants with canonical mass dimension $4$ and higher (all invariants beyond the cosmological constant and the Ricci scalar) only contribute the shift $-0.3$ and $-0.2$ to the two leading exponents, which furthermore is half as large as the shifts at $\text{FP}_4$. Once more, Riemann tensor interactions make the scaling exponents more relevant, albeit mildly. It is in this sense that $\text{FP}_3$ is as close as it gets to the Reuter fixed point. 

Turning to the large-order behaviour of universal eigenvalues we perform a linear fit as in \eq{eqn:slopefit}. 
We again find a pattern similar to the one observed in Fig.~\ref{fig:fp05522slope} for the Riemann fixed point. 
Taking the average of values for the slope and the offset over the ten highest  approximation orders $(N=62 - 72)$ we  find the estimates
\beq\label{fit-FP3}
	\begin{split}
		a |_{{\rm FP}_3}\approx& \, 1.991 \pm 0.002 \, , \\
		b |_{{\rm FP}_3} \approx& \, 4.15 \pm 0.08 \, .
	\end{split}
\eeq
We notice that the slope parameter is essentially classical with $a-a_{\rm cl}\approx 0$  in accord with the approach to near-Gaussian scaling. Also, the shift $b-b_{\rm cl}$ is smaller  then what is found at ${\rm FP}_4$ \eq{fit-FP4} and at the Riemann fixed point \eq{abfit}. These findings further underline that this fixed point is as close to the Reuter fixed point as it gets.

 \subsection{Effective Action and de Sitter Solutions}

\bfi
	\includegraphics[width=.6\linewidth]{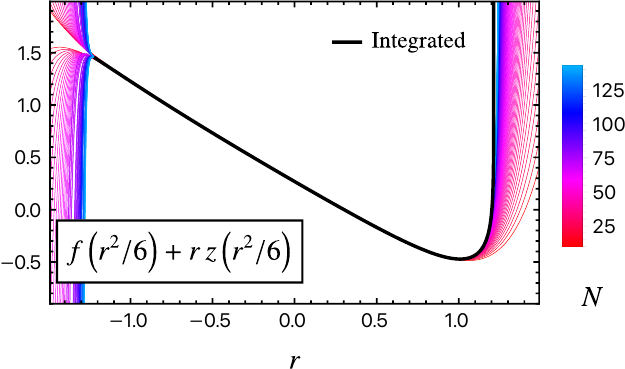}
	\caption{Shown is the effective fixed point action for the 
	secondary fixed point  FP$_4$ and the convergence with  increasing approximation orders $N$. A similar result is found for FP$_3$ (not shown).}
	\label{fig:fp027_FixPointFunc}
\efi

Next, we discuss the curvature dependence of the weakly interacting Riemann fixed points. 
In \fig{fig:fp027_FixPointFunc} we show the fixed point functional at the fixed point $\text{FP}_4$ for different orders $N$ as a function of dimensionless curvature $r=R/k^2$. We observe that the polynomial approximation shows an alternating-sign convergence on the negative real axis and  a same-sign convergence on the positive real axis. The latter is  often indicative for  the presence of a pole. We confirm this view by a full numerical integration of \eq{FlowEvenOdd} at the fixed point, which is unproblematic due to the absence of  removable singularities. Hence, it appears that the fixed point solution does not extend beyond dimensionless background curvatures of order unity in the present approximation.\footnote{Due to this divergence we can neither continue the integration on the negative nor on the positive side in \fig{fig:fp027_FixPointFunc}. This is due to how the differential equations are integrated using two functions $f$ and $z$ depending on the square of the Ricci scalar curvature. Each function on its own is independent of the sign of the scalar curvature and takes the same value for positive and negative curvature. Only the sum of them which enters the effective action may differ on the positive and the negative side.}
It is interesting to note that this pole is not generated by a pole in the differential equation. Instead, it is of the
Landau type and generated dynamically similarly to Landau poles in perturbative $\beta$-functions, \textit{e.g.} QED.
Since we expect the fixed point functional to be well-behaved for all dimensionless Ricci scalar, this result suggests that possibly new effects must arise due to further curvature invariants which ensure that the effective action extends to larger curvature (recall that our study employs the heat kernel expansion which is well-suited for small curvature). 
Similar results  and conclusions apply for the fixed point $\text{FP}_3$.

Next, we search for de Sitter solutions in the deep UV.  With increasing curvature, the fixed point functionals essentially behave linearly up to the end of the radius of convergence, see \fig{fig:fp027_FixPointFunc}. This behaviour, together with the proximity of the fixed point couplings to EH values, suggests that the equations of motion will have de Sitter solutions similar to those of the Einstein-Hilbert approximation. Quantitatively, we find one de-Sitter solution for each of the fixed points $\text{FP}_4$ and $\text{FP}_3$. Their numerical values are given by 
\beq\label{dS-FP4-FP3}
r_{\rm dS}|_{{\rm FP}_4} \approx 0.573\,,
\quad r_{\rm dS}|_{{\rm FP}_3} \approx 0.534\,,
\eeq 
respectively. Both are larger than the EH result \eq{deSitter-EH}, and all three are larger than the de Sitter solution \eq{dS1} of the strongly interacting Riemann fixed point. Retaining only the cosmological constant and the Ricci scalar terms of the fixed point action, the result would have come out as $r_{\rm dS}|_{{\rm FP}_4} \approx 0.58$
and $ r_{\rm dS}|_{{\rm FP}_3} \approx0.56$. Here we used  \eq{FP4-data}, \eq{FP3-data}, and the fact that $r_{\rm dS}=4\lambda$ in any Einstein-Hilbert theory. The result strengthens the view that the $\text{FP}_4$ and $\text{FP}_3$ fixed points are weakly coupled, and only differ mildly from the Reuter fixed point.

\bfi
	\includegraphics[width=.6\linewidth]{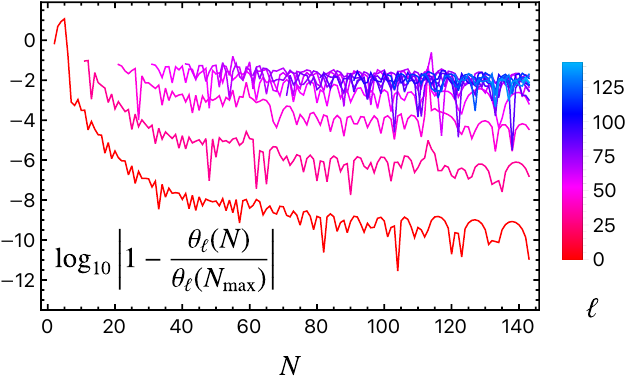}
	\caption{The accuracy of  (the real part of) eigenvalues of FP$_4$ at different orders $N$, in comparison to their values at $N_{\text{max}} = 144$. For better visualisation we only show the convergence of every $10$th eigenvalue. Unlike FP$_{4s}$ (Fig.~\ref{fig:fp05522eigenvalueconvergence}) the accuracy ceases to improve and exponents do not converge beyond a certain order.}
	\label{fig:fp027_EigenvalueConvergence}
\efi

\subsection{Convergence}
\label{sct:weakFpconvergence}
In \fig{fig:fp027_EigenvalueConvergence}, we show the convergence of eigenvalues towards their values at approximation order $N_\text{max} = 144$. For the Riemann fixed point, all eigenvalues converge algebraically fast (\fig{fig:fp05522eigenvalueconvergence}). However, this is not observed for $\text{FP}_4$. Rather, a fast and constant rate of convergence is observed for the first 20 orders in the expansion, reaching up to six significant digits in the exponents. With increasing order, however, the rate begins to flatten and the accuracy stops increasing around the order $N\approx 60-100$. 
Specifically, the findings show  that  the scaling exponent $\theta_0$ cannot be determined beyond an accuracy of $10^{-8}$, and increasingly less so for the higher order eigenvalues. 
A similar behaviour is found for the fixed point couplings of the weakly Riemann fixed points $\text{FP}_4$ and $\text{FP}_3$, and for the scaling exponents of the fixed point $\text{FP}_3$.

The  slight deterioration of convergence towards the highest orders is further corroborated in \fig{fig:fp027_GaussianApproach}, where we show  the difference of the real part of the eigenvalues and their classical scaling at different orders $N$. This should be compared with the result at the Riemann fixed point  (see \fig{fig:fp4smallgaussappro}).  Overall, eigenvalues approach Gaussian values with small deviations, as seen for the Riemann fixed point. However, in contrast to \fig{fig:fp4smallgaussappro}, this plot shows that the convergence of increasingly irrelevant eigenvalues slowly deteriorates. While up to $\ell = 30$ an apparent convergence of eigenvalues can be seen, higher eigenvalues continue to strongly fluctuate within a belt around Gaussian scaling,  another consequence of the lack of convergence at highest orders (see \fig{fig:fp027_EigenvalueConvergence}).  Similar results are found for $\text{FP}_3$.  In summary, we are lead to the conclusion that a full determination of the fixed point and its scaling exponents cannot be achieved beyond a limiting accuracy within the local curvature expansion of the effective action performed here. However, since the effect is numerically small, and most likely below systematic  errors
due to approximations, it may be neglected for the present purposes.

It is conceivable that the slight lack of convergence is a technical artifact, which can be overcome by an extended study using different momentum cutoffs, or by using spectral sums instead of heat kernel expansions. In functional RG studies of critical phenomena it has been noted that   certain technical choices may hamper the accurate determination of scaling exponents due to convergence-limiting poles in the complex plane  \cite{Morris:1994ki}. Still, the apparent lack of convergence can be overcome by taking constant field backgrounds \cite{Aoki:1998um,Litim:2001dt}, by using more suitable cutoffs \cite{Litim:2002cf}, or by dropping the polynomial approximation in the first place \cite{Morris:1994ki}. These extensions, albeit interesting, are beyond the scope of the present work.

\bfi
	\includegraphics[width=.45\linewidth]{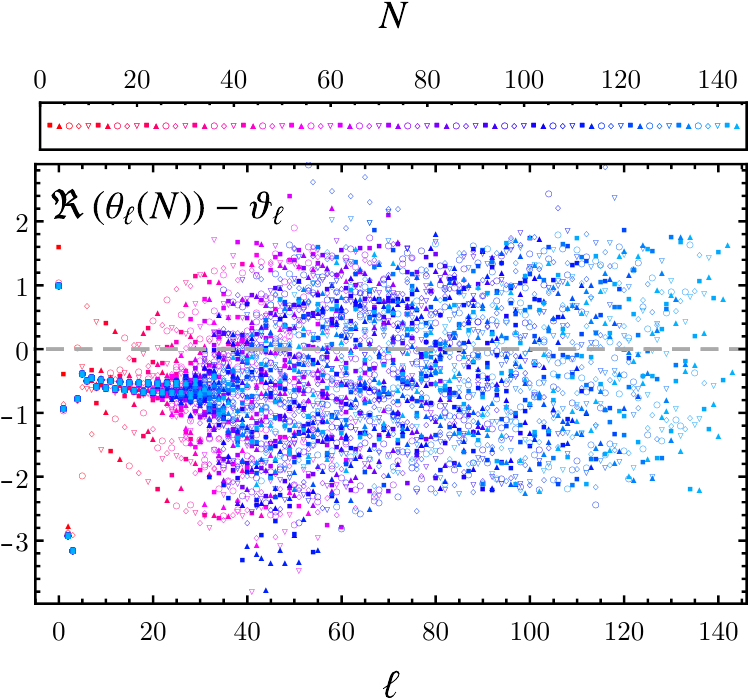}
	\includegraphics[width=.5\linewidth]{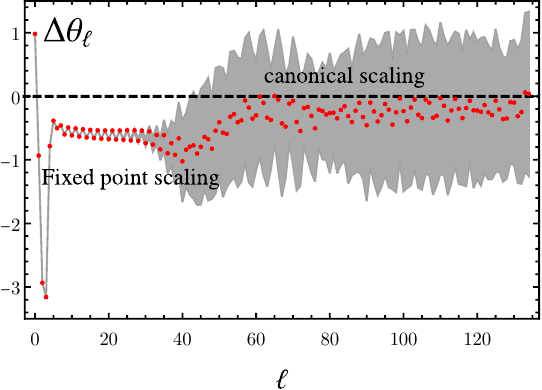}
	\caption{The difference between the real part of the universal eigenvalues $\theta_\ell (N)$ of FP$_4$ and their canonical values $\vartheta_\ell$ is shown. In the left plot, we show every second order $N$ up to $N = 144$, using different symbols and colour coding to distinguish different orders. We observe that higher order eigenvalues do not converge but fluctuate within a belt around their canonical values. This is further clarified in the plot on the right-hand side which shows averaged values for the eigenvalues of the last $10$ orders together with a band of $1\sigma$ standard deviation resulting from the averaging.}
	\label{fig:fp027_GaussianApproach}
\efi

\section{\bf Eigenvectors and Equal Weight Condition}\label{misc}
In this section, we summarise results for eigenvectors and eigenperturbations.
As a main novelty we put forward an ``equal weight'' condition which allows the identification of eigenperturbations irrespective of overall parametrisation ambiguities of couplings in the effective action.

\subsection{Eigenvectors}\label{EVs1}
The goal of analysing eigenvectors
is to clarify which invariants in the effective action relate to which eigenvalues, and to understand which interaction monomials dominate the relevant eigendirections. We expect that there will not be a one-to-one correspondence between an eigenvector and  monomials in the effective action. Rather, different operators in the effective action mix and as a result eigenvectors often do not point into the direction of an individual monomial only.

\begin{figure}
	\includegraphics[width=.9\linewidth]{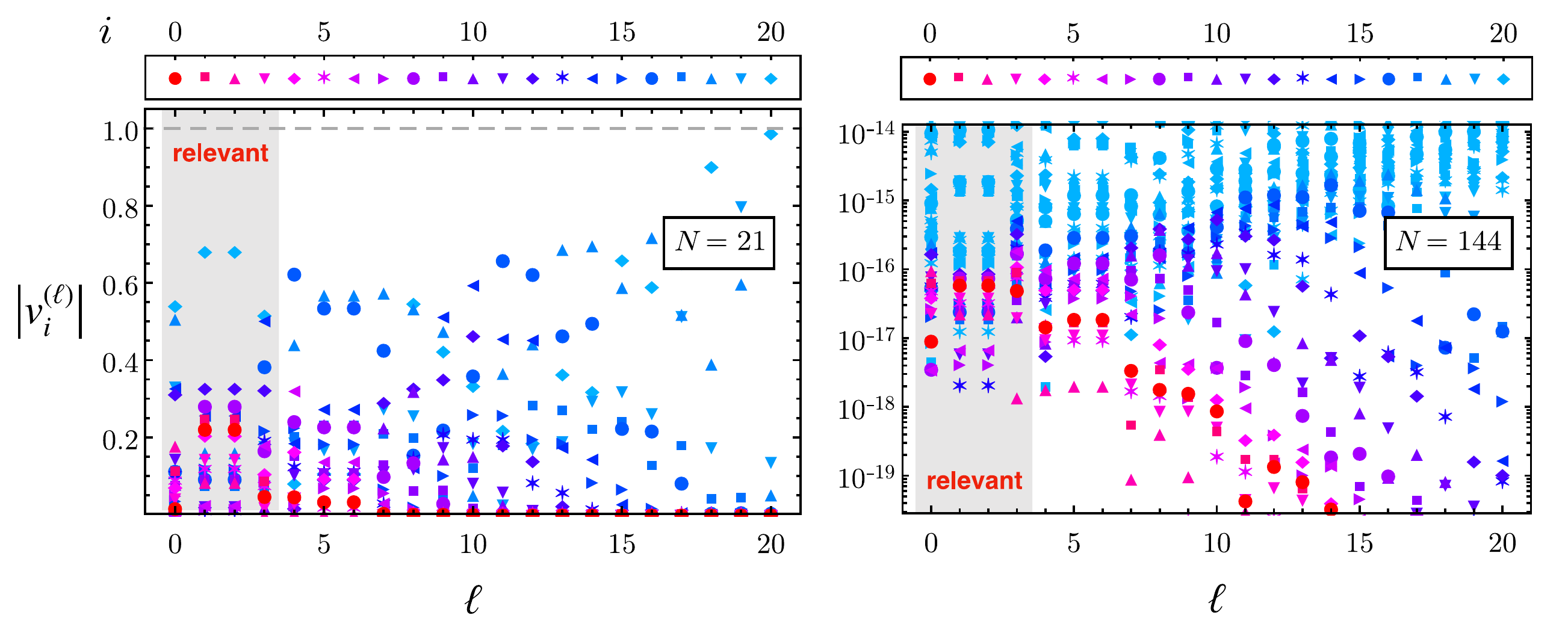}
	\caption{Shown are the first 21   ``ad hoc'' eigenvectors $\bm{v}^{(\ell)}$ $(\ell=0,\cdots, 20$) at the Riemann fixed point (absolute values of their components), normalised to equal length  \eq{eqn:evnorm}, also  comparing  polynomial approximations at order $N =21$ (left panel) with $N = N_\text{max} =144$ (right panel).  The  components $(i=0,\cdots,N - 1$) of eigenvectors representing~the cosmological constant ($i=0)$, the Ricci scalar $(i=1)$, ${\rm Riem}^2$ $(i=2)$, and higher order interaction monomials $(i\ge 2)$ are indicated by different symbols and colours according to the  legend. Gray-shaded areas indicate the set of relevant eigenvectors. The right panel also  contains symbols corresponding to invariants higher than those indicated in the legend, represented by additional light blue symbols.   The ad hoc eigenvectors are spuriously  dominated by the highest interaction monomials in any approximation which makes them unreliable.}
	\label{eigvnotrescaled}
\end{figure}

To begin with, we define the eigensystem $\{(\theta_\ell, \bm{v^{(\ell)}}),\, \ell=0,\cdots,N-1\} $ of eigenvalues $\theta_\ell$ and eigenvectors $\bm{v^{(\ell)}}$ at approximation order $N$ via the stability matrix at the fixed point, $M_{ij}=\partial\beta_i/\partial g_j |_*$,
\beq\label{eigensystem}
	\sum_j M_{ij}
	\, v_j^{(\ell)} = \theta_\ell \, v_i^{(\ell)} \, ,
\eeq
where we use $\ell$ to label different eigenvalues $\theta_\ell $ and their corresponding eigenvectors $\bm{v^{(\ell)}}$. Having calculated this eigensystem for the Riemann fixed point at different orders, we plot the absolute values of the components $i$ for eigenvectors $\bm{v^{(\ell)}}$ at $N = 21$  and $N = N_\text{max} = 144$ in \fig{eigvnotrescaled}. Starting with a discussion of the eigenvectors at $N = 21$, we observe that all eigenvectors appear  to be dominated by classically irrelevant interactions: The largest components of all eigenvectors point into the directions of the five highest included operators in the truncation. In contrast to this, other directions corresponding to operators with smaller mass dimension are suppressed in all eigenvectors. However, taking a closer look we see that the components corresponding to operators of small mass dimension are significantly enhanced in the relevant eigenvectors. Moreover, the most irrelevant eigenvectors only get significant contributions from the classically most irrelevant eigenvectors in the truncation. 

Qualitatively similar features can be seen in the eigenspectrum at $N = 144$ as well. Again,  the whole eigenspectrum appears to be dominated by the classically most irrelevant directions included in the truncation. In particular, relevant eigenvectors are dominated by the highest operators included in the truncation and classically relevant directions contribute significantly more to relevant eigenvectors than irrelevant ones. However, note that the quantitative values have changed drastically by going from $N = 21$ to $N = 144$. While the classically relevant directions had contributions of the order of $0.1$ in the first case, these contributions are suppressed to orders of $10^{-16}$ in the latter. Further, all operators which are included at $N = 21$ turn out to be strongly suppressed by higher operators which are additionally included at $N = 144$.

Some of the features seen in \fig{eigvnotrescaled}   are  reasonable ($e.g.$~the fact that classically relevant directions contribute significantly more into the directions of relevant eigenvectors) but others are clearly not ($e.g.$~the fact that all eigenvectors seem to be dominated by the classically most irrelevant operators).  Moreover, the findings do not reflect the order-by-order stability of fixed point couplings and eigenvalues.

\subsection{Equal Weight Condition}\label{EVs}
We are now going to make the case that the eigenvectors in  \fig{eigvnotrescaled} do not display universal features of the theory, while a suitably modified version of them does. To appreciate this point we note that eigenvectors, unlike eigenvalues, are not universal quantities in the sense that they change under a redefinition or rescaling of the underlying fields or couplings. For example, if we rescale polynomial couplings as
$\lambda_i \rightarrow \overline{\lambda}_i = \zeta_i\cdot  \lambda_i$ (no sum),
the corresponding $\beta$-functions are given by
$\overline{\beta}_i (\overline{\lambda}_j)=\zeta_i \cdot\beta_i(\overline{\lambda}_j/\zeta_j)$.  It then follows that the elements of the stability matrix of the rescaled system are given by
\beq
	\overline{M}_{ij}
	= {\zeta_i} \cdot M_{ij}\cdot ({\zeta_j})^{-1}\,. 
\eeq
While eigenvalues remain unchanged, eigenvectors do change, as can be seen from the relationship between eigenvectors before and after rescaling,
\beq
	\overline{v}_i^{(\ell)} = \zeta_i \,v_i^{(\ell)} \,.
	\label{eqn:evrescaling}
\eeq
In general, the normalisation of an eigenvector changes under rescaling of couplings, and we  may introduce rescaled eigenvectors,  normalised to unity, as
\beq
	\overline{\omega}_i^{(\ell)} = \frac{\overline{v}_i^{(\ell)}}{\sqrt{\sum_{j = 0}^{N - 1} \left| \overline{v}_{j}^{(\ell)} \right|^2}} = \frac{\zeta_i v_i^{(\ell)}}{\sqrt{\sum_{j = 0}^{N - 1} \left| \zeta_j v_j^{(\ell)} \right|^2}} \,.
	\label{eqn:evomegadef}
\eeq
 A remnant of this  "rescaling ambiguity" is visible in \fig{eigvnotrescaled}, where the entire eigenspectrum appears to be dominated by the highest operators retained in the approximation. As is evident from  \eq{eqn:evrescaling} with suitable choices for the rescaling parameters $\zeta_i$ any operator can be made to appear dominating the  eigenspectrum.

To remedy this ambiguity, we  lay out a simple procedure to find eigenvectors which are best qualified to describe the physical eigenperturbations without any  rescaling ambiguities. Since we are only interested in the physical information of the eigensystem, we  fix the rescaling ambiguity  such that all operators contribute with  "equal weight" to the entire eigensystem. This idea ensures that no operator is   dominating artificially in the spectrum, and should leave us directly with the physical information stored in the eigenvectors. On a quantitative level, this amounts to the "equal weight" condition for the rescaling parameters $\zeta_i$,
\beq
	\sum_{\ell = 0}^{N - 1} \left| \overline{\omega}_i^{(\ell)} \right|^2 = 1 \qquad \forall \,\, i \,.
	\label{eqn:compnorm}
\eeq
Note that this equation sums over different vectors $\bm{\overline{\omega}}^{(\ell)}$ with the components $i$ held fixed. In physical terms, the condition  \eq{eqn:compnorm} imposes that the absolute-value-squared contributions of any operator to all eigenvectors adds up to unity. In this manner, it is guaranteed that the contributions to eigenvectors of all operators retained in the approximation are weighted equally.

In addition, recall that the $\overline{\omega}_i^{(\ell)}$ are defined such that they also fulfil the conventional  "unit length" normalisation condition for eigenvectors,
\beq
	\sum_{i = 0}^{N - 1} \left| \overline{\omega}_i^{(\ell)} \right|^2 = 1 \qquad \forall \,\, \ell \, ,
	\label{eqn:evnorm}
\eeq
which sums over different components $i$. Thus, considering the matrix of eigenvectors,
\beq
	\overline{\omega} = 
	\left( \begin{matrix}
		\overline{\omega}_0^{(0)} & \overline{\omega}_0^{(1)} & \dots & \overline{\omega}_0^{(N - 1)} \\
		\overline{\omega}_1^{(0)} & \overline{\omega}_1^{(1)} & \dots & \overline{\omega}_1^{(N - 1)} \\
		\vdots & \vdots & \ddots & \vdots \\
		\overline{\omega}_{N - 1}^{(0)} & \dots & \dots & \overline{\omega}_{N - 1}^{(N - 1)}
	\end{matrix} \right) \, ,
	\label{eqn:omegamatrix}
\eeq
the normalisation conditions \eq{eqn:compnorm} and \eq{eqn:evnorm} correspond to normalising the absolute-value-squared sum of each row and each column to unity.

Next, we show that the equal weight \eq{eqn:compnorm} and unit length condition \eq{eqn:evnorm}  for the matrix \eq{eqn:omegamatrix}  can always be achieved. The condition \eq{eqn:evnorm} follows trivially from the normalisability of eigenvectors. The condition \eq{eqn:compnorm}, on the other hand, requires a special choice for the rescaling parameters $\zeta_i$, which can always be achieved.
To see this, we write \eq{eqn:compnorm} in terms of \eq{eqn:evomegadef} to get
\beq
	\sum_{\ell = 0}^{N - 1} \left| \overline{\omega}_i^{(\ell)} \right|^2 = \sum_{\ell = 0}^{N - 1} \left| \zeta_i \frac{v_i^{(\ell)}}{\sqrt{ \sum_{j = 0}^{N - 1} \left| \zeta_j v_j^{(\ell)} \right|^2}} \right|^2 = 1 \qquad \forall \,\, i \, .
	\label{eqn:compnormrescaled}
\eeq
Notice that due to the normalisation of eigenvectors, an overall rescaling factor,
\beq
	\zeta_i \rightarrow \text{const.} \,\, \zeta_i \, ,
\eeq
drops out of \eq{eqn:compnormrescaled} leaving us with only $N - 1$ open parameters. Due to the irrelevancy of this overall factor, we can always choose the rescaling factors to be of the form
\beq
	\bm{\zeta} =
	\left( \begin{matrix}
	1 \\
	\zeta_1 \\
	\zeta_2 \\
	\vdots \\
	\zeta_{N - 1}
	\end{matrix} \right) \, .
	\label{eqn:zetanormchoice}
\eeq
With this choice, we can use the $N - 1$ open parameters of \eq{eqn:zetanormchoice} to fulfil \eq{eqn:compnormrescaled} for all $i > 0$. Doing so, in the matrix of eigenvectors \eq{eqn:omegamatrix} this corresponds to having all columns and all rows except for the first row normalised to $1$. From this it follows, however, that
\beq
	\begin{split}
		\sum_{l = 0}^{N - 1} \left| \overline{\omega}_0^{(l)} \right|^2 =& \sum_{l = 0}^{N - 1} \left( 1 - \sum_{i = 1}^{N - 1} \left| \overline{\omega}_i^{(l)} \right|^2 \right) 
		=N - \sum_{i = 1}^{N - 1} 1 
		= 1 \, .
	\end{split}
\eeq
Thus, fulfilling \eq{eqn:compnormrescaled} for $i > 0$ and normalising all eigenvectors automatically ensures that \eq{eqn:compnormrescaled} is fulfilled for $i = 0$ as well. It is, therefore, always possible to rescale the eigenvectors such that the naturalness condition \eq{eqn:compnorm} is fulfilled.

\subsection{Eigenvectors from Riemann Interactions}\label{EV-Riemann}
\begin{figure}
	\includegraphics[width=.9\linewidth]{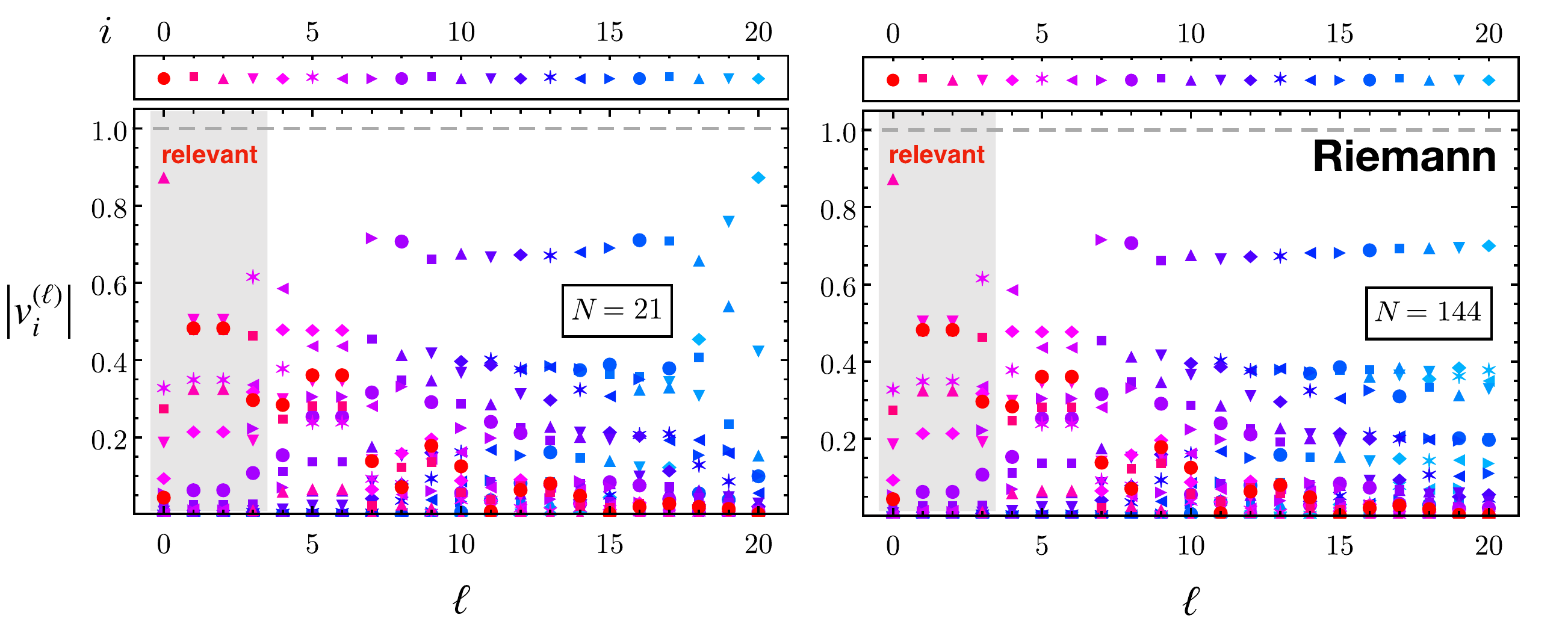}
	\caption{
	Shown are the first 21  eigenvectors at the Riemann fixed point FP${}_{4s}$ at approximation order $N =21$ (left panel) and $N = N_\text{max} =144$ (right panel),  normalised according to the equal weight    and  unit length conditions  \eq{eqn:compnorm}, \eq{eqn:evnorm}. Unlike in \fig{eigvnotrescaled}, 
	normalised eigenvectors only change mildly under extensions of  approximations, in accord with the stability and convergence of approximations (\fig{fig:fp05522couplingsstandard}), and the near-Gaussianity of scaling exponents (Figs.~\ref{fig:fp05522BootStrap} and~\ref{fig:fp4Gaussian}).
	Gray-shaded areas indicate the set of relevant eigenvectors.  The leading  eigenvectors  $(\ell\lesssim 6)$ are dominated by mixtures of the relevant interaction monomials while increasingly irrelevant eigenvectors $\bm v^{(\ell)}$ $(\ell\gtrsim 7)$ are dominated by the component $i=\ell$. }
	\label{eigvrescaled}
\end{figure}

In \fig{eigvrescaled}, we apply our procedure to the theory at hand and show  eigenvectors  rescaled according to  \eq{eqn:compnorm} and \eq{eqn:evnorm}, and for different truncation orders $N = 21$ and $N = N_\text{max} = 144$. We emphasise that our procedure has been applied 
to the full eigenspectrum at order $N = 21$ and $N = 144$, even though only the first $21$ eigenvectors are shown in either case.
The results in  \fig{eigvrescaled}  should now be compared with \fig{eigvnotrescaled}, where several points are worth noting. First, with the equal weight normalisation, eigenvectors are now  seen to be stable from order to order in the approximation, including up to the highest order. This is consistent with the order-by-order stability of fixed points and eigenvalues, but much unlike the pattern observed  in \fig{eigvnotrescaled}. Second, we observe that eigenvectors associated to the relevant eigenvalues are dominated by linear combinations of the leading few operators with small mass dimensions. The result reflects the fact that interactions induce strong correlations amongst the leading few interaction terms, implying that linear combinations of them scale with identical exponents (see $e.g.$~Tab.~\ref{tab:fp05522eigvlow}). Third, we also observe that  increasingly irrelevant eigenvalues $\theta_\ell$ become dominated by a single interaction monomial  with canonical mass dimension $2\ell-4$, with other monomials only providing subleading corrections. This result reflects the near-Gaussian scaling of operators with increasing mass dimension observed in the previous section (\sct{sec:boot}). We conclude that key features of the eigensystem have become visible thanks to a suitable normalisation of eigenvectors.

We may now take a closer look at \fig{eigvrescaled} to understand which interaction terms are responsible for the four-dimensionality of the UV critical surface. To that end, we divide the set of eigenvectors into subsets of strongly  and weakly correlated eigenvectors.  The set of weakly correlated eigenvectors contains those  where each eigenvector is approximately dominated by a single operator in the effective action. 
If so, it means that quantum corrections are small, only introducing mild mixing, and a rough correspondence between eigenvectors and field monomials can be established. On the other hand, the set of strongly correlated eigenvectors are those which show a strong mixing between  interaction monomials. Here, quantum effects are strong, and an association of an eigendirection with a unique interaction monomial is no longer possible. 

In \fig{eigvrescaled}, the first subset contains the eigenvectors $\bm{v}^{(\ell)}$ with $\ell \leq 6$ while the second one  consists of those with $\ell > 6$.  Specifically, the vacuum energy contribution to the most relevant eigenvector $\bm{v}^{(0)}$  is suppressed, despite of being the operator of lowest canonical mass dimension in the effective action. Rather, the eigenvector $\bm{v}^{(0)}$ is dominated by the classically marginal operator $\int\sqrt{g}\, {\rm Riem}^2$, modulo corrections from other leading interaction terms. Similarly, the complex conjugate pair of eigenvalues corresponding to $\bm{v}^{(1)}$ and $\bm{v}^{(2)}$ have leading contributions from the vacuum energy and the $\int\sqrt{g} R\cdot {\rm Riem}^2$ interaction, while 
 $\bm{v}^{(4)}$ gets its largest contribution from the $\int\sqrt{g} R\cdot {\rm Riem}^4$ interaction which, classically, is irrelevant  with eigenvalue $\vartheta_{5,\rm cl.} = 6$. Hence, the naive expectation that the four most relevant eigenvectors correspond to the four classically most relevant operators does not hold true. Rather, all eigenvectors in the first part of the plot (all eigenvectors with $\ell \leq 6$) get  contributions due to mixing, which, moreover, can be large quantitatively. We conclude  that the interaction terms responsible for the four-dimensional UV critical surface are not only the four classically most relevant operators. Rather,  the mixing between  operators up to and including $\int\sqrt{g}\,  {\rm Riem}^6$ is responsible for the four relevant eigenvalues. This matches the  observation that the fixed point converges poorly (with standard boundary conditions) at low approximation orders up to and including $N = 6$ (see~Tab.~\ref{tab:fp05522eigvlow}), which is now understood as a remnant of strong correlations amongst eigenperturbations.

\begin{figure}
	\includegraphics[width=.9\linewidth]{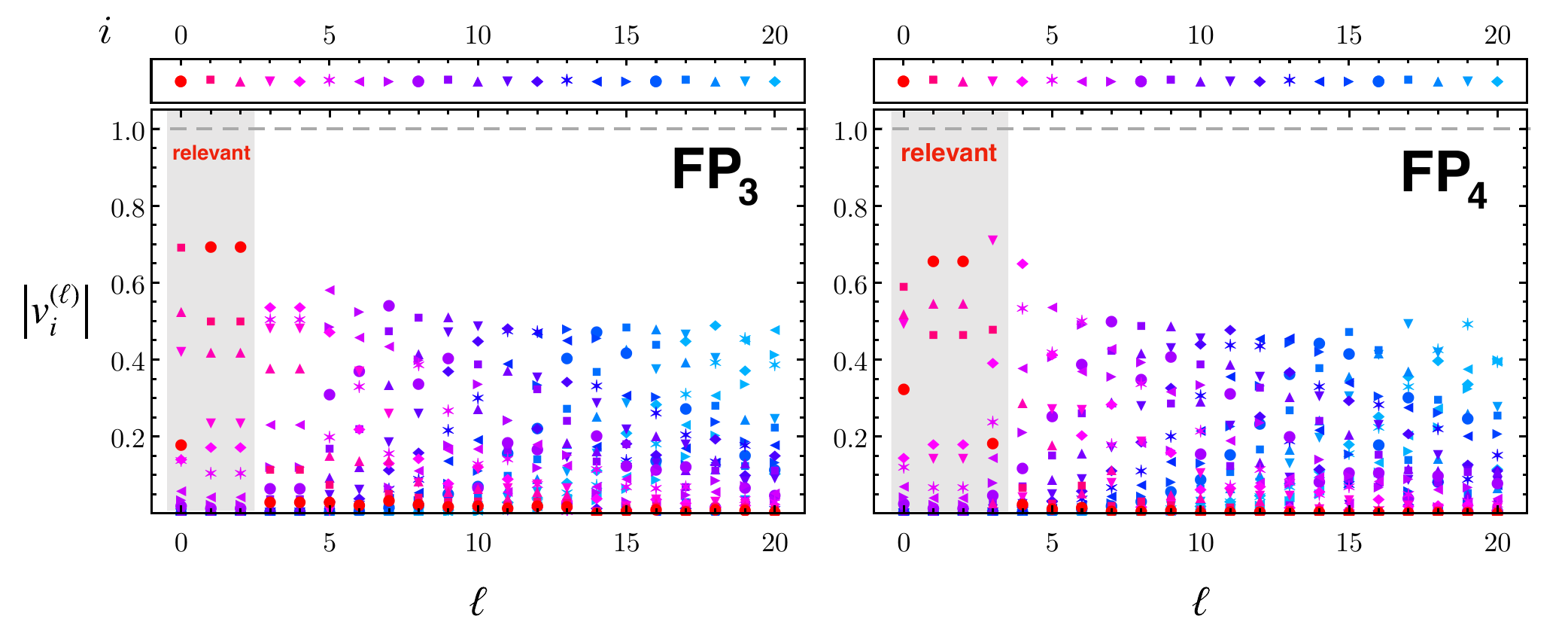}
	\caption{
	Shown are the components of first 21  eigenvectors $\bm{v}^{(\ell)}$ of the secondary Riemann fixed points $\text{FP}_3$ (left panel) and $\text{FP}_4$ (right panel) 
	normalised according to the equal weight   and unit length condition,  \eq{eqn:compnorm} and \eq{eqn:evnorm}. Gray-shaded areas indicate the set of relevant eigenvectors.}
	\label{eigvrescaled_secondary}
\end{figure}

The same analysis of eigenvectors can now be done for the secondary fixed points ${\rm FP}_4$ and ${\rm FP}_3$. Without using the equal weight condition, the resulting plots show qualitatively the same picture as \fig{eigvnotrescaled} for the primary fixed point: The ad hoc eigenvectors spuriously appear to be dominated by the canonically most irrelevant operators in the spectrum with the canonically most relevant operators being enhanced in the relevant directions. 

Using the equal weight condition, on the other hand, a stable and transparent picture becomes visible. Our results are shown in \fig{eigvrescaled_secondary} where the first $21$ eigenvectors calculated at the highest approximation order ($N_{\rm max}=72$ for $\text{FP}_3$, and $N_{\rm max}=144$ for $\text{FP}_4$) are shown. 
In either case, the relevant eigenvectors (grey shaded area)  are dominated by leading operators of small mass dimension -- the cosmological constant, the Ricci scalar, and the ${\rm Riem}^2$ interaction. For ${\rm FP}_4$, we also notice that the operator $\int\sqrt{g}\ R\cdot{\rm Riem}^2$  contributes sizeably to the relevant eigenvectors. In turn, irrelevant eigenvectors are dominated by canonically irrelevant operators. In comparison with  the primary fixed points \fig{eigvrescaled} we note that the mixing between different operators in relevant directions is smaller. This is balanced by a stronger mixing of operators in the irrelevant eigenvectors. The one-to-one correspondence observed earlier is much weaker and there is a shift between irrelevant eigenvectors and their corresponding operators. Due to the weaker mixing in relevant directions, the correspondence between relevant directions and operators is clearer and we expect that the relevant directions are created by the operators up to and including $\int\sqrt{g}\, {\rm Riem}^4$. In comparison with the primary fixed point, the canonically marginal interaction $ \sim{\rm Riem}^2$ now contributes much less to the leading eigenvalue.
The three most relevant directions have eigenvalues of roughly the same magnitude and arise from linear combinations of the cosmological constant, the Ricci scalar and the ${\rm Riem}^2$ interactions.
For all three of these fixed points, it can be seen that the second and third most relevant eigenvectors get substantially large contributions from the cosmological constant.

We also observe from  \fig{eigvrescaled_secondary} 
that the increasingly irrelevant eigenvectors $\bm{v}^{(\ell)}$ are approximately dominated by the  interaction monomial   $\sim ({\rm Riem}^2)^{i/2}$ for even $i$ and by terms $\sim R\cdot ({\rm Riem}^2)^{(i-1)/2}$ for odd $i$, with $i\approx \ell + 2$ in the range covered by   \fig{eigvrescaled_secondary}. The shift $\ell-i$ increases further with increasing $i$ beyond $i=21$.
This observation applies to both $\text{FP}_4$ and  $\text{FP}_3$, with other monomials only adding subleading corrections. 
The result indicates, once more, that the UV scaling of higher curvature interactions is non-classical despite of being near-Gaussian.

\section{\bf Discussion}
\label{sec:discus}
In this section, we compare our results from Riemann tensor interactions with earlier higher order studies involving Ricci scalar and Ricci tensor interactions. We also discuss the impact of higher order curvature invariants on fixed point couplings and a scale invariant notion for the interaction strength, and on vacuum solutions to the quantum equations of motion. 

\subsection{Ricci vs Riemann Interactions}
We begin with a  comparison  of results of  $f(R, \text{Riem}^2)$-type actions studied in  this work  with results from $f(R, \text{Ric}^2)$ and $f(R)$ actions, summarised in~\tab{tCompare}.  A first important joint feature is the  near-Gaussianity of high order eigenvalues in all three settings. Even though other aspects differ from each other, the near-Gaussian scaling of the eigenvalues seems to be hard-wired in all of them. This can be seen as an indication that the gravitational fixed point shows signatures of weak couplings as the classical eigenvalues are only corrected by small quantum corrections, especially regarding high eigenvalues. Since this feature is present in all three  settings we have reason to believe that only a finite number of eigenvalues will become relevant at an interacting gravitational fixed point as the quantum corrections are not expected to be big enough to render arbitrary classically irrelevant eigenvalues relevant.

Apart from this feature, the $f(R, \text{Riem}^2)$ and  $f(R, \text{Ric}^2)$ fixed points  show further similarities. After initial fluctuations at low approximation orders both truncations show a fast convergence in the fixed point couplings and eigenvalues. The radius of polynomial convergence is maximal in the $\text{Ric}^2$ case and almost maximal for the polynomial solution of $\text{Riem}^2$. In the latter setting, however, it becomes maximal using Pad\'e resummation or numerical integration. Further, both settings have two de Sitter solutions. The AdS solution apparent in $\text{Ric}^2$ is, however, missing in $\text{Riem}^2$. In contrast, $f(R)$-type theories are  different in these aspects. The polynomial radius of convergence is only half of the maximal possible radius, and the convergence of couplings and eigenvalues is rather slow. dS and AdS solutions can both be found in $f(R)$ quantum gravity.

Further differences appear in the spectra of eigenperturbations. Most notably, we have indications that operators including the square of the Riemann tensor induce a relevant fourth direction to the UV critical surface. This is a very important novelty, hinting at the relevancy of Riemann tensor interactions. Moreover, we also observe that only two pairs of eigenvalues in $f(R, \text{Riem}^2)$-type theories arise as complex conjugates (Fig.~\ref{fig:fp05522eigvstandard}). This is quite different from $f(R)$ and $f(R, \text{Ric}^2)$-type models where  complex conjugate pairs continue to arise with increasing approximation orders. In general, complex scaling exponents indicate a degeneracy \cite{Falls:2014tra}, meaning that the RG scaling of two sets of eigenoperators has become identical up to a phase. Including more interaction terms should lift the degeneracy, and eigenvalues in the full physical theory are expected to be real. It is noteworthy that the Riemann tensor interactions  lift the hitherto  observed degeneracy  nearly entirely, accept for a pair each of  relevant and irrelevant eigenperturbations \eq{eqn:fp05522eigvfinal}.

\begin{table*}
		\addtolength{\tabcolsep}{3pt}
		\setlength{\extrarowheight}{3pt}
\normalsize
\scalebox{0.8}{
\begin{tabular}{`cccc`} 
\toprule
\rowcolor{Yellow}
\bf Action&$\bm{f(R)}$
&{$\bm{f(R, {\rm Ric}^2)}$} 
&{$\bm{f(R, {\rm Riem}^2)}$} \\ 
\midrule
Parameter $(a,b,c)$
&$(1,0,0)$& $(0,1,0)$&(0,0,1)\\
\rowcolor{LightGray}
Bootstrap test&Yes 
&Yes&Yes\\
&3-dimensional&$ {}\quad $3-dimensional${}\quad$  &4-dimensional\\
\multirow{-2}{*}{Critical surface}
&$\ \ \{\Lambda,R,R^2\}$  &$\{\Lambda,R,{\rm Ric}^2\}$ 
&$\{\Lambda,R,{\rm Riem}^2,R\cdot{\rm Riem}^2\}$\\
\rowcolor{LightGray}
Near-Gaussianity&Yes
&Yes&Yes\\
&Moderate
&Maximal&Large\\
\multirow{-2}{*}{Radius}&$(R_c/R_{\rm max}\approx 0.41\ldots0.45)$&$(R_c/R_{\rm max}=1)$&$(R_c/R_{\rm max}\approx 0.90\ldots 0.95)$\\
\rowcolor{LightGray}
Convergence&Slow&Fast&Fast\\
Recursive relations&Yes (level 2)  \cite{Falls:2013bv,Falls:2014tra} &Yes  (level 3)&Yes (level 3)\\
\rowcolor{LightGray}
de Sitter &No \cite{Falls:2016wsa} Yes \cite{Falls:2018ylp}&Yes \cite{Falls:2017lst}&Yes\\
AdS&Yes\ \ \ \cite{Falls:2016wsa}&Yes&No\\
\bottomrule
\end{tabular}  
}
\caption{\label{tCompare} Comparison of results with  $f(R)$ quantum gravity  \cite{Falls:2013bv,Falls:2014tra,Falls:2018ylp} (2nd column) and  $f(R, {\rm Ric}^2)$-type quantum gravity  \cite{Falls:2017lst} (3rd column)  and $f(R, {\rm Riem}^2)$-type quantum gravity  (this work, 4th column) in an otherwise identical setup (same gauge fixing, momentum cutoffs, and heat kernels).}
\end{table*}

Algebraic differences between models arise at the level of recursive relations for couplings. For $f(R)$ theories, these lead to explicit  expressions for couplings in terms of two unknowns which must be determined by other means  (level 2) \cite{Falls:2013bv,Falls:2014tra}. For $f(R, \text{Ric}^2)$ and $f(R, \text{Riem}^2)$-type theories, the presence of 
propagating fourth order degrees of freedom leads to recursive relations involving three unknown parameters (level 3). For comparison, in non-gravitational quantum field theories, recursive relations for fixed point couplings of functional RG flows involve genuinely two unknown parameters (level 2) \cite{Litim:2016hlb}, except in specific limits such as infinitely many matter fields, or for suitably chosen expansion points in field space, where the recursive relations  reduce to one (level 1) or even no (level 0) free parameter; see \cite{Morris:1994ki,Litim:2002cf,Bervillier:2007tc,Marchais:2017jqc,Juttner:2017cpr} for  examples.

We conclude that 
different choices for the parameters $(a, b, c)$ share important universal properties. 
A key novelty has arisen from  Riemann interactions  in  that their dynamics generate a new fundamentally free parameter for the UV critical surface. 

 \begin{figure}
	\includegraphics[width=.5\linewidth]{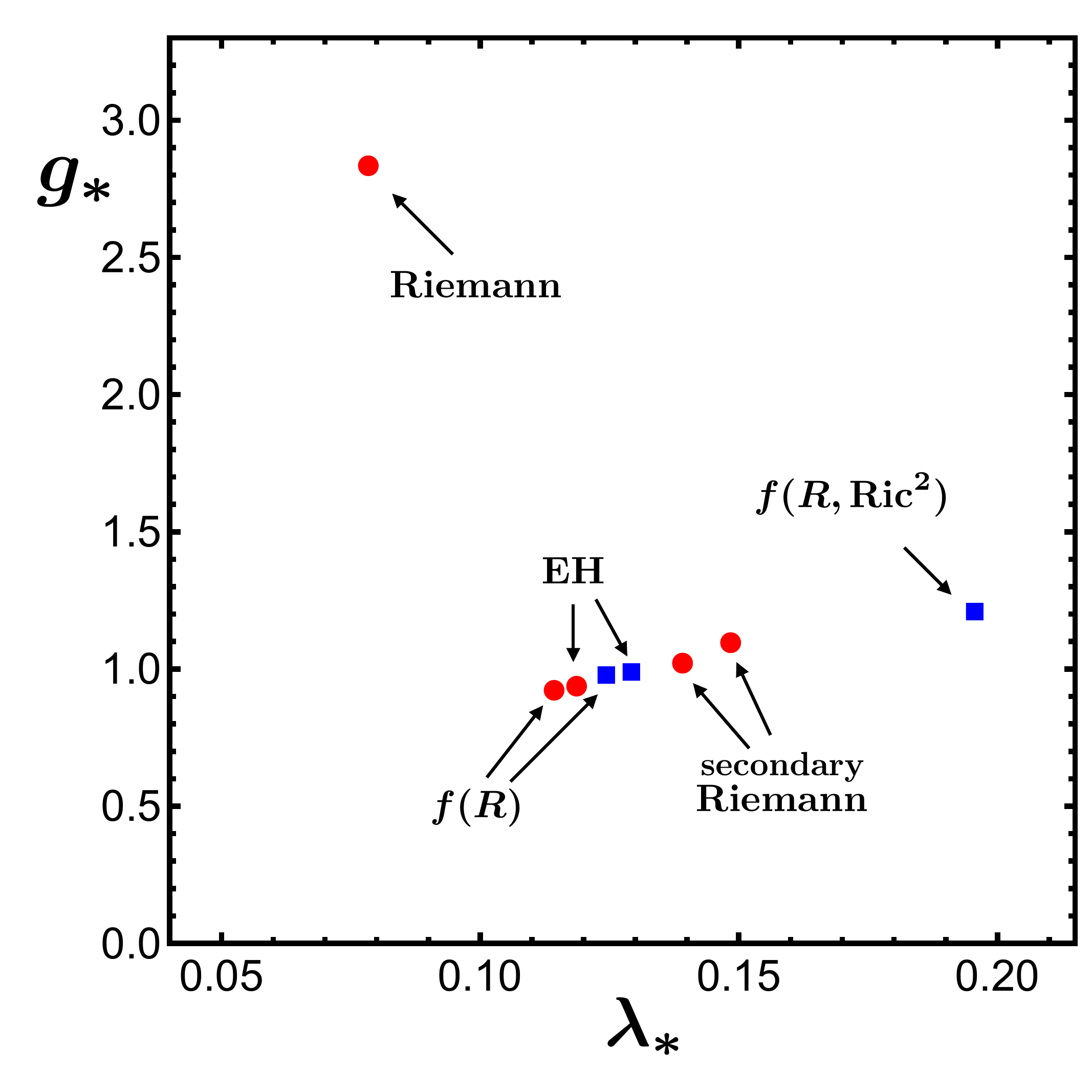}
	\caption{Shown are the values for the cosmological constant  $\lambda_*$ and Newton's coupling $g_*$, comparing the Riemann fixed points (this work) with the Einstein-Hilbert fixed point, the $f(R)$ fixed point \cite{Falls:2013bv,Falls:2014tra,Falls:2018ylp}, and the  Ricci fixed point  \cite{Falls:2017lst}. With all key technical parameters (cutoff scheme, gauge fixing, heat kernels)  the same, minor differences due to the treatment of ghost fields  \cite{Falls:2018ylp} are indicated by red circles (blue squares). We observe that Ricci and Riemann  tensor interactions shift fixed point couplings  away from the Einstein-Hilbert results, and substantially so for the primary Riemann fixed point. In contrast, higher order Ricci scalar interactions only add mild modifications. See also Fig.~\ref{pLaG}.}
	\label{pCompareLaG}
\end{figure}

\subsection{Cosmological Constant and Newton's Coupling}
Next, we discuss more specifically the impact of higher dimensional curvature interactions on the fixed points for Newton's coupling and the cosmological constant.  To that end, Fig.~\ref{pCompareLaG} zooms into Fig.~\ref{scatter} to show the values for the fixed point couplings $(\lambda_*,g_*)$ at the Riemann fixed points, and in comparison to other models and  approximations. 

The colour coding in Fig.~\ref{pCompareLaG} (blue squares vs red dots) distinguishes underlying technical choices.
Results indicated by blue squares  use the heat kernel expansion alongside optimised cutoff functions \cite{Litim:2001up,Litim:2003vp}, and gauge fixing parameters  identical to those first adopted by Codello, Percacci and Rahmede \cite{Codello:2007bd} for polynomial $f(R)$ models. The same technical setting has been used for high order $f(R)$ models in  \cite{Falls:2013bv,Falls:2014tra}. Subsequently, new flow equations have been derived for  $f(R,{\rm Ric}^2)$ models, again using the same technical choices \cite{Falls:2017lst}; see Fig.~\ref{pLaG}.

Results indicated by red dots use the heat kernel expansion with optimised cutoff functions \cite{Litim:2001up,Litim:2003vp}, and the same gauge fixing parameters as in  \cite{Codello:2007bd}, except that ghost fields are now treated differently to ensure that unphysical, convergence-limiting poles of the flow are removed  from the outset \cite{Falls:2018ylp}. The  new treatment of ghosts leads to quantitative, and also qualitative differences. Most notably, $f(R)$ models with improved ghosts display real de Sitter solutions \cite{Falls:2018ylp}. Without improved ghosts, de Sitter solutions disappear narrowly into the complex field plane \cite{Falls:2016wsa}.

A few points are worth emphasizing in Fig.~\ref{pCompareLaG}. We note that  $f(R)$  results for fixed point couplings are very close to those for the Einstein-Hilbert (EH) approximation. Higher order interactions lead to a small reduction of $\lambda_*$ while leaving $g_*$ mostly unchanged. This pattern is neatly visible irrespective of the presence or absence of ghost-induced poles. The small quantitative differences are entirely due to the additional $f(R)$ type interactions beyond the Einstein-Hilbert terms. Turning now to the models with Ricci and Riemann tensor interactions, we notice that the inclusion of Ricci tensors, leads to a more substantial increase of the cosmological constant $\lambda_*$, alongside a  moderate increase in $g_*$. The same holds true at the secondary Riemann fixed points, which also show a  moderate increase in $g_*$ over the Einstein Hilbert result.
On the other hand, the primary Riemann fixed point  shows a substantial increase of $g_*$ and a decrease in $\lambda_*$ over findings in the EH and $f(R)$ approximations. Incidentally, the strong shift in the value for $g_*$ is also at the root for the slow convergence of  the polynomial expansion at the first few orders observed previously (see Tab.~\ref{tab:fp05522couplings}). 
We conclude from Fig.~\ref{pCompareLaG} that theories with Ricci and Riemann tensor interactions lead to more substantial alterations of the fixed point couplings $\lambda_*$ and $g_*$ over the Einstein-Hilbert and $f(R)$ approximations. Hence, the pattern of alterations is quite sensitive to whether the interactions involve Ricci scalars, tensors, or Riemann tensors.

\subsection{Interaction Strength and Perturbativity}\label{ISP}
The gravitational couplings by themselves are not universal quantities at a fixed point. Still, some universal quantities of interest are given by products of couplings which remain invariant under a rescaling of the metric field
\beq\label{rescale}
g_{\mu\nu}\to \ell\,g_{\mu\nu}\,.
\eeq
In our conventions, the gravitational couplings \eq{series} transform as $\lambda_n\to \ell^{4-2n}\lambda_n$ under the rescaling   \eq{rescale}
in four dimensions. Clearly, many  scale invariant products can be formed  out of the fixed point couplings \cite{Falls:2014tra}. Kawai and Ninomiya \cite{Kawai:1989yh} have explained that the scale-invariant product  of the lowest-order couplings $\lambda_0/(2\lambda_1^2)$  is of particular interest in that it serves as an indicator for the effective gravitational coupling strength.  It will then be useful to compare different fixed point theories from the viewpoint of their effective coupling strength. 

\begin{figure}[t]
	\includegraphics[width=.5\linewidth]{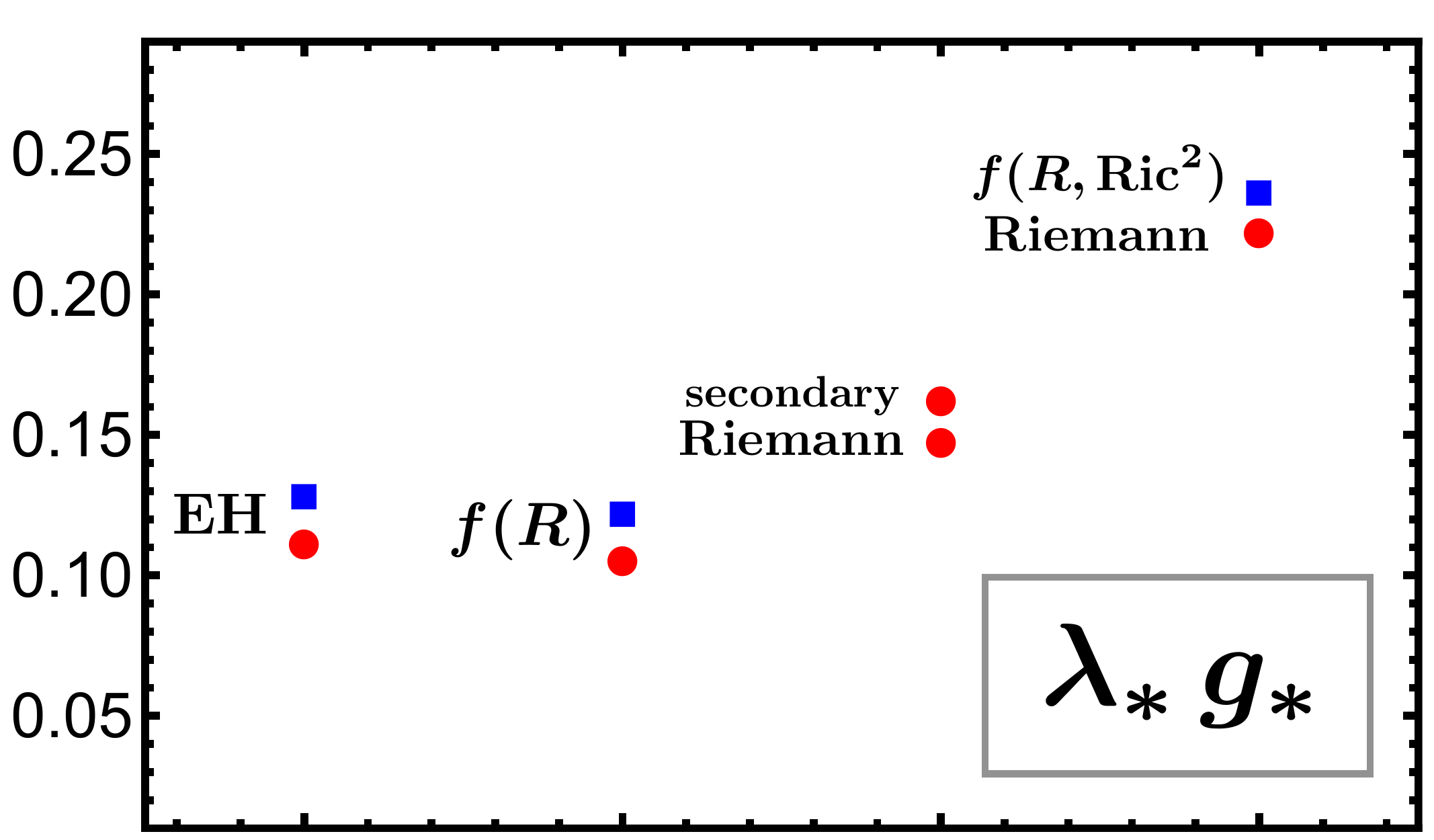}
	\caption{Shown is the effective coupling strength  
	at ultraviolet fixed points, comparing  the Riemann fixed points with the Einstein-Hilbert, the $f(R)$  and the  $f(R,{\rm Ric}^2)$ fixed points;  same colour coding as in Fig.~\ref{pCompareLaG}. Notice the enhancement due to Ricci or Riemann tensor interactions.}
	\label{pCompareLG}
\end{figure}

In our settings, the effective coupling strength  is given by the product $\lambda_*\, g_*$ in terms of the dimensionless cosmological constant and Newton's coupling, see  \eq{gla}. In Fig.~\ref{pCompareLG}, we show our results for the scale-invariant product at the fixed point, comparing different approximations and using the same colour-coding as in Fig.~\ref{pCompareLaG}. In the Einstein-Hilbert approximation, we find 
\beq\label{lag_EH}
(\lambda_*\, g_*)\big|_{{}_{\rm EH}}\approx 0.11 - 0.13\,.
\eeq 
Here, the range indicates the small variations due to different treatment of ghosts.  Moreover, it is also well-known that this product is rather stable under alterations of the RG scheme, the shape of the Wilsonian momentum cutoff, and under changes of gauge fixing parameters \cite{Lauscher:2001ya,Fischer:2006fz,Falls:2014zba}. The smallness of the result may be taken as an indicator for the weakness of gravitational interactions at an asymptotically safe UV fixed point. 
Within $f(R)$ models for quantum gravity,
findings for the effective coupling strength are in the range of
\beq\label{lag_fR}
( \lambda_* \,g_*)\big|_{f(R)}\approx 0.10 - 0.12\,.
\eeq 
The rather small decrease of the effective coupling strength  over \eq{lag_EH} is due to the combined small decrease in $\lambda_*$ and $g_*$.  This is in accord with the picture laid out above, in particular Fig.~\ref{pCompareLaG}. We conclude that the presence of higher order Ricci scalar interactions has only a very mild impact on this universal parameter. This result also confirms the view that the fixed point in $f(R)$ type models of quantum gravity can be interpreted as the extension of the fixed point already observed in the much simpler Einstein-Hilbert approximation.

We now turn to models with Ricci tensor or Riemann tensor interactions. Unlike in models with Ricci scalar interactions only,
these models also feature fourth order propagating degrees of freedom. In comparison with \eq{lag_EH}, \eq{lag_fR}, we observe that the scale-invariant product of couplings is  larger  by a factor of two for $f(R,{\rm Ric}^2)$ models studied in \cite{Falls:2017lst}, 
\beq\label{lag_fRic}
(\lambda_* \,g_*)\big|_{f(R,{\rm Ric}^2)}\approx 0.24\,.
\eeq
This is due to a strong increase in $\lambda_*$ and a small increase in $g_*$, as can be seen from Fig.~\ref{pCompareLaG}. In combination, these alterations  make the effective coupling strength twice as large as in models with Ricci scalar interactions.
Finally, for $f(R,{\rm Riem}^2)$ models studied in this work, we have
\beq\label{lag_fRiem}
(\lambda_* \,g_*)\big|_{\rm Riemann}\approx 0.22\,,
\eeq
which once more is   larger than \eq{lag_EH}, \eq{lag_fR} by a factor of two. Here, the enhancement is due to a substantial  increase in $g_*$ in combination with a small decrease in $\lambda_*$, see  Fig.~\ref{pCompareLaG}.  It is quite intriguing that fixed point theories from Ricci tensor or Riemann tensor interactions both lead to very similar values for the scale-invariant interaction strength (Fig.~\ref{pCompareLG}). Finally, for the secondary Riemann fixed points we have
\beq\label{lag_2nd}
(\lambda_* \,g_*)\big|_{{\rm FP}_3-{\rm FP}_4}\approx 0.14 - 0.17\,.
\eeq
see \eq{FP4-data} and \eq{FP3-data}. Quantitatively, this is closer to the EH result \eq{lag_EH}  than to \eq{lag_fRiem}, owing to the fact that the fixed point coordinates only differ mildly from the EH result. Still, for either of these the additional Riemann interactions increase the scale-invariant product of couplings over the Einstein Hilbert values, much unlike in $f(R)$ models \eq{lag_fR}.

The comparison of results  in Fig.~\ref{pCompareLG} shows that Ricci and Riemann tensor interactions increase, while Ricci scalar decrease the quantity $\lambda_*\,g_*$. This  indicates that the additional degrees of freedom in theories with Ricci tensor or Riemann tensor interactions make the theory more strongly coupled, roughly by up to a factor of two, and irrespective of the finer details for the higher dimensional interaction terms.  
In this light, the closeness of \eq{lag_fRic} and \eq{lag_fRiem} can be seen as an indication for universality. At the same time, the quantitative smallness of   \eq{lag_fRic} and \eq{lag_fRiem}, albeit twice as large as \eq{lag_EH}, \eq{lag_fR}, and the near-Gaussianity of scaling exponents with increasing mass dimension (see Sec.~\ref{sec:weakcoup}) still indicates that the fixed point theory is largely weakly coupled. All in all, this further substantiates  the view that quantum gravity remains ''as Gaussian as it gets" to accommodate non-perturbative renormalisability \cite{Falls:2013bv,Falls:2014tra,Falls:2017lst,Falls:2018ylp}.

\subsection{Quantum Vacuum}

Finally, we compare solutions to the quantum equations of motion $E(R_{\rm vac})=0$, the `vacuum' solutions, which in our models correspond to field configurations of constant Ricci scalar curvature.  
In our study, vacuum solutions  have been found both at the primary Riemann fixed point \eq{dS1}, \eq{dS2}, and at the secondary ones \eq{dS-FP4-FP3},  in accord with expectations \cite{Dietz:2013sba}.

\begin{figure}[t]
	\includegraphics[width=.5\linewidth]{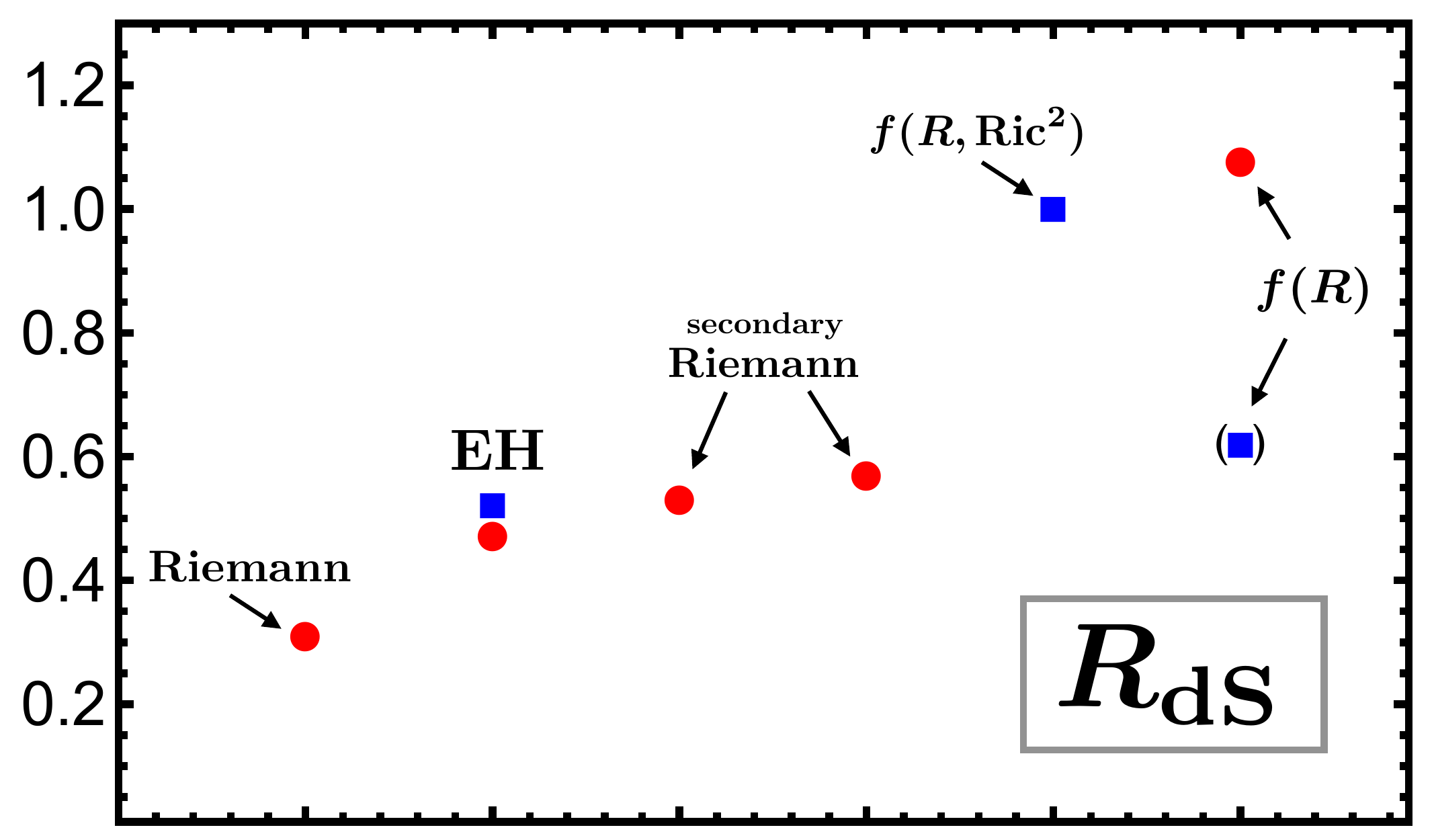}
	\caption{Shown are the vacuum solutions $R_{\rm dS}$  of the quantum equations of motion at different gravitational UV fixed points. Data is provided for  Riemann fixed points (this work), the Einstein-Hilbert fixed point, and the $f(R)$  and   $f(R,{\rm Ric}^2)$ fixed points;  same colour coding as in Fig.~\ref{pCompareLaG}. Taking the EH results as a point of reference, we note a strong enhancement (decrease) of $R_{\rm dS}$ due to Ricci  (Riemann) interactions, while de Sitter solutions of the secondary  Riemann fixed points 
	are only mildly enhanced.}
	\label{pComparedS}
\end{figure}

We compare the vacuum solutions with those from  the Einstein Hilbert approximation \eq{deSitter-EH}, and the $f(R)$ \cite{Falls:2018ylp} and $f(R,{\rm Ric}^2)$ fixed point \cite{Falls:2017lst}, shown in Fig.~\ref{pComparedS}, 
with $R_{\rm dS}$ denoting the background Ricci scalar curvature in units of the RG scale. We use once more the same colour-coding as in Fig.~\ref{pCompareLaG}.
Taking the Einstein Hilbert approximation with the Reuter fixed point as a point of reference, we see that the treatment of ghost only makes a minor difference, with roughly $R_{\rm dS}|_{\rm EH}\approx 0.5$ for either of the settings.  

For the Riemann fixed points, we see that the secondary ones have nearby vacuum solutions with $R_{\rm dS}\approx 0.55$. This is in accord with the view that these fixed points can be considered as natural extensions of the Reuter fixed point. The vacuum solution for the primary Riemann fixed point  is much smaller, around $R_{\rm dS}\approx 0.32$. This effect  is entirely due to the coupling $g_*$ being much larger and $\lambda_*$ much smaller than their EH counterparts. Moreover, all three Riemann fixed points share the feature that their vacuum solutions are controlled by the cosmological constant $\lambda_*$, and by Newton's coupling $g_*$ and the Ricci scalar term in the action. Consequently, all higher order Riemann interaction terms only add a small  subleading contribution to $R_{\rm dS}$.
Turning to $f(R)$ and $f(R,{\rm Ric}^2)$ models, 
we note that both have vacuum solutions in the range of $R_{\rm dS}\approx 1.0-1.1$, roughly twice as large as those at the Reuter fixed point. Here, the enhancement is entirely due to higher order interaction terms in the effective action, rather than to a softening of $g_*$ or an enhancement of $\lambda_*$. Also, for $f(R)$ models the improved treatment of ghosts (red dot) makes an important difference \cite{Falls:2018ylp}. Without it, the vacuum solution has a small complex component  \cite{Falls:2016wsa} as  indicated by brackets around the corresponding data point (blue square) in Fig.~\ref{pComparedS}.

We conclude that higher order Ricci scalar, Ricci tensor or Riemann tensor  interactions affect the  location of vacuum solutions in vastly different manners. One may speculate that those vacuum solutions related to models with a larger UV critical surface are more likely to be realised in the full (un-approximated) theory of asymptotically safe quantum gravity. However, this remains to be clarified in future works.

\section{\bf Summary and Conclusions}\label{sec:conclusions}

Starting from   gravitational  actions involving Ricci scalars, tensors, and Riemann tensors, 
we have derived a new  family of functional renormalisation group equations  for the running gravitational couplings. 
This completes a line of work initiated in \cite{Falls:2014tra,Falls:2017lst,Falls:2018ylp}, and allows systematic studies of quantum gravity including a large variety of higher order curvature invariants. 
Different parameter choices correspond to different rays in the space of gravitational actions after projection onto the sphere, much in the spirit of a local potential approximation \cite{Benedetti:2012dx}.  The setting is  versatile and, together with the known heat kernel coefficients   \cite{Kluth:2019vkg}, can  be used to  explore   quantum effects for higher curvature extensions of general relativity \cite{Sotiriou:2008rp,DeFelice:2010aj,Clifton:2011jh,Capozziello:2011et} or holography   \cite{Hung:2011xb}, the role of higher order curvature interactions for the asymptotic safety conjecture as done here,  and the inclusion of matter, or more general gauge fixing conditions and   cutoff profiles.

We have applied the formalism to  models of quantum gravity with Riemann interactions. Using polynomial expansions of the action to high order 
a  primary Riemann fixed point has been found satisfying all standard  tests including a fast convergence of couplings (\fig{fig:fp05522couplingsstandard}) and scaling exponents  (\fig{fig:fp05522eigenvalueconvergence}), the bootstrap test  (\fig{fig:fp05522BootStrap}),  near-Gaussian behaviour at  large orders  (\fig{fig:fp4Gaussian}), and a maximal radius of convergence (\fig{fig:fp4smallfixpointfunc} and \fig{fig:EOM_FP4}). High order polynomial expansions of the action up to 144 powers in curvature have been used to establish convergence and stability. Together with numerical integration, they have also been used to obtain solutions beyond any polynomial order.  Findings  differ  noticeably from those in Einstein-Hilbert and $f(R)$ approximations (Fig.~\ref{pCompareLaG}) and come out more strongly coupled   (Fig.~\ref{pCompareLG}). 
We have also found a  stable pair of  weakly coupled Riemann fixed points including de Sitter solutions and 
well-defined UV-IR connecting RG trajectories with General Relativity at low energies. For these, the leading couplings and scaling exponents are  close to those of the Einstein-Hilbert  fixed point  (Fig.~\ref{scatter}) with elsewise mild quantum corrections. For these reasons, they can be viewed as higher curvature extensions of the Reuter fixed point.

An important novelty is the dimensionality of the critical surface which is found to be four (Figs.~\ref{fig:fp05522Gap} and \ref{fig:FP4-UV-Surface}). This is  rather different from 
models without Riemann tensor interactions, where the critical surface -- with all other technical choices  the same -- comes out as three-dimensional.
Our study highlights  that quantum-induced shifts   in the scaling dimensions of the order of a few (\fig{fig:fp05522slope}) may  well turn canonically irrelevant dim-6 operators into relevant ones.
The smallness of  the scale-invariant product of couplings $\lambda_* \,g_*$
and the near-Gaussianity of high order scaling dimensions further substantiates  the view that quantum gravity remains ''as Gaussian as it gets".
Future work should investigate  the relevancy of other dim-6 curvature invariants (Tab.~\ref{tab:dim246operators}),  and  other technical choices, to further substantiate the number of fundamentally free parameters of  quantum gravity.

Another novelty of this work is the detailed study of eigenvectors and eigenperturbations in the ultraviolet. 
Eigenvectors are nonuniversal quantities which change under rescalings of  couplings
 (\fig{eigvnotrescaled}). We have shown that  a simple  equal weight condition  
 leads to order-by-order stable eigenvectors
 which represent the relevant physics correctly  (Figs.~\ref{eigvrescaled},~\ref{eigvrescaled_secondary}).
 Results further substantiate  that  relevant eigenperturbations are linear combinations of the leading handful of interaction terms, with irrelevant ones  dominated by   subleading interactions.  Anomalous dimensions are found to be of order unity, 
 having the largest impact on interaction monomials with low mass dimension. Eigenvectors  are seen to align  accordingly.
It will be interesting to  exploit these ideas for other theories at criticality, with or without gravity. 
 \\[3ex]

\centerline{\bf Acknowledgments}
\noindent
YK is supported by the Science Technology and Facilities Council (STFC) under grant number [ST/S505766/1]. Some  of our results have been presented at the workshops {\it Asymptotic Safety Meets Particle Physics V} (U Dortmund, Dec 2018), {\it Beyond the Standard Model XXXI} (Bad Honnef, Mar 2019), and at the workshop {\it Quantum and Gravity} (U Okinawa, Jul 2019). YK thanks the organisers for hospitality and participants  for discussions.

\appendix
\renewcommand{\thesubsection}{\Alph{subsection}}
\renewcommand{\theequation}{\thesubsection.\arabic{equation}}

\counterwithin*{equation}{subsection}

\section*{\bf Appendices}

In the appendix, we summarise some of the technical expressions and formul\ae\ required in the main text, including   first and second variations of the basic field monomials in the action (App.~\ref{AppA}), the Hessians of the effective action (App.~\ref{AppB}), and
the general form of the functional flow equations (App.~\ref{sec:floweq}).  
\subsection{First and Second Variations}\label{AppA}
In the main text, we require the first two variations of 
\beq
X = a\,R^2 + b \,\text{Ric}^2 + c\, \text{Riem}^2
\eeq
subject to the linear split \eq{eqn:linsplit} in general dimension. The first variation is given by
\beq
		\delta \left( X \right) = \left(a+\frac{b}{d}+\frac{2 c}{(d-1) d}\right) \left(-\frac{2 R^2}{d} h + 2 R \nabla^{\mu } \nabla^{\nu } h_{\mu  \nu } - 2 R \nabla^2 h \right) \, ,
	\label{eqn:xvar1}
\eeq
while the second variation gives
\beq
	\begin{split}
		\delta^{(2)} \left( X \right) =& R^2 h_{\mu  \nu } \Bigg[ \left( \frac{b}{2} + 2 c \right) \nabla^4 + \left( a + \frac{d-3}{(d-1) d} b - \frac{2}{d-1} c  \right) R \nabla^2 \\
		& \quad + \left( \frac{2 (d-2)}{(d-1) d} a + \frac{2 (d-3) d+6}{(d-1)^2 d^2} b + \frac{4 (2 d-3)}{(d-1)^2 d^2} c  \right) R^2 \Bigg] h^{\mu  \nu } \\
		& + h \Bigg[\left( 2a + \frac{b}{2} \right) \nabla^4 + \left( \frac{d+4}{d} a + \frac{1 + 3 d}{2 d (d - 1)} b + \frac{6}{(d-1) d} c \right) R \nabla^2 \\
		& \quad + \left( \frac{4 d-2}{(d-1) d^2} a + \frac{4 d-6}{(d-1)^2 d^2} b + \frac{4}{(d-1)^2 d^2} c \right) R^2 \Bigg] h \\
		& + h \Bigg[ (-4a -b) \nabla^2 + \left( -\frac{4 a}{d} + \frac{8 c}{(d-1) d} \right) R \Bigg] \nabla^{\mu }\nabla^{\nu }h_{\mu  \nu } \\
		& + h_{\mu  \rho } \nabla ^{\rho }\Bigg[ \left( -2 a - \frac{3}{d} b - \frac{12}{d (d - 1)} c \right) R - (b + 4 c) \nabla^2 \Bigg] \nabla _{\nu } h^{\mu  \nu } \\
		& +(2 a+b+2 c) h_{\mu  \nu } \nabla ^{\mu } \nabla ^{\nu } \nabla ^{\rho } \nabla ^{\sigma } h_{\rho  \sigma
   } \, .
	\end{split}
	\label{eqn:xvar2}
\eeq
\subsection{Hessians}\label{AppB}
The Hessians of $h^{T \mu \nu} h^T_{\mu \nu}$ and $hh$ for actions \eq{eqn:effansatz} in general dimension with  $X$ as defined in \eq{X} are required for the derivation of the flow equation. We find
\beq
	\begin{split}
		\left. \frac{\delta^2 \overline{\Gamma}_k}{\delta h^T_{\mu \nu} \delta h^T_{\rho \sigma}} \right|_{\phi_i = 0} =& \left( g^{\mu (\rho} g^{\sigma) \nu} - \frac{1}{d} \, g^{\mu \nu} g^{\rho \sigma} \right) \Bigg\{ \left(F_k'+R Z_k'\right) \Bigg[ \left( \frac{1}{2} b + 2 c \right) \nabla^4 \\
		& \qquad + \left( a + \frac{d-3}{d (d - 1)} b - \frac{2}{d - 1} c \right) R \nabla^2 \\
		& \qquad +  2 \left( \frac{(d-2)}{(d-1) d} a + \frac{d^2-3d+3}{(d-1)^2 d^2} b + \frac{4 d-6}{(d-1)^2 d^2} c \right) R^2 \Bigg] \\
		& \quad + Z_k \left( \frac{1}{2}\nabla^2 -\frac{d^2-3d+4}{2 (d-1) d} R \right)-\frac{F_k}{2} \Bigg\} \, ,
	\end{split}
	\label{eqn:hessianhtht}
\eeq
and
\beq
	\begin{split}
		\left. \frac{\delta^2 \overline{\Gamma}_k}{\delta h \delta h} \right|_{\phi_i = 0} =& \, 4 \left(F_k''+R Z_k''\right) R^2 \left(\frac{d-1}{d} a + \frac{d-1}{d^2} b + \frac{2}{d^2} c \right)^2 \left( \nabla^4 + \frac{2R}{d-1}  \nabla^2 + \frac{R^2}{(d-1)^2}  \right) \\
		& + F_k' \Bigg[ \frac{d - 1}{d} \left(\frac{2 (d-1)}{d} a + \frac{1}{2} b + \frac{2}{d} c \right) \nabla^4 \\
		& \quad - \left(\frac{(d-6) (d-1)}{d^2} a + \frac{((d-10) d+8)}{2 d^3} b - \frac{8}{d^3} c \right) R \nabla^2 \\
		& \quad + R^2 \frac{d - 3}{d^2} \left(-2a - \frac{2}{d} b - \frac{4}{(d-1) d} c \right) \Bigg] \\
		& +R Z_k' \Bigg[ \frac{d - 1}{d^2} \left(6 (d-1) a + \frac{d^2 + 8 d - 8}{2 d} b + \frac{2 (d+4)}{d} c \right) \nabla^4 \\
		& \quad - \left(\frac{(d-14) (d-1)}{d^2} a + \frac{(d-26) d+24}{2 d^3} b - \frac{24}{d^3} c \right) R \nabla^2 \\
		& \quad+ \frac{d - 5}{d^2} \left(-2 a-\frac{2}{d} b - \frac{4}{(d-1)d} c \right) R^2 \Bigg] \\
		& +Z_k \frac{d - 2}{d^2} \left( - \frac{d-1}{2} \nabla^2 + \frac{d-4}{4} R \right) + F_k \frac{d-2}{4 d} \, .
	\end{split}
	\label{eqn:hessianhh}
\eeq

\subsection{Flow Equation and Fluctuation Integrals}
\label{sec:floweq}

The central flow equation \eq{fRG} of this work  arises from \eq{eqn:effansatz} with \eq{X}, \eq{eqn:fullgamma}. Adopting the cutoff  \eq{opt} and evaluating all operator traces, the explicit flow
takes the form
\begin{equation}
	4 f + 2 rz  + \partial_t (f + r z) - \frac{12 a + 3 b + 2 c}{3} r^2 (f' + rz') = \frac{1}{\kappa} I [f, z; a, b, c] (r) \,,
	\label{floweqnabcApp}
\end{equation}
with $\kappa = 24 \pi$. Here, we specify the fluctuation-induced terms contained in the fluctuation integrals $I [f, z; a, b, c]$ on the RHS. Adopting conventions and notation of \cite{Falls:2014tra,Falls:2017lst}, we find
\begin{equation}
	\begin{split}
		I [f, z] (r) =& I_0 [f, z] (r) + I_1 [f, z] (r) \cdot \partial_t z + I_2 [f, z] (r)\cdot \partial_t f' \\[1ex]
		& + I_3 [f, z] (r) \cdot\partial_t z'+ I_4 [f, z] (r) \cdot(\partial_t f'' + r \partial_t z'') \,,
	\end{split}
\end{equation}
where we recall that the flow terms arise due to couplings introduced via the regularisation.
 All five terms $I_i [f, z](r)$ $(i=0,\cdots,4)$ arise from tracing over the fluctuations of the metric field for which we have adopted a transverse traceless decomposition. They do not contain any renormalisation scale derivative. The term $I_0[f,z]$ also receives $f$- and $z$-independent contributions from the ghosts and from the Jacobians originating from the split of the metric fluctuations into tensor, vector and scalar parts. To indicate the origin of the various contributions in the expressions below, we use superscipts $T$, $V$, and $S$ to refer to the transverse traceless tensorial, vectorial, and scalar origin. The specific form of the functions $I_i[f, z]$ also depends on the choice of gauge fixing and the details for the regulator function. In our setting, they 
are given by
\begin{equation}\label{In}
	\begin{aligned}
		I_0 [f, z] (r) =& P^V + P^S + \frac{P_0^T}{D^T} + \frac{P_0^S}{D^S} \, , \\ \qquad I_1 [f, z] (r) =& \frac{P_1^T}{D^T} + \frac{P_1^S}{D^S} \, , \\
		I_2 [f, z] (r) =& \frac{P_2^T}{D^T} + \frac{P_2^S}{D^S} \, , & \qquad \\ I_3 [f, z] (r) =& \frac{P_3^T}{D^T} + \frac{P_3^S}{D^S} \, , \qquad \\
		I_4 [f, z] (r) =& \frac{P_4^S}{D^S} 
 \, .
	\end{aligned}
\end{equation}
The denominators $D^T$ and $D^S$ in \eq{In} arise from the tensorial (T) and scalar (S) metric fluctuations, and are defined as
\bea
		D^T &=& \, [24 a (r-3) r+b (r (7 r-6)+36)+2 c (r (5 r+24)+72)] \left(f'+r z'\right)\nonumber \\
		&& -36 f-12 (2 r+3) z \, ,\label{floweqnI0}  \\
		D^S &=& \, r \big(r (r-3)^2 (12 a+3 b+2 c)^2 \left(f''+r z''\right)+6 z' (12 a ((r-15) r+27)+3 b ((r-16) r+30) \nonumber \\
			&& +2 c ((r-18) r+36))\big)-6 f' (12 a (r (r+3)-9)+(r+6) (3 b (r-2)+2 c r)-36 c) \\
			&&+72 f+108 z \,. \nonumber
\eea
The various terms $P$ in \eq{In} are polynomials in curvature. Specifically, the terms $P^V$ and $P^S$ are given by
\bea
P^S&=&\frac{271}{90}r^2-12r-12\\[1ex]
P^V&=&\frac{191}{30}r^2-24r-36\,.
\eea
The   numerators  $P_0$ appearing  in \eq{In}  can be written as
\bea
		P_0^T &=& P_0^{T z0} z + P_0^{T f1} f' + P_0^{T z1} z' + P_0^{T 2} (f'' + r z'') \, , \\[1ex]
		P_0^S &=& P_0^{S z0} z + P_0^{S f1} f' + P_0^{S z1} z' + P_0^{S f2} f'' + P_0^{S z2} z'' + P_0^{S3}( f^{(3)} + r z^{(3)}) \, ,
\eea
with coefficient functions
\bea
		P_0^{T z0} &=& \, -\frac{311}{63} r^3+4 r^2+1080 r-2880 \, , \\
		P_0^{T f1} &=& \, \frac{1}{3} (r (r+360)-1080) (12 a r+b (r-12)-8 c (r+6)) \, , \\
		P_0^{T z1}& =& \, \frac{14928 a+2597 b-3296 c}{756} r^5-\frac{12 a+3 b+2 c}{3} r^4-120 (12 a+7 (b+2 c)) r^2 \nonumber\\
			&& \qquad-2 (61 b+304 c) r^3 +3240 (b+4 c) r \, , \\
		P_0^{T2} &=& \, (12 a+3 b+2 c) \bigg[\frac{7464 a+731 b-4540 c}{4536} r^6+\frac{-12 a-b+8 c}{18} r^5\nonumber \\ 
		&& \qquad
		+\frac{-360 a-29 b+244 c}{3} r^4 
			 + (240 a+100 b+160 c)r^3 + (-180 b-720 c) r^2\bigg] \, , \\
		P_0^{S z0} &=& \, \frac{37}{21} r^3+\frac{348}{5} r^2+648 r+1728 \, , \\
		P_0^{S f1} &=& \, -\frac{4 (3 a+b+c)}{5} \left(29 r^3+186 r^2-1080 r-6480\right) \, , \\
		P_0^{S z1} &=& \, \frac{20442 a+5555 b+4296 c}{1260} r^5 -\frac{29}{5} (12 a+3 b+2 c) r^4 -\frac{6}{5} (1638 a+425 b+304 c) r^3 \nonumber\\
			&& \qquad +72 (12 a+5 b+6 c) r^2 +1944 (18 a+5 b+4 c) r \, , \\
		P_0^{S f2} &=& \, (12 a+3 b+2 c) \bigg[\frac{24882 a+6665 b+5036 c}{7560} r^6 + \frac{29 (18 a+5 b+4 c)}{15} r^5 \nonumber \\
			&& \qquad+\frac{62 (3 a+b+c)}{5} r^4 +r^3 (-144 a-48 b-48 c)+r^2 (648 a+108 b)\bigg] \, , \\
		P_0^{S z2} &=& \, (12 a+3 b+2 c) \bigg[\frac{108408 a+27991 b+19846 c}{15120} r^7 +\frac{58 (27 a+7 b+5 c)}{15} r^6 \nonumber \\
			&& \quad + \frac{3648 a+943 b+670 c}{10} r^5 - (144 a+48 b+48 c) r^4 - (1296 a+378 b+324 c) r^3 \bigg] \, , \\
		P_0^{S3} &=& \, (12 a+3 b+2 c)^3 \left[\frac{181}{10080} r^8 + \frac{29}{90} r^7 + \frac{91}{60} r^6 - 9 r^4\right] \,.
\eea
The  numerators  $P_{1,2,3,4}$ appearing  in \eq{In}  take the form
 \begin{equation}\label{P1}
	\begin{split}
		P_1^T =& \, -\frac{311}{126} r^3+r^2+180 r-360 \, , \\
		P_1^S =& \, \frac{37}{42} r^3+\frac{87}{5} r^2+108 r+216 \,,
		\end{split}
\eeq
\begin{equation}\label{P2}
	\begin{split}
		P_2^T =& \, -\frac{7464 a+731 b-4540 c}{1512} r^4 + \frac{12 a+b-8 c}{6} r^3 + (360 a+29 b-244 c) r^2 \\
			& \qquad -60 (12 a+5 b+8 c) r + 540 (b+4 c) \, , \\
		P_2^S =& \, - (3 a+b+c) \left[\frac{127}{180} r^4+\frac{58}{5} r^3+\frac{186}{5} r^2-144 r-648\right] \,,\end{split}
\eeq
\begin{equation}\label{P3}
	\begin{split}
		P_3^T =& \, -\frac{7464 a+731 b-4540 c}{1512} r^5 + \frac{12 a+b-8 c}{6} r^4 + (360 a+29 b-244 c) r^3 \\
			& \qquad - 60 (12 a+5 b+8 c) r^2 + 540 (b+4 c) r \, , \\
		P_3^S =& \, -\frac{24882 a+6665 b+5036 c}{2520} r^5 - \frac{58 (15 a+4 b+3 c)}{5} r^4 -\frac{3 (1278 a+335 b+244 c)}{5} r^3 \\
			& \qquad +144 (3 a+b+c) r^2 + 324 (18 a+5 b+4 c) r \,,\end{split}
\eeq
\begin{equation}\label{P4}
	\begin{split}
		P_4^S =& \, (12 a+3 b+2 c)^2 \left[\frac{181 r^6}{3360}+\frac{29 r^5}{30}+\frac{91 r^4}{20}-27 r^2\right] \, .
	\end{split}
\end{equation}

\bibliography{bibtex}
\bibliographystyle{mystyle}

\end{document}